\documentclass{article} %
\usepackage{microtype,times}
\usepackage{graphicx}
\usepackage{booktabs} %

\usepackage{hyperref}

\usepackage[accepted]{icml2023}
\usepackage{blindtext}

\definecolor{navyblue}{rgb}{0.0, 0.0, 0.5}

\definecolor{darkraspberry}{rgb}{0.53, 0.15, 0.34}
\definecolor{pansypurple}{rgb}{0.47, 0.09, 0.29}
\definecolor{richmaroon}{rgb}{0.69, 0.19, 0.38}

\usepackage[utf8]{inputenc} %
\usepackage[T1]{fontenc}    %
\usepackage{url}            %
\usepackage{booktabs}       %
\usepackage{amsfonts}       %
\usepackage{nicefrac}       %
\usepackage{microtype}      %
\usepackage{tabu}
\usepackage{multicol}
\usepackage{soul}
\usepackage{bbm}
\usepackage{lipsum}
\usepackage{kantlipsum}
\usepackage{subfigure}

\usepackage{cancel}

\usepackage{amsmath,amssymb,amsfonts,amsthm, dsfont, color}
\usepackage{algorithm}
\usepackage{mathtools}
\usepackage{graphicx}
\usepackage{textcomp}
\usepackage{natbib}
\usepackage{xcolor, fancyhdr}
\usepackage{enumerate}
\usepackage{float}

\usepackage{mathtools}
\usepackage{cuted}

\definecolor{MyGreen1}{RGB}{20,180,40}
\usepackage{multirow, tikz,float}

\usepackage{nicefrac,color,mathrsfs,float}
\usepackage{multirow,caption,tikz}
\usetikzlibrary{shapes.misc, positioning}
\usetikzlibrary{decorations.pathreplacing}

\tikzset{
  block/.style    = {draw, thick, rectangle, minimum width = 2em},
sblock/.style      = {draw, thick, rectangle, minimum height = 2em,
minimum width = 2em}, 
}

\newcommand{\HS}{\hspace{\fontdimen2\font}}

\newcommand{\polar}{\text{Polar}}
\newcommand{\pac}{\text{PAC}}

\newcommand{\mhat}{\hat{m}}
\newcommand{\mpast}{\hat{\bma}_{<i}}

\newcommand{\crisp}{\text{CRISP}}
\newcommand{\plotkin}{\mathrm{Plotkin}}

\usepackage{comment}

\usepackage{thmtools}

\usepackage{prettyref,xspace}
\usepackage{tikz}

\newrefformat{cond}{Condition~\ref{#1}}
\newrefformat{eq}{Eq.~\ref{#1}}
\newrefformat{thm}{Theorem~\ref{#1}}
\newrefformat{th}{Theorem~\ref{#1}}
\newrefformat{chap}{Chapter~\ref{#1}}
\newrefformat{sec}{Sec.~\ref{#1}}
\newrefformat{algo}{Algorithm~\ref{#1}}
\newrefformat{fig}{Fig.~\ref{#1}}
\newrefformat{tab}{Table~\ref{#1}}
\newrefformat{rmk}{Remark~\ref{#1}}
\newrefformat{clm}{Claim~\ref{#1}}
\newrefformat{def}{Definition~\ref{#1}}
\newrefformat{cor}{Corollary~\ref{#1}}
\newrefformat{lmm}{Lemma~\ref{#1}}
\newrefformat{prop}{Proposition~\ref{#1}}
\newrefformat{pr}{Proposition~\ref{#1}}
\newrefformat{app}{App.~\ref{#1}}
\newrefformat{prob}{Problem~\ref{#1}}
\newrefformat{ques}{Question~\ref{#1}}
\newrefformat{notee}{Note~\ref{#1}}
\newrefformat{assump}{Assumption~\ref{#1}}
\newrefformat{issuee}{Issue ~\ref{#1}}
\newrefformat{fix}{Fix ~\ref{#1}}

\newcommand{\reals}{\mathbb{R}}
\newcommand{\naturals}{\mathbb{N}}

\newcommand{\pprob}[1]{\mathbb{P}[#1]}

\newcommand{\calN}{\mathcal{N}}

\newcommand{\vect}[1]{\boldsymbol{#1}}

\newcommand{\bc}{\vect{c}}

\newcommand{\bh}{\vect{h}}

\newcommand{\bma}{\vect{m}}

\newcommand{\bu}{\vect{u}}
\newcommand{\bv}{\vect{v}}

\newcommand{\bx}{\vect{x}}
\newcommand{\by}{\vect{y}}
\newcommand{\bz}{\vect{z}}

\newcommand{\bI}{\vect{I}}

\newcommand{\bL}{\vect{L}}

\newcommand{\abs}[1]{\left|{#1} \right|}

\newcommand{\binary}{\{0,1\}}

\newcommand{\define}{\triangleq}

\newcommand{\pth}[1]{\left( #1 \right)}

\newcommand{\ie}{i.e.\xspace}
\newcommand{\iid}{i.i.d.\xspace}

\mathchardef\mhyphen="2D

\definecolor{MyGreen1}{RGB}{20,180,40}

\usepackage{siunitx}

\newcommand{\Cref}[1]{Co\-ro\-lla\-ry\,\ref{#1}}

\newcommand{\walk}[1]{\xrightarrow{#1}}

\icmltitlerunning{CRISP: Curriculum based Sequential neural decoders for Polar code family}

\begin{document}

\twocolumn[
\icmltitle{CRISP: Curriculum based Sequential neural decoders for Polar code family}

\icmlsetsymbol{equal}{\dag}

\begin{icmlauthorlist}
\icmlauthor{S. Ashwin Hebbar}{equal,princeton}
\icmlauthor{Viraj Nadkarni}{equal,princeton}
\icmlauthor{Ashok Vardhan Makkuva}{epfl}
\icmlauthor{Suma Bhat}{princeton}
\icmlauthor{Sewoong Oh}{wash}
\icmlauthor{Pramod Viswanath}{princeton}
\end{icmlauthorlist}

\icmlaffiliation{princeton}{Princeton University}
\icmlaffiliation{wash}{University of Washington}
\icmlaffiliation{epfl}{EPFL}

\icmlcorrespondingauthor{Ashwin Hebbar}{hebbar@princeton.edu}
\icmlcorrespondingauthor{Viraj Nadkarni}{viraj@princeton.edu}
\icmlcorrespondingauthor{Ashok Makkuva}{ashok.makkuva@epfl.ch}

\icmlkeywords{Machine Learning, ICML}

\vskip 0.3in
]
\printAffiliationsAndNotice

\def\thefootnote{\dag}\footnotetext{Equal contribution}
\def\thefootnote{*}

\begin{abstract}
Polar codes are widely used state-of-the-art codes for reliable communication that have recently been included in the $5^{\text{th}}$ generation wireless standards ($5$G). However, there remains room for the design of polar decoders that are both efficient and reliable in the short blocklength regime. Motivated by recent successes of data-driven channel decoders, we introduce a novel {\bf C}ur{\bf RI}culum based {\bf S}equential neural decoder for {\bf P}olar codes (CRISP)\footnote{Source code available at the following \href{https://github.com/hebbarashwin/neural_polar_decoder}{link}.}. 
We design a principled curriculum, guided by information-theoretic insights, to train CRISP and show that it outperforms the successive-cancellation (SC) decoder and attains near-optimal reliability performance on the $\polar(32,16)$ and $\polar(64,22)$ codes. 
The choice of the proposed curriculum is critical in achieving the accuracy gains of CRISP, as we show by comparing against other curricula. More notably, CRISP can be readily extended to  Polarization-Adjusted-Convolutional (PAC) codes, where existing SC decoders are significantly less reliable. To the best of our knowledge, CRISP constructs the first data-driven decoder for PAC codes and attains near-optimal performance on the $\text{PAC}(32,16)$ code. 
\end{abstract}

\section{Introduction}
\label{sec:intro}

Error-correcting codes (codes) are the backbone of modern digital communication. Codes, composed of (encoder, decoder) pairs, ensure reliable data transmission even under noisy conditions. Since the groundbreaking work of \cite{shannon1948mathematical}, several landmark codes have been proposed: Convolutional codes, low-density parity-check (LDPC) codes, Turbo codes, Polar codes, and more recently, Polarization-Adjusted-Convolutional (PAC) codes \citep{richardson2008modern}. 
In particular, polar codes, introduced by \cite{Arikan_2009}, are widely used in practice owing to their reliable performance in the short blocklength regime. 
 A family of variants of polar codes known as PAC codes further improves performance, nearly achieving the fundamental lower bound on the performance
of any code at finite lengths, albeit at a higher decoding complexity \citep{arikan2019sequential}.
In this paper, we focus on the {\em decoding} of these two classes of codes, jointly termed the ``Polar code family''.

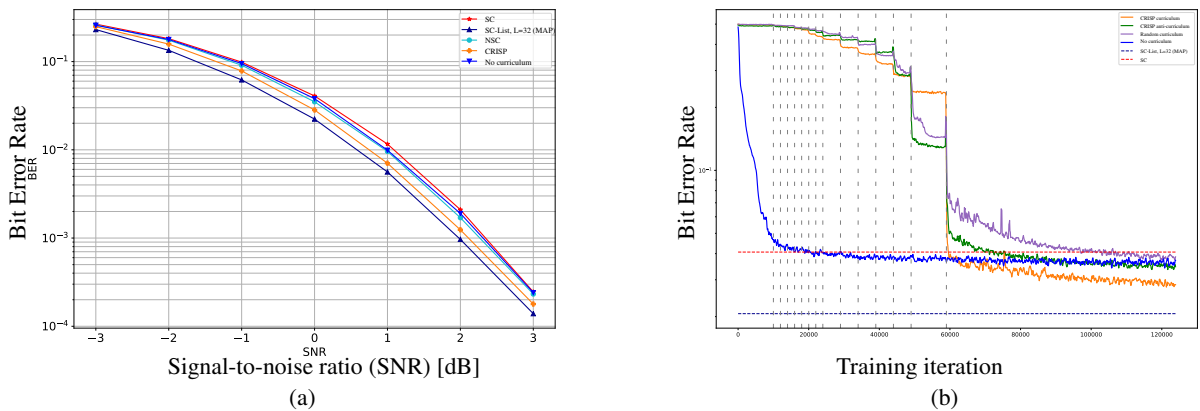
\begin{figure*}[ht]
\centerline{
\subfigure[]
{
  \centering
  \includegraphics[width=\columnwidth]{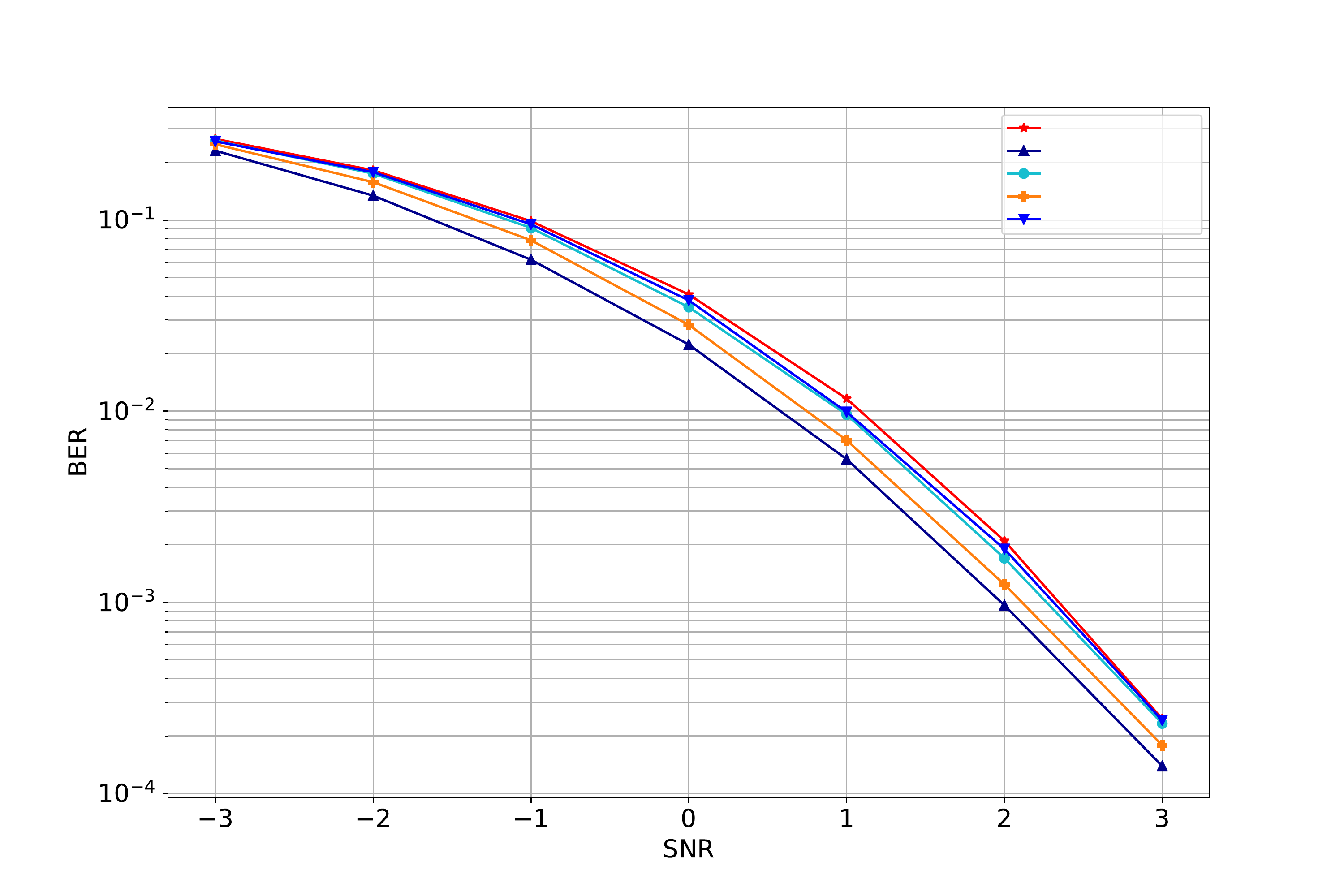}
    \put(-50,133){\fontsize{3}{5}\selectfont SC}
    \put(-50,129){\fontsize{3}{5}\selectfont SC-List, L=32 (MAP)}
    \put(-50,125){\fontsize{3}{5}\selectfont NSC}
    \put(-50,121){\fontsize{3}{5}\selectfont CRISP}
    \put(-50,117){\fontsize{3}{5}\selectfont No curriculum}
    \put(-170,0){\footnotesize Signal-to-noise ratio (SNR) [dB]}
    \put(-230,70){\rotatebox[origin=t]{90}{\footnotesize Bit Error Rate}}
  \label{fig:2264}
}
\hfill
\subfigure[]
{
  \includegraphics[width=\columnwidth]{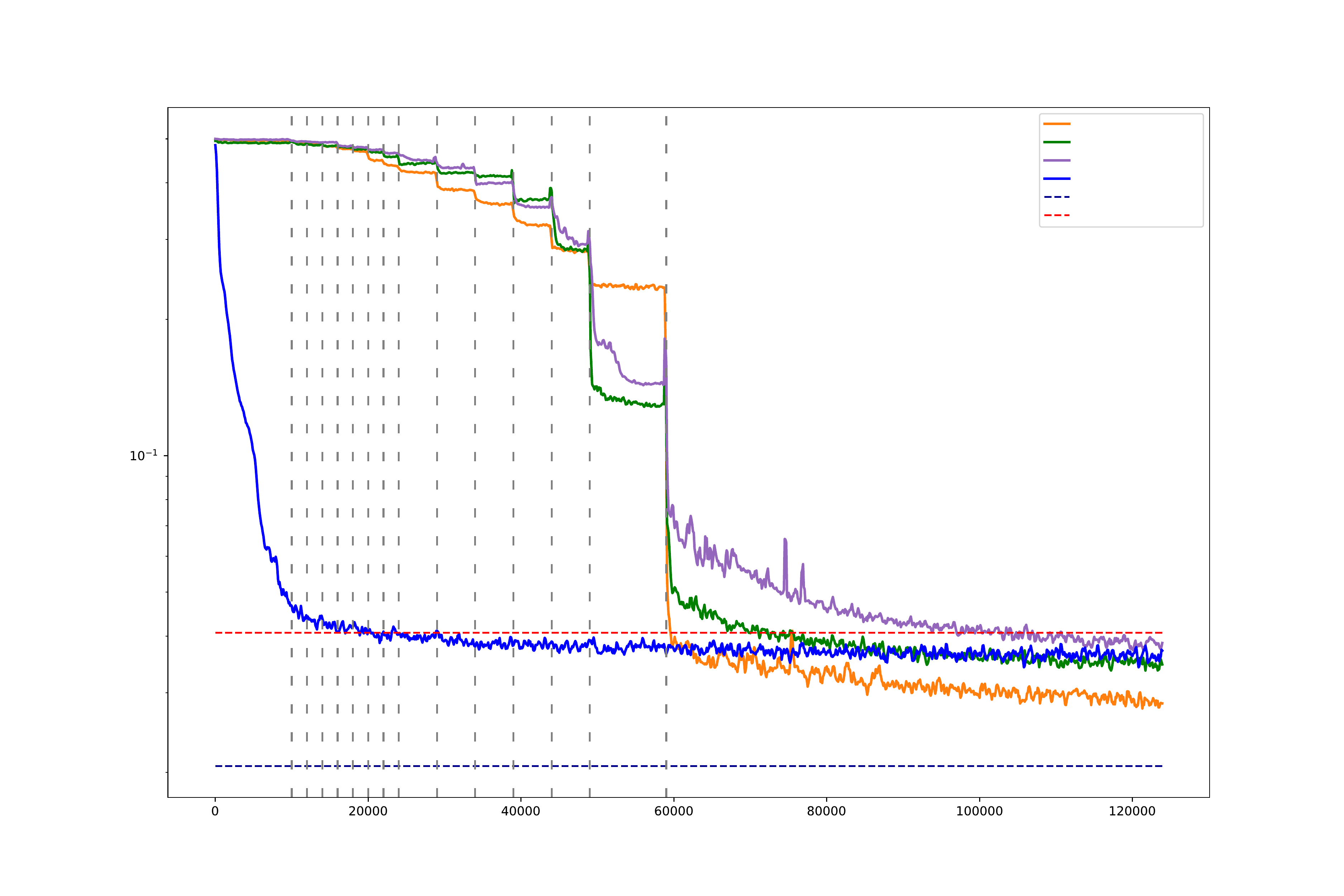}
    \put(-45,134){\fontsize{2}{5}\selectfont CRISP curriculum}
    \put(-45,131){\fontsize{2}{5}\selectfont CRISP anti-curriculum}
    \put(-45,128){\fontsize{2}{5}\selectfont Random curriculum}
    \put(-45,125){\fontsize{2}{5}\selectfont No curriculum}
    \put(-45,121.5){\fontsize{2}{5}\selectfont SC-List, L=32 (MAP)}
    \put(-45,118){\fontsize{2}{5}\selectfont SC}
    \put(-160,0){\footnotesize Training iteration}
    \put(-220,70){\rotatebox[origin=t]{90}{\footnotesize Bit Error Rate}}

  \label{fig:2264_progressive}
}
}
\caption{(a) CRISP achieves near-MAP reliability for $\polar(64, 22)$ code on the  AWGN channel. (b) Our proposed curriculum is crucial for the gains CRISP attains over the baselines;   details  in \prettyref{sec:main_results}. }

\label{fig:main_figure}
\end{figure*}

The polar family exhibits several crucial information-theoretic properties; practical finite-length performance, however, depends on high complexity decoders. This search for the design of efficient and reliable decoders for the Polar family is the focus of substantial research in the past decade. {\bf (a) Polar codes:} The classical successive cancellation (SC) decoder achieves information-theoretic capacity asymptotically, but performs poorly at finite blocklengths compared to the optimal maximum a posteriori (MAP) decoder \citep{arikan2019sequential}. To improve upon the reliability of SC, several polar decoders have been proposed in the literature  (\prettyref{sec:related}). One such notable result is the celebrated Successive-Cancellation-with-List (SCL) decoder \citep{tal2015list}. SCL improves upon the reliability of SC and approaches that of the MAP with increasing list size (and complexity). {\bf (b) PAC codes:} The sequential ``Fano decoder'' \citep{fano1963heuristic} allows PAC codes to perform information-theoretically near-optimally; however, the decoding time is long and variable \citep{rowshan2020complexity}. Although SC is efficient, $O(n \log n)$, its performance with PAC codes is significantly worse than that of the Fano decoder. Several works \citep{yao2021list, rowshan2020polarization, zhu2020fast, rowshan2021list, rowshan2021convolutional, sun2021optimized} propose ameliorations; it is safe to say that constructing efficient and reliable decoders for the Polar family is an active area of research and of utmost practical interest given the advent of Polar codes in 5G wireless cellular standards. The design of efficient and reliable decoders for the Polar family is the focus of this paper.

In this paper, we  introduce a novel {\bf C}ur{\bf RI}culum based {\bf S}equential neural decoder for {\bf P}olar code family (CRISP). When the proposed curriculum is applied to neural network decoder training, 
thus trained decoders  outperform existing baselines and attain near-MAP reliabilty on $\polar(64,22)$, $\polar(32,16)$ and PAC$(32,16)$ codes while maintaining low   computational  complexity  (Figs.~\ref{fig:main_figure}, \ref{fig:awgn_restplots}, \prettyref{tab:complexity}).
CRISP builds upon an inherent nested hierarchy of polar codes; 
 a $\polar(n,k)$ code subsumes all the codewords of lower-rate subcodes $\polar(n,i), 1 \leq i \leq k$ (\prettyref{sec:all_enc}). 
 We provide principled curriculum of training on examples from a sequence  of sub-codes along this hierarchy, 
and demonstrate that the proposed curriculum is critical in attaining  near-optimal performance (\prettyref{sec:main_results}).

Curriculum-learning (CL) is a training strategy to train machine learning models, starting with easier subtasks and then gradually increasing the difficulty of the tasks \citep{wang2021survey}.  \citep{elman1993learning}, a seminal work, was one of the first to employ CL for supervised tasks,  highlighting the importance of ``starting small". Later,  \cite{bengio2009curriculum} formalized the notion of CL and studied when and why CL helps in the context of visual and language learning \citep{wu2020curricula, wang2021survey}. In recent years, many empirical studies have shown that CL improves generalization and convergence rate of various models in domains such as computer vision \citep{pentina2015curriculum, guo2018curriculumnet, wang2019dynamic}, natural language processing \citep{cirik2016visualizing, platanios2019competence}, speech processing \citep{amodei2016deep, gao2016snr}, generative modeling \citep{karras2017progressive, wang2018fully}, and neural program generation \citep{zaremba2014learning, reed2015neural}. Viewed from this context, our results add decoding of algebraic codes (of the Polar family) to the domain of successes 
of supervised CL. In summary, we make the following contributions:

\begin{itemize}
    \item We introduce CRISP, a novel curriculum-based sequential neural decoder for the Polar code family. Guided by information-theoretic insights, we propose CL-based techniques to train CRISP, that are crucial for its superior performance (\prettyref{sec:crisp}).
    \item CRISP attains near-optimal reliability performance on $\polar(64,22)$ and $\polar(32,16)$ codes whilst achieving improved throughput (\prettyref{sec:awgn} and \prettyref{sec:complexity_compar}).\\
    \item CRISP further achieves near-MAP reliability for the PAC$(32,16)$ code with significantly higher throughput compared to the Fano decoder. To the best of our knowledge, this is the first learning-based PAC decoder to achieve this performance (\prettyref{sec:pac}).
\end{itemize}

\section{Problem formulation}\label{sec:probForm}
In this section we formally define the channel decoding problem and provide background on the Polar code family. Our notation is the following: we denote Euclidean vectors by small bold face letters $\bx, \by$, etc. $[n] \define \{1,2,\ldots, n \}$.  For $\bma \in \reals^n, \bma_{<i} \define (m_1,\ldots,m_{i-1})$. $\calN(0, \bI_n)$ denotes a standard Gaussian distribution in $\reals^n$. $\bu \oplus \bv$ denotes the bitwise XOR of two binary vectors $\bu, \bv \in \binary^\ell$.

\subsection{Channel decoding}
\label{sec:channel_dec}

\begin{figure}[ht]

  \centering
  \includegraphics[width=1\columnwidth]{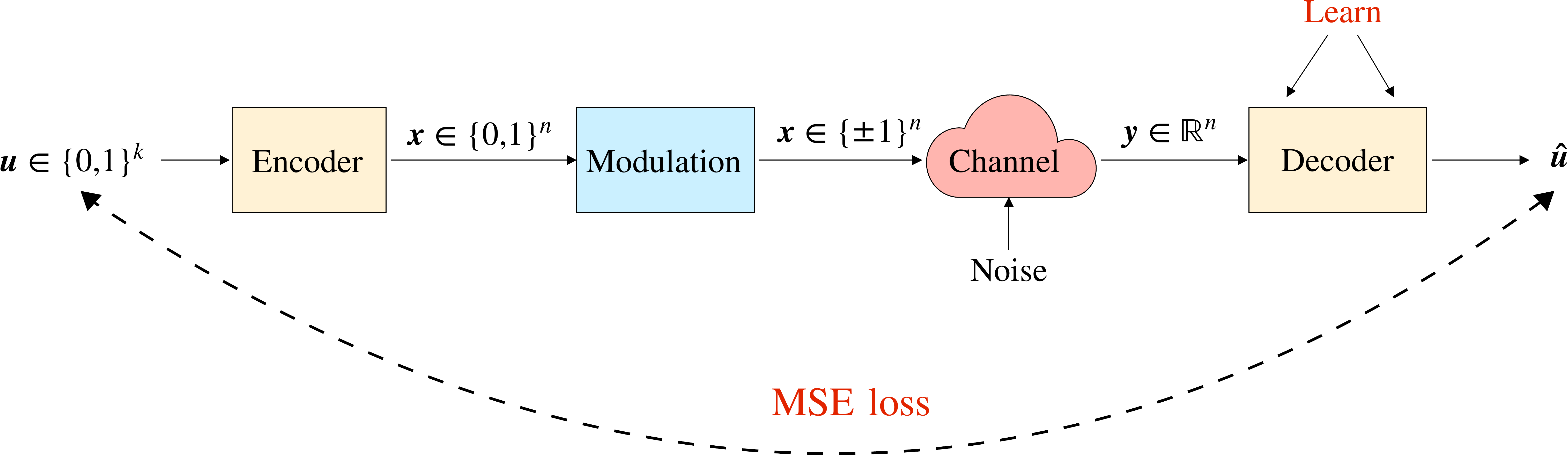}
  \caption{Channel decoding problem.}
  \label{fig:channel_dec}

\end{figure}

\begin{figure*}[h]
\centerline{
\subfigure[Polar encoder]{
  \includegraphics[width=0.35\textwidth]{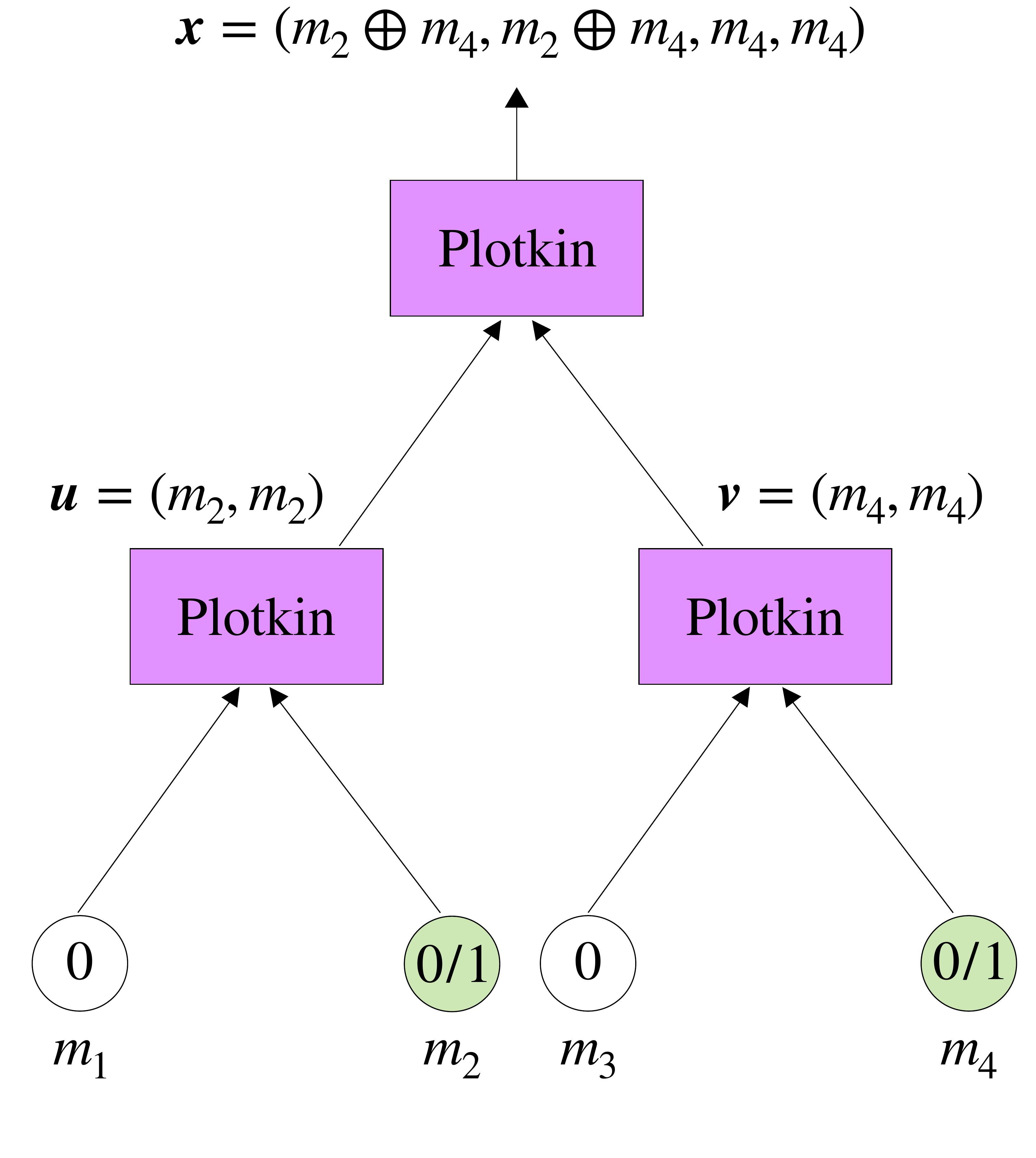}
  \label{fig:polar_enc}
}
\hspace{0.2in}
\subfigure[Successive cancellation decoder]{
  \includegraphics[width=0.35\textwidth]{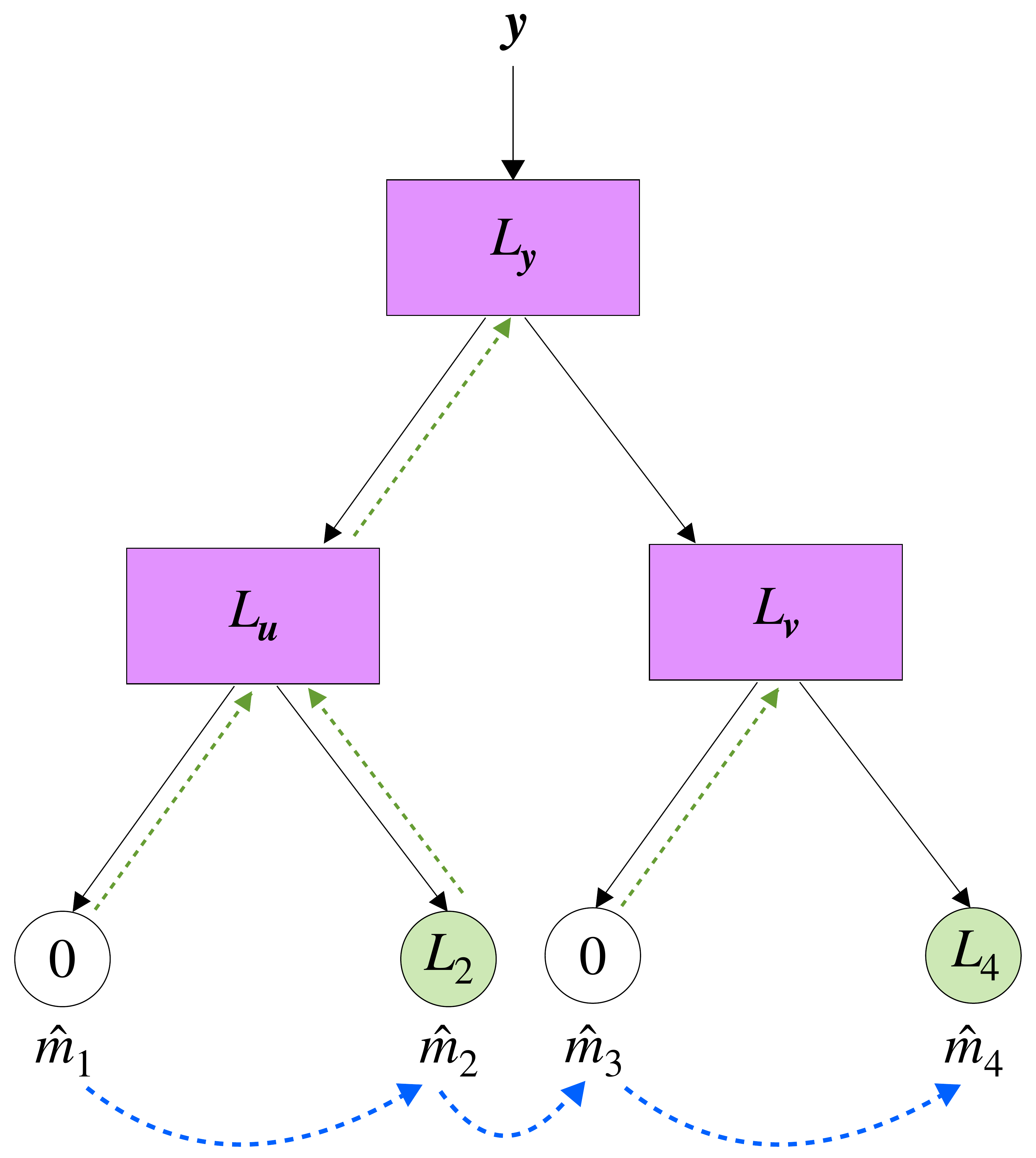}
  \label{fig:polar_dec}
}
}
\caption{$\polar(4,2)$: (a) Polar encoding via Plotkin tree; (b) Blue arrows indicate the decoding order. }
\end{figure*}

The primary goal of channel decoding is to design efficient decoders that can correctly recover the message bits upon receiving codewords corrupted by noise (\prettyref{fig:channel_dec}). More precisely, let $\bu =(u_1,\ldots, u_k) \in \{0,1\}^k$ denote a block of \emph{information/message} bits that we wish to transmit. An encoder $g:\binary^k \to \binary^n$ maps these message bits into a binary codeword $\bx$ of length $n$, \ie $\bx=g(\bu)$. The encoded bits $\bx$ are modulated via Binary Phase Shift Keying (BPSK), \ie $\bx \mapsto 1-2\bx \in \{\pm 1\}^n$, and are transmitted across the channel. We denote both the modulated and unmodulated codewords as $\bx$. The channel, denoted as $P_{Y|X}(\cdot|\cdot)$, corrupts the codeword $\bx$ to its noisy version $\by \in \reals^n$. Upon receiving the corrupted codeword, the decoder $f_\theta$ estimates the message bits as $\hat{\bu}=f_\theta(\by)$. The performance of the decoder is measured using standard error metrics such as Bit-Error-Rate (BER) or Block-Error-Rate (BLER): $\mathrm{BER}(f_\theta) \define (1/k) \sum_i \pprob{\hat{\bu}_i \neq \bu_i}$, whereas $\mathrm{BLER}(f_\theta) \define \pprob{\hat{\bu} \neq \bu}$. 
\begin{figure*}[htbp]
\centerline{
\subfigure[t][CRISP decoder]
{
\includegraphics[width=\columnwidth]{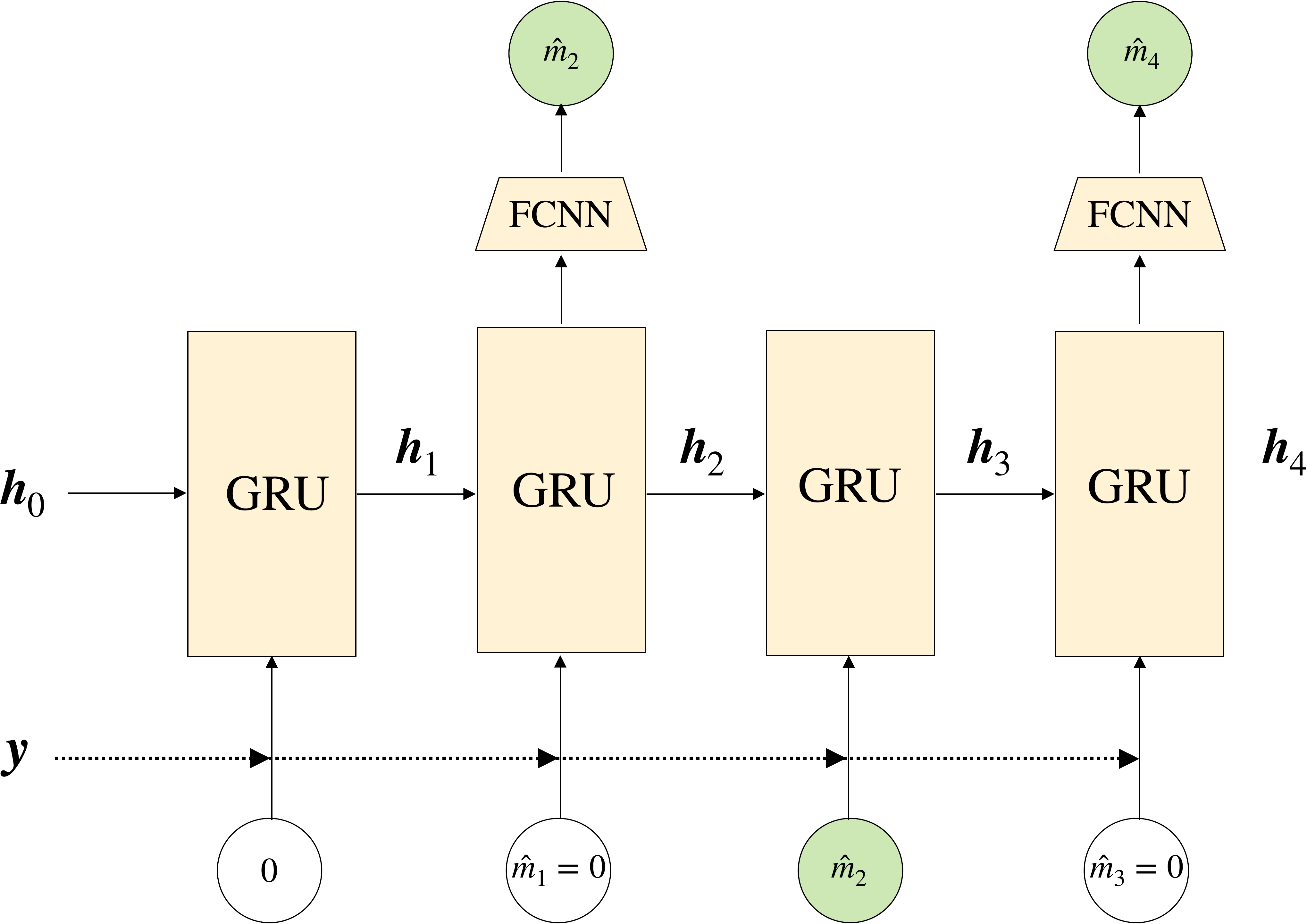}
  \label{fig:crisp_arch}
}
\hfill
\subfigure[Curriculum to train CRISP]
{
  \includegraphics[width=\columnwidth]{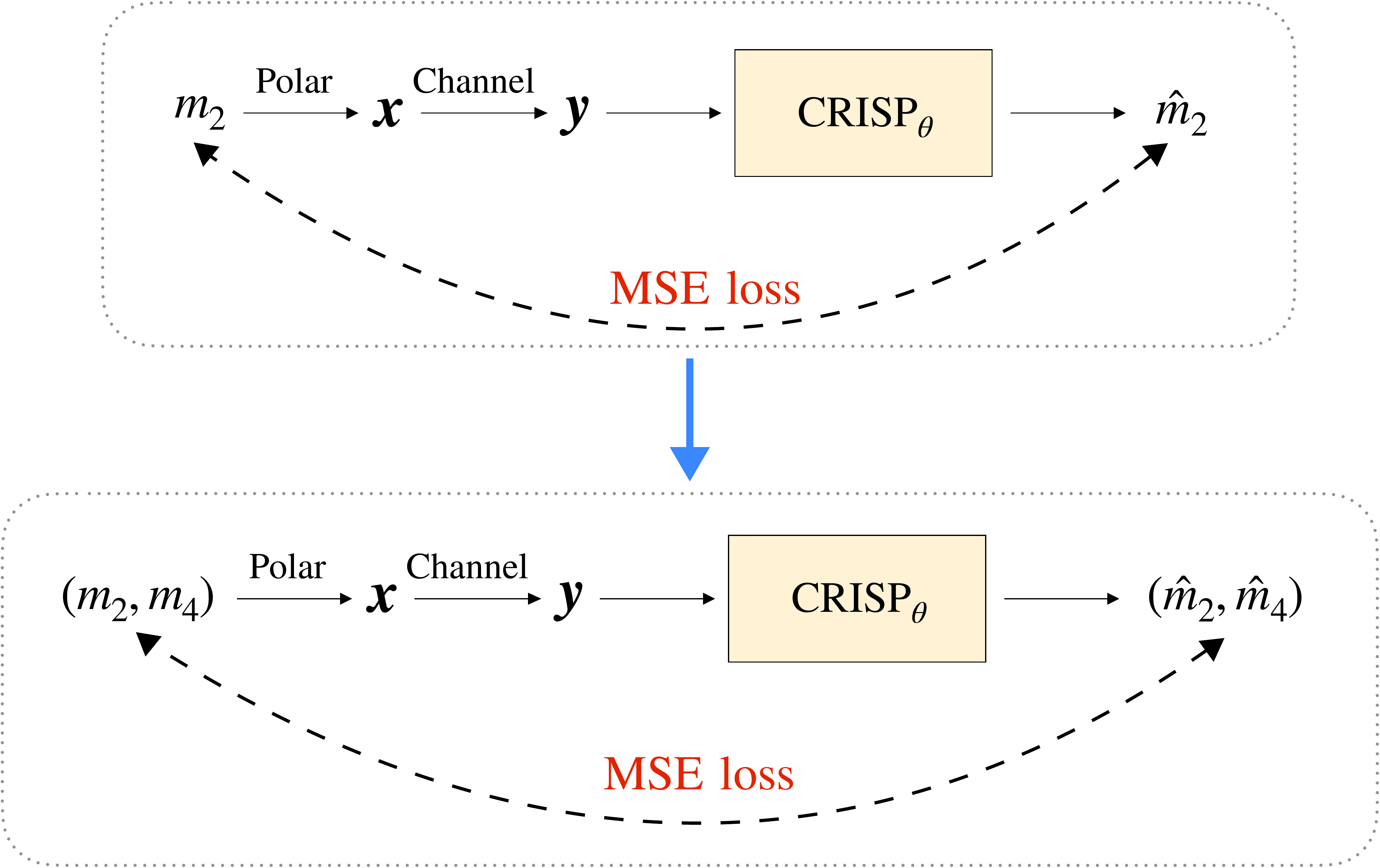}
  \label{fig:crisp_curr}
}
}
  \caption{CRISP decoder and its training by curriculum-learning for $\polar(2,4)$.}
  \label{fig:crisp_dec}
\end{figure*}

Given an encoder $g$ with code parameters $(n,k)$ and a channel $P_{Y|X}$, the channel decoding problem can be mathematically formulated as:
\begin{align}
    \theta \in \arg \min_\theta \mathrm{BER}(f_\theta),
    \label{eq:channel_dec}
\end{align}
which is a joint classification of $k$ binary classes. To train the parameters $\theta$, we use the mean-square-error (MSE) loss as a differentiable surrogate to the objective in \prettyref{eq:channel_dec}. It is well known in the literature that naively parametrizing $f_\theta$ by general-purpose neural networks does not work well and they perform poorly even for small blocklengths like $n=16$ \citep{gruber2017deep}. Hence it is essential to use efficient decoding architectures that capitalize on the structure of the encoder $g$ \citep{kim2018communication,chen2021cyclically}. To this end, we focus on a popular class of codes, the \emph{Polar code family}, that comprises two state-of-the-art codes: Polar codes \citep{Arikan_2009} and Polarization-Adjusted-Convolutional (PAC) codes \citep{arikan2019sequential}. Both these codes are closely related and hence we first focus on polar codes in \prettyref{sec:all_enc}. In \prettyref{sec:crisp}, we present CRISP, our novel curriculum-learning based neural decoder to decode polar codes. In \prettyref{sec:pac} we detail PAC codes.

\subsection{Polar codes}
\label{sec:all_enc}

{\bf Encoding.} Polar codes, introduced in \cite{Arikan_2009}, were the first codes to be theoretically proven to achieve capacity for any binary-input discrete memoryless channel. Their encoding is defined as follows: let $(n,k)$ be the code parameters with $n=2^p, 1 \leq k \leq n$. In order to encode a block of message bits $\bu= (u_1,\ldots, u_k) \in \binary^k$, we first embed them into a source message vector $\bma \define (m_1,\ldots,m_n) = (0, \ldots, u_1, 0,  \ldots, u_2, 0, \ldots, u_k, 0, \ldots) \in \binary^n$, where $\bma_{I_k}= \bu$ and $\bma_{I_k^C}=0$ for some $I_k \subseteq [n]$.  Since the message block $\bma$ contains the information bits $\bu$ only at the indices pertaining to $I_k$, the set $I_k$ is called the \emph{information set}, and its complement $I_k^C$ the \emph{frozen set}. For the set $I_k$, we first compute the capacities of the $n$ individual polar bit channels and rank them in their increasing order \citep{tal2013construct}. Then $I_k$ picks the top $k$ out of them. For example, $\polar(4,2)$ has the ordering $m_1 < m_2 = m_3 < m_4$ and $I_k=\{2,4\}$, and thus $\bma=(0,m_2,0,m_4)$. Similarly, $\polar(8,4)$ has $m_1 < m_2 < m_3< m_5<m_4 <m_6 <m_7<m_8$, $I_4=\{4, 6, 7, 8 \}$ and $\bma=(0,0,0,m_4,0,m_6,m_7,m_8)$.

Finally, we obtain the polar codeword $\bx = \mathrm{PlotkinTree}(\bma)$, where the mapping $\mathrm{PlotkinTree} :\binary^n \to \binary^n$ is given by a complete binary tree, known as Plotkin tree \citep{plotkin60}.
\prettyref{fig:polar_enc} details the Plotkin tree for $\polar(4,2)$. Plotkin tree takes the input message block $\bma \in \binary^n$ at the leaves and applies the ``$\plotkin$" function at each of its internal nodes recursively to obtain the codeword $\bx \in \binary^n$ at the root. The function $\plotkin:\binary^\ell \times \binary^\ell \to \binary^{2\ell}$, $\ell \in \naturals$, is defined as 
\begin{align*}
    \plotkin(\bu, \bv) \define (\bu \oplus \bv, \bv).
\end{align*}
For example, in \prettyref{fig:polar_enc}, starting with the message block $\bma= (0, m_2,0, m_4)$ at the leaves, we first obtain $\bu=\plotkin(0,m_2)=(m_2, m_2)$ and $\bv=\plotkin(0,m_4)=(m_4, m_4)$. Applying the function once more, we obtain the codeword $\bx=\plotkin(\bu,\bv)=(m_2 \oplus m_4,m_2 \oplus m_4, m_4, m_4 )$.

{\bf Decoding.} The successive-cancellation (SC) algorithm is one of the most efficient decoders for polar codes, with a decoding complexity of $O(n \log n)$. The basic principle behind the SC algorithm is to  sequentially decode one message bit $m_i$ at a time according to the conditional log-likelihood ratio (LLR), $L_i \define \log (\pprob{m_i=0|\by, \hat{\bma}_{<i} }/ \pprob{m_i=1|\by, \hat{\bma}_{<i} }) $, given the corrupted codeword $\by$ and previous decoded bits $\hat{\bma}_{<i}$ for $i \in I_k$. \prettyref{fig:polar_dec} illustrates this for the $\polar(4,2)$ code: for both the message bits $m_2$ and $m_4$, we compute these conditional LLRs and decode them via $\mhat_2= \mathbbm{1}\{L_2 < 0 \}$ and $\mhat_4=\mathbbm{1}\{L_4 < 0 \}$. Given the Plotkin tree structure, these LLRs can be efficiently computed sequentially using a depth-first-search based algorithm (\prettyref{app:sc}).

As discussed in \prettyref{sec:intro}, SC achieves the theoretically optimal performance only asymptotically, and its reliability is sub-optimal at finite blocklengths. SC-list (SCL) decoding improves upon its error-correction performance by maintaining a list of $L$ candidate paths at any time step and choosing the best among them in the end. In fact, for a reasonably large list size $L$, SCL achieves MAP performance at the cost of increased complexity $O(L n \log n)$, as highlighted in \prettyref{tab:complexity}.

\section{CRISP: Curriculum based sequential neural decoder for Polar family} 
\label{sec:crisp}

We design CRISP, a curriculum-learning-based sequential neural decoder for polar codes that strictly outperforms the SC algorithm and existing baselines. CRISP uses a sequential RNN decoder, powered by gated recurrent units (GRU) \citep{cho2014properties, empiricalgru}, to decode one bit at a time. Instead of standard training techniques, we design a novel curriculum, guided by information-theoretic insights, to train the RNN to learn good decoders. \prettyref{fig:crisp_dec} illustrates our approach.

{\bf CRISP decoder.} We use the $\polar(4,2)$ code as a guiding example to illustrate our CRISP decoder (\prettyref{fig:crisp_arch}). This code has two message bits $(m_2, m_4)$ and the message block is $\bma=(0, m_2, 0, m_4)$. Upon encoding it to the polar codeword $\bx \in \{\pm 1\}^4$ and receiving its noisy version $\by \in \reals^4$, the decoder estimates the message as $\hat{\bma}=(0, \hat{m}_2,0,\hat{m}_4)$. Similar to SC, CRISP uses the sequential paradigm of decoding one bit $\mhat_i$ at a time by capitalizing on the previous decoded bits $\mpast$ and $\by$. To that end, we parametrize the bit estimate $\mhat_i$ conditioned on the past as a fully connected neural network (FCNN) that takes the hidden state $\bh_i$ as its input. Here $\bh_i$ denotes the hidden state of the GRU that implicitly encodes this past information $(\mpast, \by)$ via GRU's recurrence equation, \ie
\begin{align}
    \bh_i & = \text{GRU}_\theta(\bh_{i-1}, \mhat_{i-1}, \by), \quad i \in \{1,2,3,4 \}, \label{eq:rnn_recursion}\\
    \mhat_i |\by,\mpast & = \text{FCNN}_\theta(\bh_i), \quad i \in \{2,4\},
    \label{eq:bitestimate}
\end{align}

where $\theta$ denotes the FCNN and GRU parameters jointly. Henceforth we refer to our decoder as either CRISP or $\crisp_\theta$. Note that while the RNN is unrolled for $n=4$ time steps (\prettyref{eq:rnn_recursion}), we only estimate bits at $k=2$ information indices, \ie $\mhat_2$ and $\mhat_4$ (\prettyref{eq:bitestimate}). A key drawback of SC is that a bit error at a position $i$ can contribute to the future bit errors ($>i$), and it does not have a feedback mechanism to correct these error events. On the other hand, owing to the RNN’s recurrence relation (\prettyref{eq:rnn_recursion}), through the gradient it receives during training, CRISP can learn to better predict the bits.

{\bf Curriculum-training of CRISP.} Given the decoding architecture of CRISP in \prettyref{fig:crisp_arch}, a natural approach to train its parameters via supervised learning is to use a joint MSE loss function for both the bits $(\mhat_2,\mhat_4)$: $\mathrm{MSE}(\mhat_2, \mhat_4)= (\mhat_2(\theta)-m_2)^2+(\mhat_4(\theta)-m_4)^2$. However, as we highlight in \prettyref{sec:awgn} such an approach learns to fail better decoders than SC and gets stuck at local minima. To address this issue, we propose a curriculum-learning based approach to train the RNN parameters.

The key idea behind our curriculum training of CRISP is to decompose the problem of joint estimation of bits $(\mhat_2,\mhat_4)$ into a sequence of sub-problems with increasing difficulty: start with learning to estimate only the first bit ($\mhat_2$) and progressively add one new message bit at each curriculum step ($\mhat_4$) until we estimate the full message block $\bma=(\mhat_2,\mhat_4)$. We freeze all the non-trainable message bits to zero during any curriculum step. In other words, in the first step, we freeze the bit $m_4$ and train the decoder only to estimate the bit $\mhat_2$ (\ie the subcode corresponding to $k=1$): 
\begin{align}
(m_2, m_4=0) \rightarrow \bma= (0, m_2, 0, 0) \walk{\polar} \bx \nonumber\\ \bx \walk{\text{Channel}} \by \walk{\crisp_\theta} \mhat_2.
\label{eq:m2_train}
\end{align}

We use this trained $\theta$ as an initialization for the next task of estimating both the bits $(\mhat_2, \mhat_4)$: 
\begin{align}
(m_2, m_4) \rightarrow \bma=(0, m_2, 0, m_4) \walk{\polar} \bx \nonumber\\ \bx \walk{\text{Channel}} \by \walk{\crisp_\theta} (\mhat_2, \mhat_4).
\label{eq:m24_train}
\end{align}
\prettyref{fig:crisp_curr} illustrates this curriculum-learning approach. We note that the knowledge of decoding $\mhat_2$ when $m_4=0$ (\prettyref{eq:m2_train}) serves as a good initialization when we learn to decode $\mhat_2$ for a general $m_4 \in \binary$ (\prettyref{eq:m24_train}). With such a curriculum aided training, we show in \prettyref{sec:awgn} (Figs.~\ref{fig:main_figure}, \ref{fig:awgn_restplots}) that the CRISP decoder outperforms the existing baselines and attains near-optimal performance for a variety of blocklengths and codes. We interpret this in \prettyref{sec:interpret}. We defer the training details to \prettyref{app:exp_details}.

{\bf Left-to-Right (L2R) curriculum for $\polar(n,k)$.} For a general $\polar(n,k)$ code, we follow a similar curriculum to train $\crisp_\theta$. Denoting the index set by $I_k=\{i_1, i_2, \ldots, i_k\} \subseteq [n]$ in the increasing order of indices $i_1 < i_2 < \ldots < i_k$, our curriculum is given by: Train $\theta$ on $\mhat_{i_1} \rightarrow$ Train $\theta$ on $(\mhat_{i_1},\mhat_{i_2}) \rightarrow \ldots \rightarrow$ Train $\theta$ on $(\mhat_{i_1},\ldots, \mhat_{i_k})$. We term this curriculum \emph{Left-to-Right (L2R)}. The anti-curriculum \emph{R2L} refers to progressively training in the decreasing order of the indices in $I_k$.

\section{Main results} 
\label{sec:main_results} 
In this section, we present numerical results for the CRISP decoder on the Polar code family. 

\subsection{AWGN channel}
\label{sec:awgn}

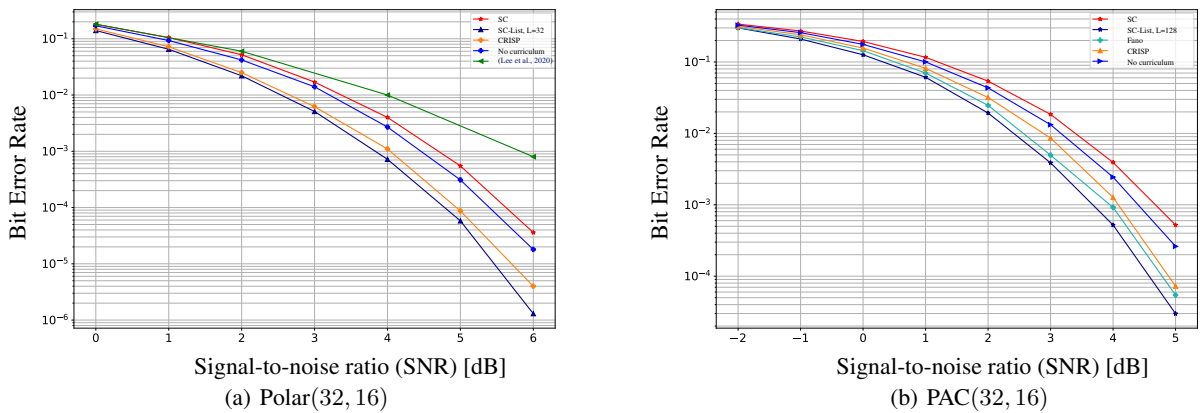
\begin{figure*}[ht]
\centerline{
\subfigure[Polar$(32,16)$]
{
 \includegraphics[width=\columnwidth]{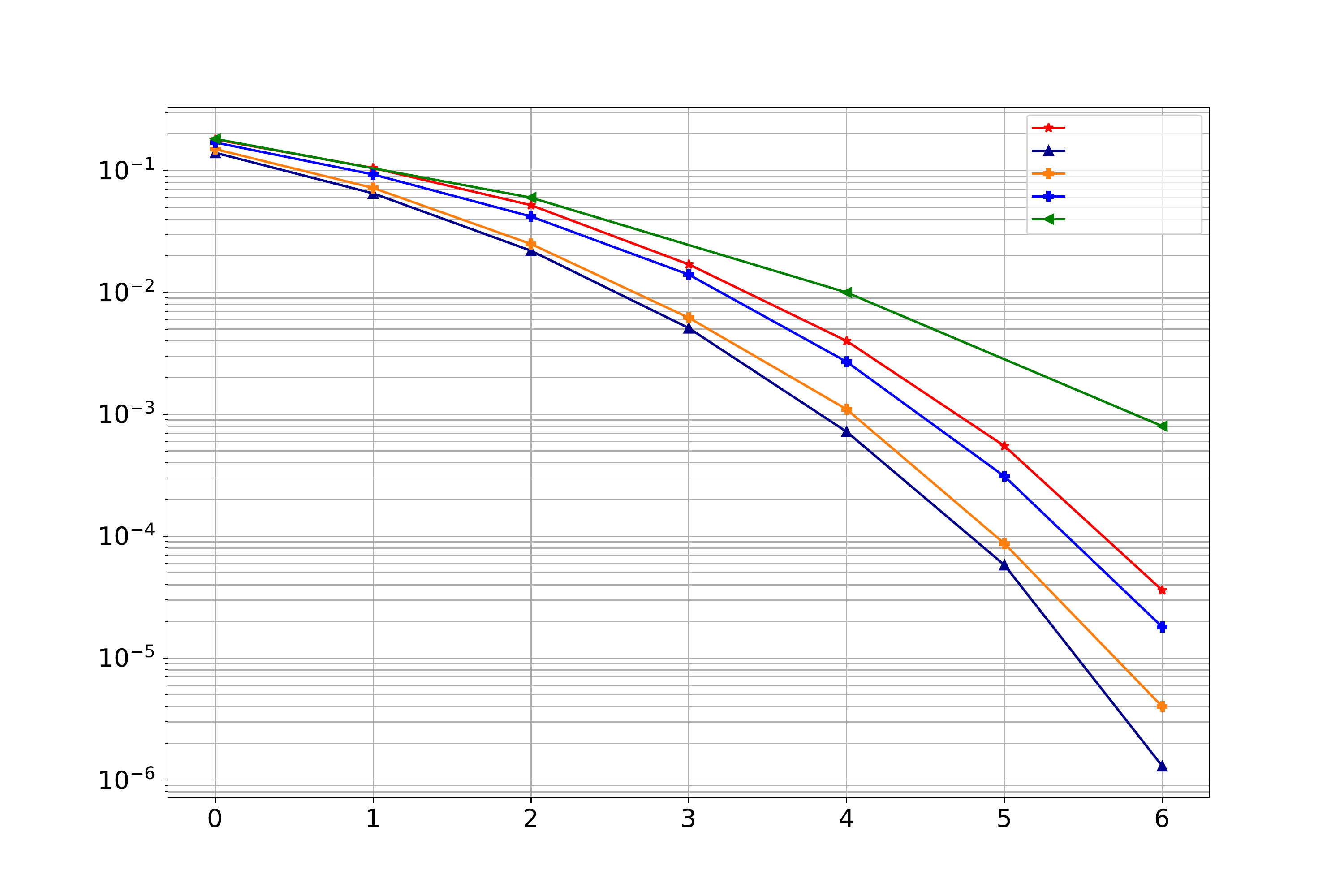}
    \put(-45,133){\fontsize{3}{5}\selectfont SC}
    \put(-45,129){\fontsize{3}{5}\selectfont SC-List, L=32}
    \put(-45,125){\fontsize{3}{5}\selectfont CRISP}
    \put(-45,121){\fontsize{3}{5}\selectfont No curriculum}
    \put(-45,117.5){\fontsize{3}{5}\selectfont \cite{lee2020training}}
    \put(-160,0){\footnotesize\selectfont Signal-to-noise ratio (SNR) [dB]}
    \put(-230,70){\rotatebox[origin=t]{90}{\footnotesize \selectfont Bit Error Rate}}
  \label{fig:1632}
}
\hfill
\subfigure[$\text{PAC}(32,16)$]
{  \hspace*{-0.25in}
 \includegraphics[width=\columnwidth]{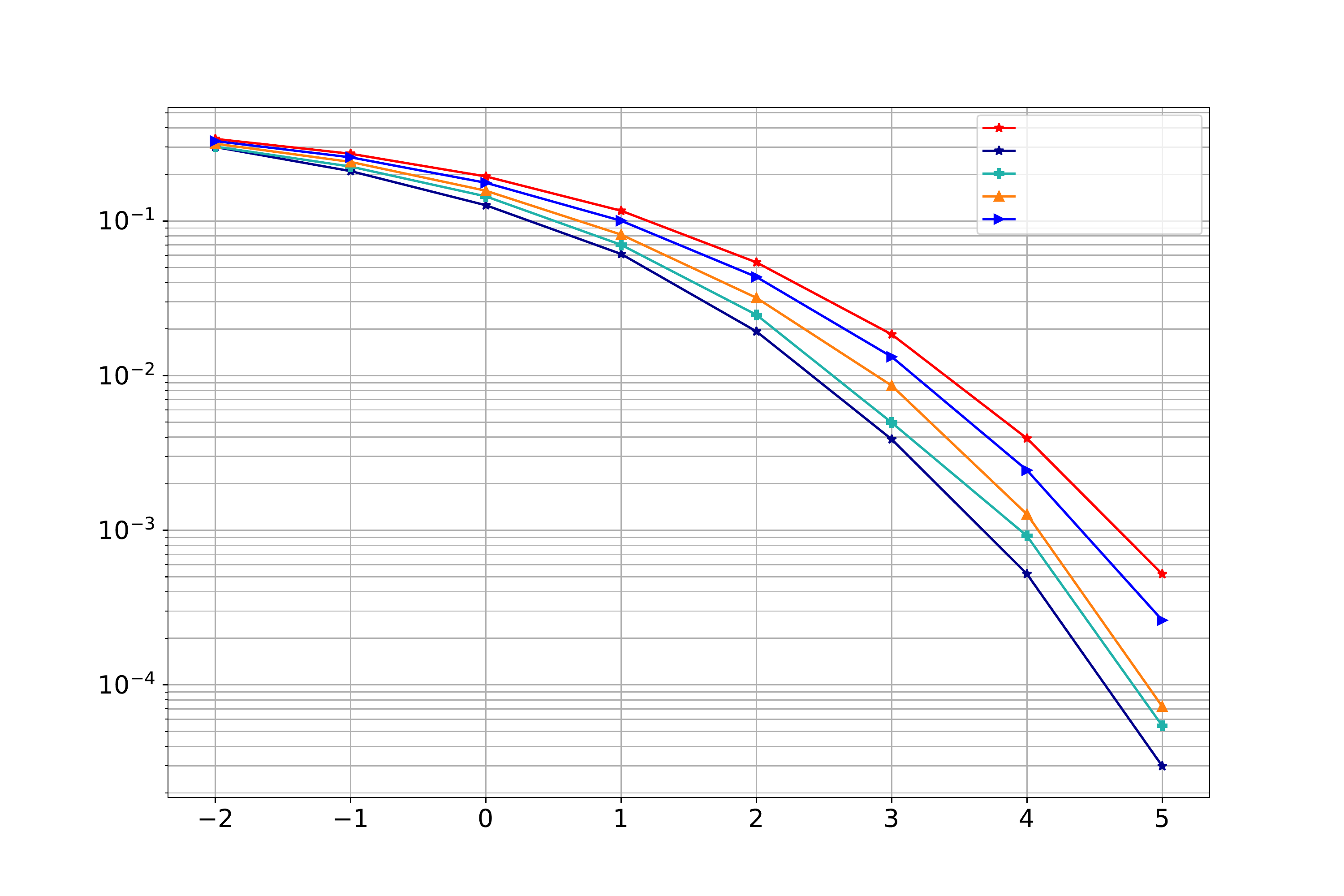} 
    \put(-50,133){\fontsize{3}{5}\selectfont SC}
    \put(-50,129){\fontsize{3}{5}\selectfont SC-List, L=128}
    \put(-50,125){\fontsize{3}{5}\selectfont Fano}
    \put(-50,121){\fontsize{3}{5}\selectfont CRISP}
    \put(-50,117){\fontsize{3}{5}\selectfont No curriculum}
    \put(-160,0){\footnotesize Signal-to-noise ratio (SNR) [dB]}
    \put(-230,70){\rotatebox[origin=t]{90}{\footnotesize\selectfont Bit Error Rate}}
  \label{fig:pac1632}
 }
 
}
\caption{CRISP outperforms baselines and attains near-MAP performance for $\polar(32,16)$ and PAC$(32,16)$ codes on the AWGN channel.}
\label{fig:awgn_restplots}

\end{figure*}

{\bf Data generation}. The input message $\bu \in \binary^k$ is randomly drawn uniformly from the boolean hypercube and encoded as a polar codeword $\bx \in \{\pm 1 \}^n$. The classical additive white Gaussian noise (AWGN) channel, $\by = \bx + \bz$, $\bz \sim \mathcal{N}(0, \sigma^2 \bI_n)$, generates the training/test data $(\by, \bu)$ for the decoder. The signal-to-noise ratio, \ie $\text{SNR} = -10\log_{10} \sigma^2$, characterizes the noise level in the channel. Here we fix the channel to be AWGN in all our experiments, as per the standard convention \citep{kim2018communication}, and refer to \prettyref{app:additional} for additional results on fading and t-distributed channels.  \prettyref{app:exp_details} details the training procedure. Once trained, we use the CRISP models for comparison against the baselines.

{\bf Baselines}. The optimal channel decoder is the MAP estimator: $\hat \bu = \arg\max_{\bu \in \binary^k} \pprob{\bu|\by}$, whose complexity grows exponentially in $k$ and is computationally infeasible. Given this, we compare our CRISP decoder with the SCL \citep{tal2015list}, which has near-MAP performance for a large $L$, along with the classical SC. Among learning-based decoders, we choose the state-of-the-art Neural-Successive-Cancellation (NSC) as our baseline \citep{doan2018neural}. NSC replaces sub-components of the existing successive cancellation decoder with NNs to scale decoding to block lengths longer than 32. Each of these neural networks are trained with the LLR outputs of the SC algorithm. Since this training procedure with SC probabilities as the target is sub-optimal, we consider an improved version with end-to-end training (\prettyref{fig:channel_dec}) for a fair comparison. We also include the performance of CRISP trained directly without the curriculum. We also compare with the curriculum training procedure of \cite{lee2020training} (the original work achieves a reliability worse than SC decoding for block length 32). 
All these baselines have the same number of parameters as CRISP.

{\bf Results.} \prettyref{fig:2264} highlights that the CRISP decoder outperforms the existing baselines and attains near-MAP performance over a wide range of SNRs for the $\polar(64, 22)$ code. \prettyref{fig:2264_progressive} illustrates the mechanism behind these gains at 0dB: the curriculum-guided CRISP slowly improves upon the overall BER (over the $22$ bits) during the training and eventually achieves much better performance than SC and other baselines. In contrast, the decoder trained from scratch makes a big initial gain but gets stuck at local minima and only achieves a marginal improvement over SC. Moreover, we see that decoders trained using other curricula, e.g. R2L, also fail to show significant improvements over SC (Figs. \ref{fig:2264_progressive}, \ref{fig:plot_all_curricula}). We observe a similar trend for $\polar(32,16)$ code in \prettyref{fig:1632}, where CRISP achieves near-MAP performance. We posit that aided by a good curriculum, CRISP avoids getting stuck at bad local minima and converges to better minima in the optimization landscape.
Further, CRISP is robust to deviations from the AWGN channel, while attaining similar performance gains over SC on fading and T-distributed channels (\prettyref{app:additional}). For additional results, we refer to
\prettyref{app:additional} which highlights similar reliability gains for other blocklengths and rates, \prettyref{app:ablation} for the ablation analysis, and \prettyref{app:exp_details} for the training hyperparameters and architectures.

\textbf{Sequential vs Block decoding.} We note that the sequential RNN architecture for CRISP is inspired in part by the sequential SC algorithm. Notwithstanding, we also design block decoders that estimate all the information bits $m_i$ in one shot given $\by$. We choose $1$D Convolutional Neural Networks (CNNs) to parameterize this block decoder, {\it CRISP\_CNN}. CRISP\_CNN, trained with the L2R curriculum, achieves similar BER performance as CRISP (\prettyref{app:block_decoding}).

\subsection{Reliability-throughput comparison}
\label{sec:complexity_compar}

\begin{table}[ht]
\caption{Throughput and reliability comparison of various decoders on $\polar(n,k)$.}
\label{tab:complexity}
\centering
\resizebox{\columnwidth}{!}{
\begin{tabular}{l cccc cc}
\toprule
\multirow{3}{*}{\bf Decoder} &
     \multicolumn{4}{c}{ \bf Throughput (in Mbps)} & \multicolumn{2}{c}{ \bf Gap to SCL, L=32 (in dB)} \\ 
     \cmidrule(lr){2-5} \cmidrule(lr){6-7} 
 &     \multicolumn{2}{c}{$(32,16)$}           &  \multicolumn{2}{c}{{$(64,22)$}}  &   $(32,16)$     &   $(64,22)$ \\ 
 \cmidrule(lr){2-3} 
 \cmidrule(lr){4-5}
 & GPU & CPU & GPU & CPU\\
   \midrule
    SC    &    $0.17$     & $27$  &   $0.08$       &    $15$    &     $0.80$    &    $0.40$         \\
    FastSC    &    N/A    & {\bf 47} &   N/A      &    {\bf 40}    &     $0.80$    &    0.40         \\
    SCL, L=4    &   $0.01$  & $8.5$  &     $0.02$    &    $6.27$    &       $0.05$  &      $0.10$         \\
    FastSCL, L=4   &   N/A &  {\bf 30}  &    N/A    &     {\bf 24}     &    $0.05$      &      $0.10$         \\
    SCL, L=32 (MAP)   &    5e-3     &  $0.81$ & 2e-3     &     $0.60$     &    {\bf 0.00}      &      {\bf 0.00}         \\

    FastSCL, L=32   &    N/A  & $7.7$   &    N/A     &     $5.5$     &    { 0.00}      &      { 0.00}         \\
    NSC   &     N/A     &  N/A   & $32.6$    &    $0.02$      &     N/A     &  $0.35$      \\
    {CRISP\_GRU (Ours)}    &    {\bf 80}      & 0.04  &  {\bf 38.7}     &    $0.02$      &    {\bf 0.15}     &    {\bf 0.20}            \\ 
    CRISP\_CNN (Ours)   &    {\bf 250}      & 0.02 &    {\bf 133}     &    $0.13$      &     $0.15$     &    $0.20$             \\ 
    CRISP\_GRU - No curriculum    &    ${80}$    &     0.04 &${38.7}$    &    $0.02$      &     $0.60$     &    $0.35$             \\ 

    \bottomrule

\end{tabular}
}
\end{table}

In the previous section, we demonstrated that CRISP achieves better reliability than the baselines. Here we analyze these gains through the lens of \emph{decoding complexity}.
To quantitatively compare the complexities of these decoders, we evaluate their throughput %
on a single GTX 1080 Ti GPU as well as a CPU (Intel i7-6850K, 12 threads). %
For the GPU version, we use our implementation of SC/SCL owing to the lack of publicly available implementations. 
On the other hand,  for the CPU column we use an optimized multithreaded implementation of SC/FastSC, SCL/FastSCL \cite{leonardon2019fast} in C++ by \cite{Cassagne2019a}.
As \prettyref{tab:complexity} highlights, CRISP exploits the GPUs' inherent optimization towards NNs to achieve excellent throughput, whilst achieving near-SCL BER performance. We note that CRISP\_CNN (\prettyref{app:block_decoding}) attains better throughput than CRISP\_GRU, while maintaining gains in BER. We posit that  further improvement in throughput can be realized using techniques like pruning and knowledge distillation. This is beyond the scope of this paper and is an important and separate direction of future research. Note that we use BER$=10^{-3}$ to compute the gap to SCL (Figs. \ref{fig:2264}, \ref{fig:1632}).
We refer to \prettyref{app:complexity} for further discussion. 

\subsection{Computational complexity}
Running CRISP on suitable hardware architectures allows it to attain significant throughput gains. Nevertheless, to provide a more comprehensive performance evaluation, it's important to also consider other metrics such as power consumption. This necessitates an analysis of the computational complexity of the algorithm, which we present below.

The decoding complexity of SCL is $O(L n \log{n})$, where $L$ represents the list size. On the other hand, CRISP employs a 2-layer GRU neural network, the computational complexity of which is $n(2h(n+1)+6h^2)$, where $h$ denotes the dimension of the GRU's hidden state. The bulk of this computational complexity involves matrix-vector multiplications; modern hardware like GPUs which allow for significant speedups in these operations. This, in turn, allows for an improved performance and efficiency of CRISP on such platforms.

\subsection{Interpretation}
\label{sec:interpret}

\begin{figure*}[ht]
\centerline{
\subfigure[L2R vs. R2L for decoding $m_1$]
{
 \includegraphics[width=0.4\linewidth]{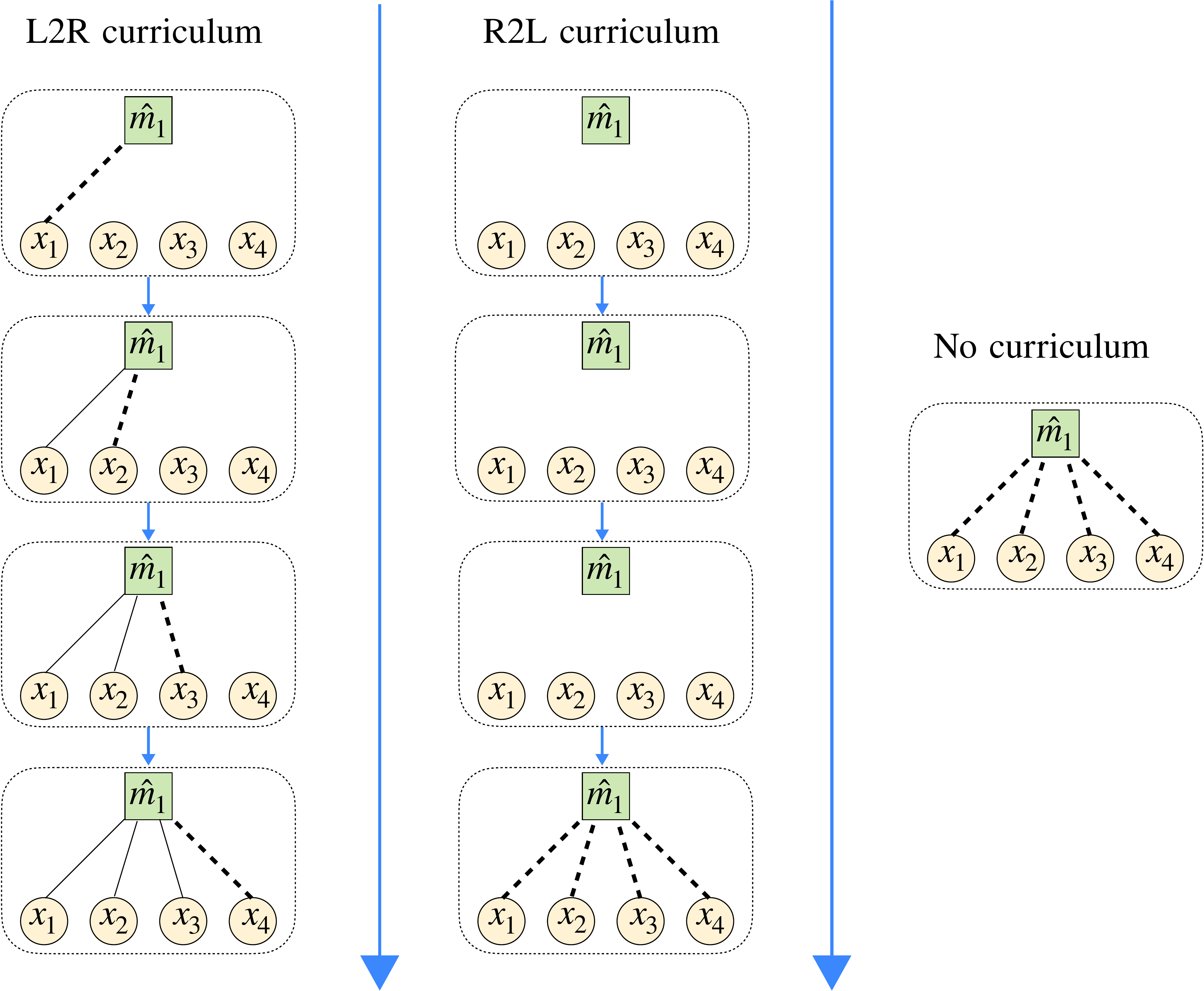}
  \label{fig:hardbit_picture}
}
\subfigure[Evolution of learning difficulty]
{
\hspace*{0.3in}
  \includegraphics[width=0.4\linewidth]{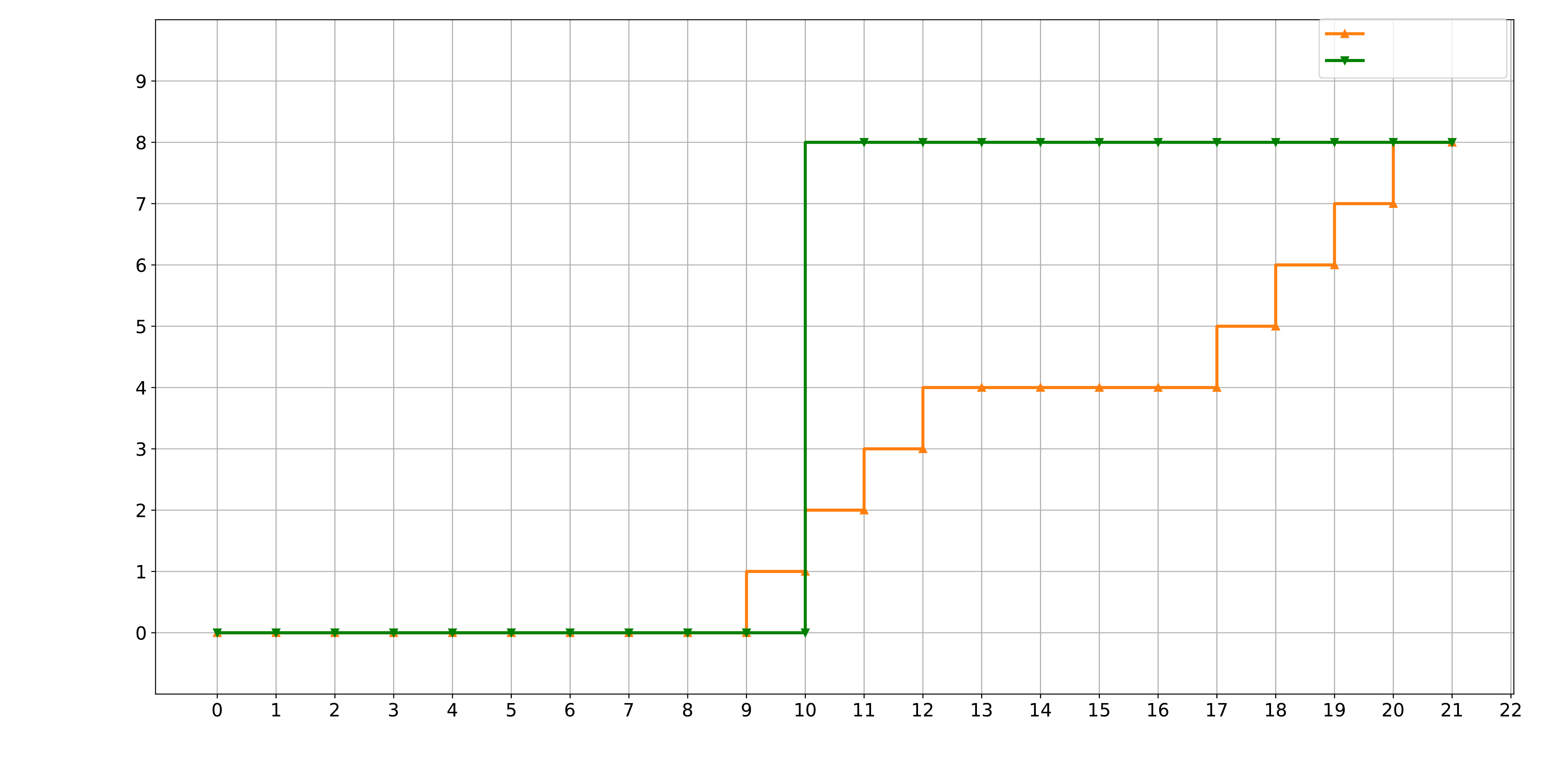}
    \put(-23,94){\fontsize{3}{5}\selectfont L2R curriculum}
    \put(-23,90){\fontsize{3}{5}\selectfont R2L curriculum}
    \put(-128,0){\footnotesize Curriculum Step}
    \put(-190,50){\rotatebox[origin=t]{90}{\footnotesize Learning Difficulty}}
  \label{fig:graph_learndifficulty}
}
}
\caption{L2R vs. R2L: (a) Bit estimates evolve more smoothly under L2R than R2L for $\polar(4,4)$, (b) Learning difficulty increases more gracefully for L2R than R2L for $\polar(64,22)$. }
\end{figure*}

This section describes why L2R is a better curriculum than others. To this end, we first claim that learning to decode uncorrupted codewords ($\by=\bx$) is critical to learning a reliable decoder. This claim follows from the following key observation: while training our model (sequential or block) at a specific SNR, we observe that whenever our model reaches SC or better performance, its BER on uncorrupted codewords, aka the noiseless BER, drops to zero very early in the training (\prettyref{app:interpret}, \prettyref{fig:goodtraining}). On the other hand, when the model gets stuck at bad minima even after a lot of training, its noiseless BER is high (\prettyref{fig:stucktraining}). Hence, without loss of generality, we focus on the setting $\by = \bx$. Under this noiseless scenario, we analyze how the optimal bit decoding rules evolve for different curricula. In particular, we focus on the least reliable bits as they contribute the largest to noiseless BER (\prettyref{fig:bitwiseNoisyVsClean} and \prettyref{fig:condBitwise}).

For the $\polar(4,4)$ code, \prettyref{fig:hardbit_picture} illustrates how the optimal rule evolves for its least reliable bit $m_1$. In this case, the MAP decoding rule for $m_1$ is: $\mhat_1 = x_1 x_2 x_3 x_4$. Under the L2R curriculum, we arrive at this expression via  $x_1 \rightarrow x_1 x_2 \rightarrow x_1 x_2 x_3 \rightarrow x_1 x_2 x_3 x_4 $, whereas R2L follows $1 \rightarrow 1 \rightarrow 1 \rightarrow x_1 x_2 x_3 x_4 $. This highlights that L2R reaches the optimal rule more gracefully by learning to include one coordinate $x_i$ at a time while this change for R2L (and no-curriculum) is abrupt, making it harder to learn.  \prettyref{fig:bitEvolving44} illustrates a similar evolution for the remaining bits $(m_2, m_3, m_4)$.

 More concretely, we define the notion of \textit{learning difficulty} for a bit: the number of bits multiplied in its optimal decoding rule. This metric roughly captures the number of operations a model has to learn at any curriculum step. \prettyref{fig:graph_learndifficulty} illustrates how it evolves over the L2R and R2L curricula for the least reliable bit in $\polar(64,22)$. If we take the maximum learning difficulty over all bits, we obtain a similar plot (\prettyref{fig:graph_maxlearndifficulty}). Note that in both the plots, the jumps in learning difficulty are larger for R2L, thus indicating a harder transfer than L2R, where it increases smoothly (at most one bit per step).

\subsection{PAC codes}
\label{sec:pac}
A recent breakthrough work \cite{arikan2019sequential} introduces a new class of codes called Polarization-Adjusted-Convolutional (PAC) codes that match the fundamental lower bound on the performance of any code under the MAP decoding at finite-lengths \citep{moradi2020performance}. The motivating idea behind PAC codes is to overcome two key limitations of polar codes at finite blocklengths: the poor minimum distance properties of the code and the sub-optimality of SC compared to the MAP \citep{mondelli2014polar}. This is addressed by adding a \emph{convolutional outer code}, with an appropriate indexing $I_k$, before polar encoding to improve the distance properties of the resulting code. More formally, the message block $\bu \in \binary^k$ is embedded into the source vector $\bma \in \binary^n$ according to the Reed-Muller (RM) indices $I_k^{(\text{RM})}$: compute the Hamming weights of integers ${0,1,\ldots, n-1}$ and choose the top $k$. Now we encode the message $\bma$ via a rate-$1$ convolutional code, \ie $\bv = \bc * \bma \in \binary^n \Leftrightarrow v_i = \sum_{j} c_j m_{i-j} $, for some $1$D convolutional kernel $\bc \in \binary^\ell$. Finally we obtain the PAC codeword $\bx$ by polar encoding $\bv$: $\bx = \mathrm{PlotkinTree}(\bv)$.

PAC codes can be decoded using the classical Fano algorithm \citep{fano1963heuristic}, a sequential decoding algorithm that uses backtracking. Coupled with the Fano decoder, PAC codes achieve impressive results outperforming polar codes (with SCL decoder) and matching the finite-length capacity bound \citep{polyanskiy2010channel}. However, the Fano decoder has significant drawbacks like variable running time, large time complexity at low-SNRs \citep{rowshan2020polarization}, and sensitivity to the choice of hyperparameters \citep{moradi2020metric}. To overcome these issues, several non-learning techniques, such as stack/list decoding, adaptive path metrics, etc., have been proposed in the literature \citep{yao2021list,zhu2020fast,rowshan2021list,rowshan2021convolutional,sun2021optimized}. In contrast, we design a curriculum-learning-based CRISP decoder for PAC codes trained directly from the data. We use the same L2R curriculum to decode PAC codes. 

\prettyref{fig:pac1632} highlights that the CRISP decoder achieves near-MAP performance for the PAC$(32,16)$ code. While Fano decoding achieves similar reliability, it is inherently non-parallelizable. In contrast, CRISP allows for batching, and achieves a higher throughput, as highlighted in \prettyref{tab:pac_complexity}. Here we measure the throughput of Fano \citep{rowshan2020complexity} at $\text{SNR}=1 ~ \text{dB}$. We note that the existing implementation of Fano is not supported on GPUs. These preliminary results suggest that curriculum-based training holds a great promise for designing efficient PAC decoders, especially for longer blocklengths, which is an interesting topic of future research (\prettyref{app:additional_pac}).
\begin{table}[ht]
\caption{Throughput and reliability comparison of various decoders on $\pac(32, 16)$.}
\label{tab:pac_complexity}
\centering
\resizebox{\columnwidth}{!}{
\begin{tabular}{l cc c}
\toprule
\multirow{2}{*}{\bf Decoder} &
     \multicolumn{2}{c}{ \bf Throughput (in Mbps)} & \multirow{2}{*}{ \bf Gap to SCL, L=128 (in dB)} \\ 
     \cmidrule(lr){2-3}  
 &     GPU     &   CPU       \\ 
   \midrule
    SC    &    N/A      &   {\bf 27}     &     $1.0$             \\
    SCL, L=128   &    N/A     &        $0.02$       &    {\bf 0.0}               \\
    Fano  &     N/A     &         4e-3       &     $0.1$         \\
    {CRISP\_GRU (Ours)}    &    {\bf 80}      &     { $0.03$}           &    $0.4$                \\ 
    CRISP\_CNN (Ours)    &    {\bf 250}      &     { $0.15$}           &     $0.4$                 \\ 
    CRISP\_GRU - No curriculum    &    {$80$}      &     {$0.03$}           &     $0.8$                 \\ 

    \bottomrule

\end{tabular}
}
\end{table}

\section{Information theory guided curricula}
\label{sec:info_theory_curr}
In \prettyref{sec:main_results}, we demonstrated the superiority of L2R curriculum over other schemes. Here we introduce an alternate curriculum, \textit{Noisy-to-Clean (N2C)}, that slightly bests the L2R, inspired by the polarization property of polar codes. The key idea behind N2C curriculum is to capitalize on the polar index set $I_k$. Recall that the set $I_k$ is obtained by ranking the $n$ polar bit channels (under SC decoding) in the increasing order of their reliabilities (from noisy to clean) and choosing the top $k$ indices. Formally, given $I_k=\{i_{r1}, i_{r2}, \ldots, i_{rk}\} \subseteq [n]$ in the increasing order of reliabilities, our \textit{N2C} curriculum is given by: Train $\theta$ on $\mhat_{i_{r1}} \rightarrow$ Train $\theta$ on $(\mhat_{i_{r1}},\mhat_{i_{r2}}) \rightarrow \ldots \rightarrow$ Train $\theta$ on $(\mhat_{i_{r1}},\ldots, \mhat_{i_{rk}})$. For both the  sequential and block decoders, we observe that N2C is the best curriculum and we have $\text{N2C} \approx \text{L2R} > \text{C2N} \approx \text{R2L} $ (\prettyref{fig:plot_all_curricula}).
This ordering is consistent with our interpretation in \prettyref{sec:interpret} of how the learning difficulty evolves over a curriculum (\prettyref{fig:graph_learndifficulty_all}). For both N2C and L2R, the learning difficulty evolves smoothly but is abrupt for C2N and R2L, thus making transfer harder in these curricula. 
Note that the \emph{C2N} curriculum refers to progressively training on subcodes of $\polar(n,k)$: $\polar(n,1) \rightarrow \ldots \rightarrow \polar(n,k)$ \citep{lee2020training}.

\section{Related work} 
\label{sec:related}

To address the sub-optimality of SC at finite lengths, a popular technique is to use list decoding \citep{tal2015list, balatsoukas2015llr}, aided by cyclic redundancy checks (CRC) \citep{li2012adaptive, niu2012crc, miloslavskaya2014sequential}. Several alternate decoding methods have also been proposed such as stack decoding \citep{niu2012stack, trifonov2018score}, belief propagation decoding \citep{yuan2014early, elkelesh2018belief}. Deep learning for communication \citep{qin2019deep, kim2020physical} has been an active field in the recent years and has seen success in many problems including the design of neural decoders for existing linear codes \citep{nachmani2016learning,  o2017introduction, lugosch2017neural, vasic2018learning, liang2018iterative, bennatan2018deep, jiang2019deepturbo, nachmani2019hyper, buchberger2020prunin, turbonet}, and jointly learning channel encoder-decoder pairs. \citep{o2016learning, kim2018deepcode, jiang2019turbo, Makkuva2021, jamali2021productae, chahine2021turbo, chahine2021deepic}.

Earlier works on designing neural polar decoders \citep{gross2020deep} used off-the-shelf neural architectures. These were only able to decode codes of small blocklength ($\leq 16$) \citep{lyu2018performance, cao2020learning}. Later works augmented belief propagation decoding \citep{xu2018polar,doan2019neural}, with neural components and improved performance. In \cite{cammerer2017scaling} and \cite{doan2018neural}, the authors replace sub-components of the existing SC decoder with NNs to scale decoding to longer lengths. However, these methods fail to give reasonable reliability gains compared to SC. In contrast, we use curriculum learning to train neural decoders, and show non-trivial gains over SC performance. \cite{lee2020training} consider a curriculum training of polar decoder, but do not achieve SC reliability for block length 32. This is owing to the sub-optimality of both the architecture and training curriculum (the C2N scheme). In contrast, we design a principled curriculum guided by information theoretic insights, and a neural architecture that fully capitalizes on the sequential polar decoding. \prettyref{fig:plot_all_curricula} and \prettyref{fig:1632} show that these design choices are essential for achieving the reliability gains over SC.

Recent research by Choukroun and Wolf \cite{choukroun2022error, choukroun2022denoising} introduces transformer-based neural decoders for block channel codes. A distinctive feature of their approach is the use of a sparse attention mask, which harnesses the structure of the parity check matrix. The application of a similar curriculum training procedure, as used in our work with CRISP, to these transformer-based architectures might potentially expedite the convergence process. Furthermore, such enhancements in the training procedure could potentially close the gap to MAP performance for higher block lengths.

\section{Conclusion}
\label{sec:conclusion}
We introduce a novel curriculum based neural decoder, CRISP, that attains near-optimal reliability on the Polar code family in the short blocklength regime. We design a principled curriculum to train CRISP, which is crucial to achieve reliability gains for both the Polar and PAC codes. To the best of our
knowledge, this is the first learning-based PAC decoder
to achieve near-MAP reliability with significantly better throughput than the Fano decoder. Extending our results to medium blocklengths ($100$-$1000$) and codes outside the Polar family are interesting future directions. While optimizing the decoder complexity is not the primary focus of this paper, our preliminary results already show gains in throughput over standard methods. Further improvement in decoding complexity whilst maintaining reliability gains is another exciting future research direction. 

\section*{Acknowledgement}
Ashok would like to thank his colleague Unnat Jain for a crucial advice about the project to switch from a RL based approach to a supervised one, which turned out to be the game changer for this paper. This work is supported by ONR grants W911NF-18-1-0332, N00014-21-1-2379, and NSF grants CNS-2002664, CNS-2002932, CCF-2312753 and CNS-2112471 as a part of NSF AI Institute for Future Edge Networks and Distributed Intelligence (AI-EDGE).

\newpage

\bibliographystyle{icml2023}
\bibliography{all_references}

\clearpage

\appendix

\section{Successive Cancellation decoder}
\label{app:sc}

Here we detail the successive-cancellation (SC) algorithm for decoding polar codes. As a motivating example, let's consider the $\polar(2,2)$ code. Let the two information bits be denoted by $u$ and $v$, where $u,v \in \binary$. The codeword $\bx \in \binary^2$ is given by $\bx = (x_1, x_2) =  (u \oplus v, v)$. Let $\by \in \reals^2$ be the corresponding noisy codeword received by the decoder. First we convert the received $\by$ into a vector of log-likelihood-ratios (LLRs), $\bL_{\by} \in \reals^2$, which contains the soft-information about coded bits $x_1$ and $x_2$, \ie
\begin{align*}
    \bL_{\by} = (\bL_{\by}^{(1)}, \bL_{\by}^{(2)}) \define \pth{\log \frac{\pprob{y_1|x_1 = 0}}{\pprob{y_1|x_1 = 1}} , \log \frac{\pprob{y_2|x_2 = 0}}{\pprob{y_2|x_2 = 1}} } \in \reals^2. 
\end{align*}
Once we have the soft-information about the codeword $\bx$, the goal is to now obtain the same for the message bits $u$ and $v$. To compute the LLRs for these information bits, SC uses the following principle: first, compute the soft-information for the left bit $u$ to estimate $\hat{u}$. Use the decoded $\hat{u}$ to compute the soft-information for the right bit $v$ and decode it. More concretely, we compute the LLR for the bit $u$ as:
\begin{align}
    L_u = \mathrm{LSE}(\bL_{\by}^{(1)}, \bL_{\by}^{(2)}) = \log \frac{1+e^{\bL_{\by}^{(1)}+ \bL_{\by}^{(2)}}}{e^{\bL_{\by}^{(1)}} + e^{ \bL_{\by}^{(2)}} } \in \reals,
    \label{eq:soft_u}
\end{align}
where $\mathrm{LSE}(a,b) \define \log(1+e^{a+b})/(e^a + e^b)$ for $ a,b \in \reals$. The expression in \prettyref{eq:soft_u} follows from the fact that  $u = (u \oplus v)\oplus v = x_1 \oplus x_2$ and hence the soft-information $L_u$ can be accordingly derived from that of $x_1$ and $x_2$, \ie $\bL_{\by}$. Now we estimate the bit as $\hat u = \mathbbm{1}\{L_u < 0\}$. Assuming that we know the bit $u = \hat u$, we observe that the codeword $\bx = (\hat u \oplus v, v)$ can be viewed as a two-repitition of $v$. Hence its LLR $L_v$ is given by
\begin{align}
    L_v = \bL_{\by}^{(1)} \cdot (-1)^{\hat u} + \bL_{\by}^{(2)} \in \reals.
    \label{eq:soft_v}
\end{align}
Finally we decode the bit as $\hat v = \mathbbm{1}\{L_v < 0\}$. To summarize, given the LLR vector $\bL_{\by}$ we first compute the LLR for the bit $u$, $L_u$, using \prettyref{eq:soft_u} and decode it. Utilizing the decoded version $\hat u$, we compute the LLR $L_v$ according to \prettyref{eq:soft_v} and decode it. 

For a more generic $\polar(n,k)$, the underlying principle is the same: to decode a polar codeword $\bx = (\bu \oplus \bv, \bv)$, first decode the left child $\bu$ and utilize this to decode the right child $\bv$. This principle is recursively applied at each node of the Plotkin tree until we reach the leaves of the tree where the decoding is trivial. In view of this principle, the SC algorithm for $\polar(2,4)$, illustrated in \prettyref{fig:polar_dec}, can be mathematically expressed as (in the sequence of steps):
\begin{align*}
    \by \in \reals^4 \longrightarrow  \bL_{\by} &= (\bL_{\by}^{(1)}, \bL_{\by}^{(2)}, \bL_{\by}^{(3)}, \bL_{\by}^{(4)}) \in \reals^4, \\
    \bL_{\bu} & = (\mathrm{LSE}(\bL_{\by}^{(1)}, \bL_{\by}^{(3)}), \mathrm{LSE}(\bL_{\by}^{(2)}, \bL_{\by}^{(4)})) \in \reals^2, \\
    \text{frozen bit} \longrightarrow \mhat_1 &= 0, \\
    L_2 &= \mathrm{LSE}(\bL_{\by}^{(1)}, \bL_{\by}^{(3)}) + \mathrm{LSE}(\bL_{\by}^{(2)}, \bL_{\by}^{(4)}) \in \reals, \\
    \mhat_2 &= \mathbbm{1}\{L_2 < 0\} \in \binary, \\
    \hat{\bu} &= (\mhat_2, \mhat_2) \in \binary^2, \\
    \bL_{\bv} &= (\bL_{\by}^{(1)}, \bL_{\by}^{(2)}) \cdot (-1)^{\hat{\bu}} + (\bL_{\by}^{(3)}, \bL_{\by}^{(4)})  \in \reals^2, \\
    \text{frozen bit} \longrightarrow \mhat_3 &= 0, \\
    L_4 &= \bL_{\bv}^{(1)}+ \bL_{\bv}^{(2)} \in \reals,\\
    \mhat_4 &= \mathbbm{1}\{L_4 <0 \} \in \binary.
\end{align*}

In \prettyref{fig:polar_dec}, the above equations are succinctly represented by two set of arrows: the black solid arrows represent the flow of soft-information from the parent node to the children whereas the green dotted arrows represent the flow of the decoded bit information from the children to the parent. We note that we use a simpler min-sum approximation for the function $\mathrm{LSE}$ that is often used in practice, \ie 
\begin{align*}
    \mathrm{LSE} (a,b) \approx \min(\abs{a}, \abs{b}) \mathrm{sign}(a) \mathrm{sign}(b), \quad a,b \in \reals.
\end{align*}

\section{Interpretation}
\label{app:interpret}

\begin{figure*}[ht]
\centerline{
\subfigure[Noiseless BER goes to zero when the model is better than SC]{
 \includegraphics[width=\columnwidth]{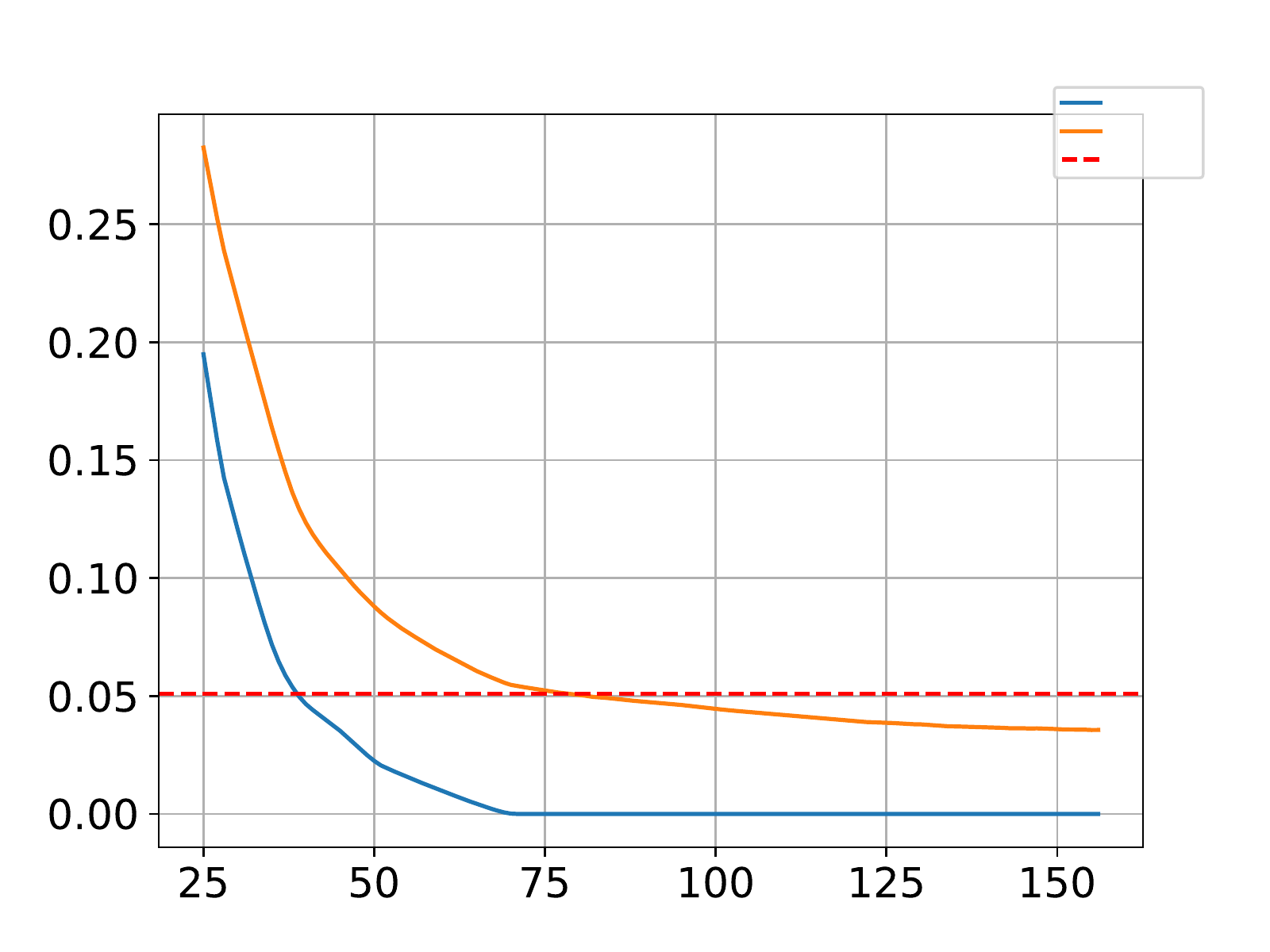}
    \put(-29,156){\fontsize{3.5}{10}\selectfont Noiseless}
 \put(-29,151){\fontsize{3.5}{10}\selectfont CRISP-1dB}
    \put(-29,146){\fontsize{3.5}{10}\selectfont SC-1dB}
    \put(-150,0){\footnotesize Training iterations}
    \put(-240,70){\rotatebox[origin=t]{90}{\footnotesize Bit Error Rate}}
  \label{fig:goodtraining}
}
\hfill
\subfigure[Noiseless BER is high when the model is worse than SC]{
 \includegraphics[width=\columnwidth]{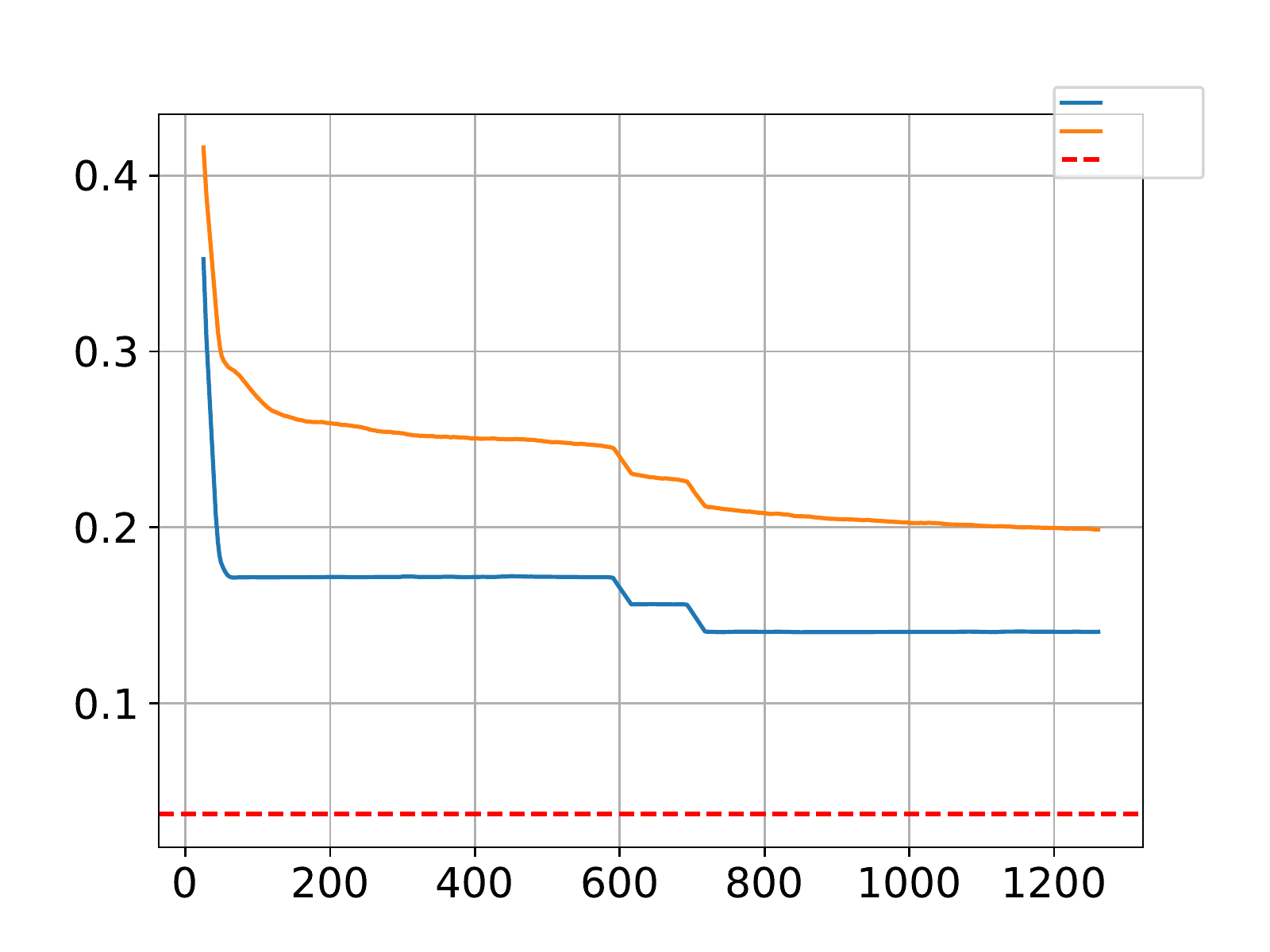}
 \put(-29,156){\fontsize{3.5}{10}\selectfont Noiseless}
 \put(-29,151){\fontsize{3.5}{10}\selectfont CRISP-1dB}
    \put(-29,146){\fontsize{3.5}{10}\selectfont SC-1dB}
    \put(-150,0){\footnotesize Training iterations}
    \put(-230,70){\rotatebox[origin=t]{90}{\footnotesize Bit Error Rate}}
  \label{fig:stucktraining}
}
}
\caption{Evolution of training BER at $1$dB and noiseless BER for CRISP. }
\end{figure*}
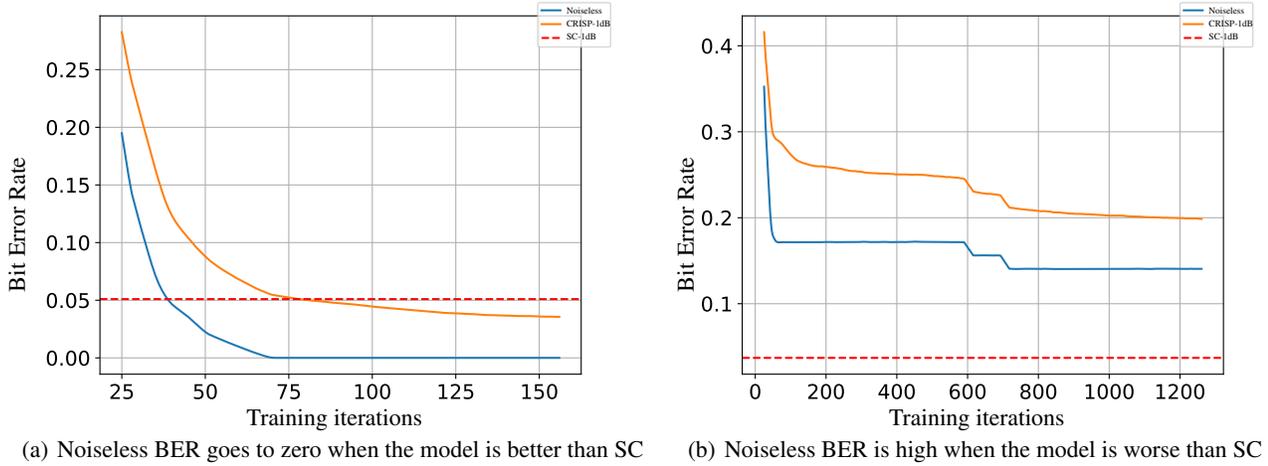

\begin{figure*}[ht]

\subfigure[Bitwise BER for clean and noisy bits]{
  \centering
 \includegraphics[width=\columnwidth]{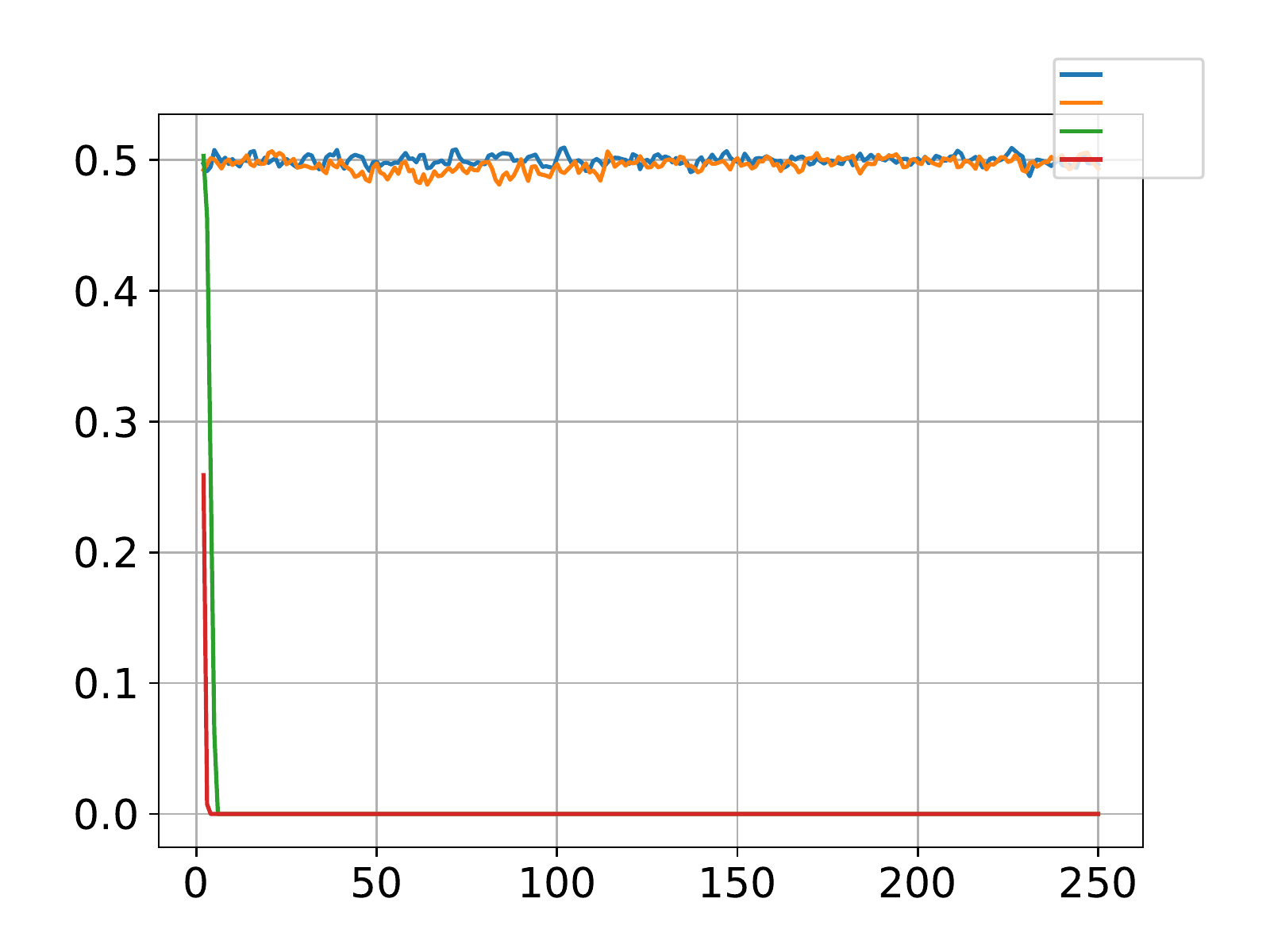}
 \put(-29,161){\fontsize{3.5}{10}\selectfont Noisy bit 1}
  \put(-29,156){\fontsize{3.5}{10}\selectfont Noisy bit 2}
 \put(-29,151){\fontsize{3.5}{10}\selectfont Clean bit 1}
    \put(-29,146){\fontsize{3.5}{10}\selectfont Clean bit 2}
 \put(-140,1){\footnotesize Training iterations}
    \put(-230,70){\rotatebox[origin=t]{90}{\footnotesize Bit Error Rate}}
  \label{fig:bitwiseNoisyVsClean}
}
\hfill
\subfigure[Bitwise contribution to the total BLER]{
  \centering
  \hspace*{-0.3in}
  \includegraphics[width=\columnwidth]{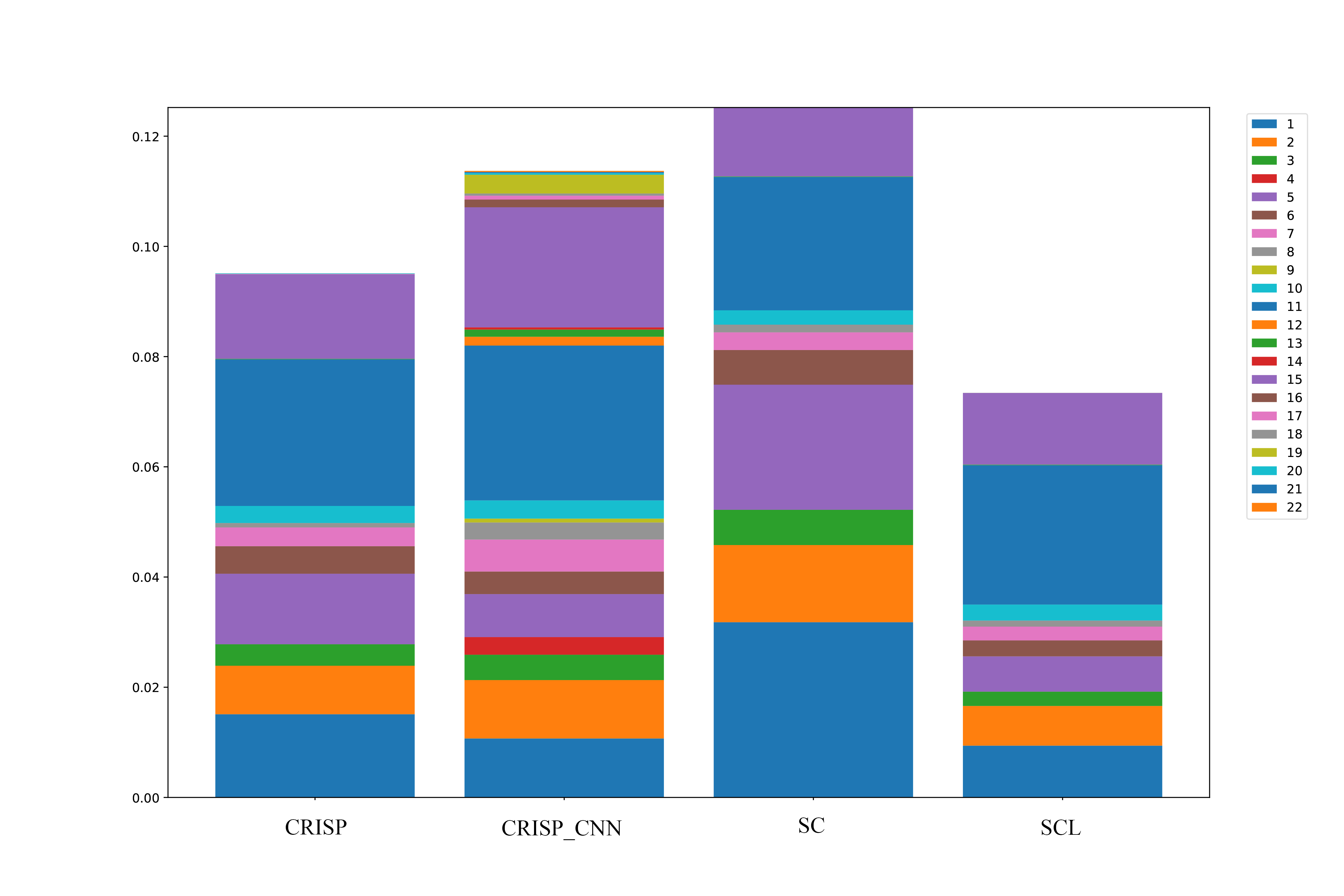}
  \label{fig:condBitwise}
}
\caption{Error analysis for $\polar(64,22)$ : (a) Noiseless BER for the two least reliable bits gets stuck at $0.5$ whereas it converges to $0$ for the two most reliable bits, (b) Contribution of each bit (conditioned on no previous errors) to the BLER. }
\end{figure*}

As discussed in \prettyref{sec:interpret}, we observe that whenever our decoder reaches SC or better performance eventually when training at a specific SNR, its BER (over all the bits) on uncorrupted codewords, noiseless BER, drops to $0$ early on in the training. \prettyref{fig:goodtraining} illustrates this for $\polar(32,16)$. Conversely, if the model gets stuck at a BER worse than that of SC, then we observe that its noiseless BER is also stuck at a non-zero value. This is highlighted in \prettyref{fig:stucktraining} for $\polar(64,32)$. In particular, we notice that the least reliable bits contribute the most to the noiseless BER, while a majority of the cleaner bits have zero individual BER (\prettyref{fig:bitwiseNoisyVsClean}). Viewed from this context, we focus on the noiseless scenario, \ie $\by=\bx$. 
\looseness=-1

As a motivating example, we first consider the $\polar(4,4)$ code. Let $\bma=(m_1,m_2,m_3,m_4) \in \binary^4$ be the block of message bits and $\bx \in \binary^4$ be the codeword. Hence under the L2R curriculum, the subcodes evolve as
\begin{itemize}
    \item $k=1$ : $m_1 \mapsto (m_1,0,0,0) \mapsto \bx = (m_1,0,0,0)$,
    \item $k=2$ : $(m_1,m_2) \mapsto (m_1,m_2,0,0) \mapsto \bx = (m_1\oplus m_2,m_2,0,0)$,
    \item $k=3$ : $(m_1,m_2,m_3) \mapsto (m_1,m_2,m_3,0) \mapsto \bx = (m_1\oplus m_2 \oplus m_3,m_2,m_3,0)$,
    \item $k=4$ : $(m_1,m_2,m_3,m_4) \mapsto (m_1,m_2,m_3,m_4) \mapsto \bx = (m_1\oplus m_2 \oplus m_3 \oplus m_4,m_2\oplus m_4,m_3\oplus m_4,m_4)$.
\end{itemize}
Correspondingly, their optimal bit decoding rules under the MAP evolve as 
\begin{itemize}
    \item $k=1$ : $\by = \bx = (m_1,0,0,0)\mapsto \mhat_1 = x_1$,
    \item $k=2$ : $\by = \bx = (m_1\oplus m_2,m_2,0,0)\mapsto (\mhat_1,\mhat_2) = (x_1\oplus x_2,x_2)$,
    \item $k=3$ : $\by = \bx = (m_1\oplus m_2 \oplus m_3,m_2,m_3,0)\mapsto (\mhat_1,\mhat_2,\mhat_3) = (x_1\oplus x_2 \oplus x_3,x_2,x_3)$,
    \item$k=4$ : $\by = \bx = (m_1\oplus m_2 \oplus m_3 \oplus m_4,m_2\oplus m_4,m_3\oplus m_4,m_4) \mapsto (\mhat_1,\mhat_2,\mhat_3,\mhat_4) = (x_1\oplus x_2 \oplus x_3 \oplus x_4,x_2\oplus x_4,x_3\oplus x_4,x_4)$.
\end{itemize}

\begin{figure*}[ht]

\subfigure[L2R curriculum]{
  \centering
 \includegraphics[width=\columnwidth]{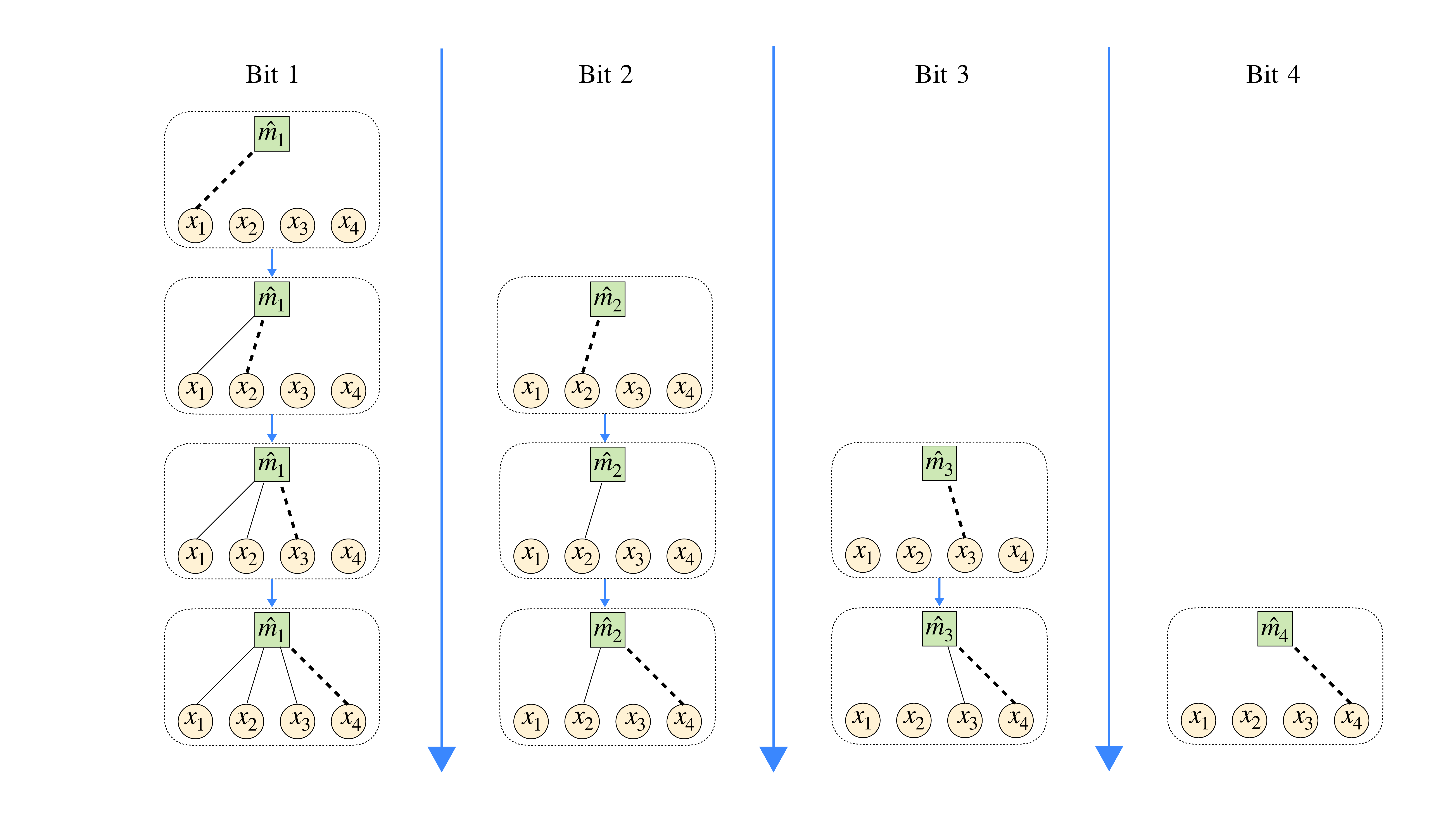}
  \label{fig:bitEvolving44L2R}
}
\hfill
\subfigure[R2L curriculum]{
  \centering
 \includegraphics[width=\columnwidth]{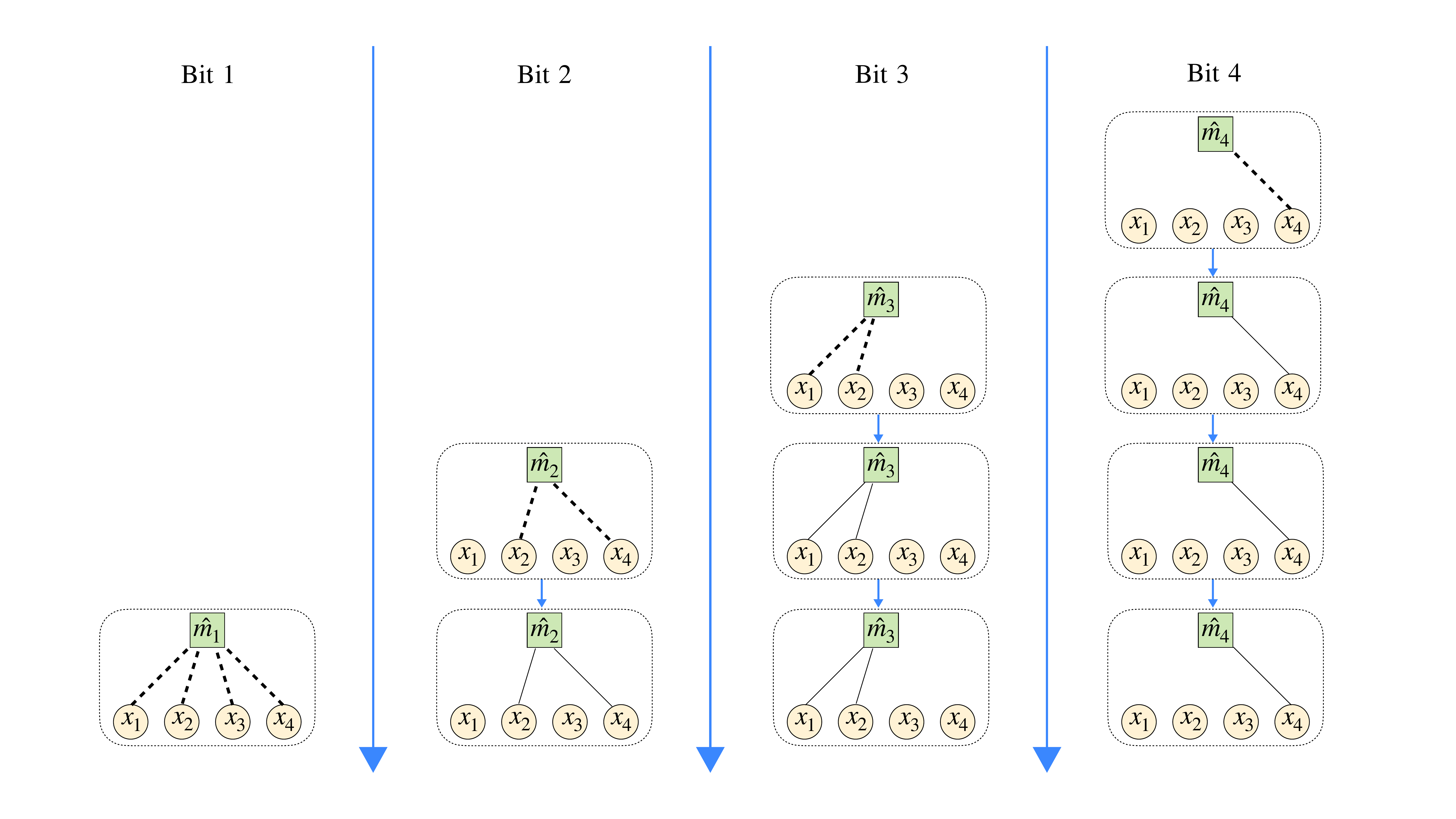}
  \label{fig:bitEvolving44R2L}
}
\caption{Evolution of the MAP decoding rules for L2R and R2L for $\polar(4,4)$. Dotted lines indicate new coded bits being introduced into the decoding rule at each curriculum step.}\label{fig:bitEvolving44}
\end{figure*}

Similarly, we can compute the subcodes and their corresponding decision rules under the R2L curriculum. \prettyref{fig:bitEvolving44} illustrates this evolution for both L2R and R2L. For the least reliable bit $m_1$, we observe that the L2R curriculum reaches the optimal rule more gracefully by including one coordinate $x_i$ at a time while this change for R2L (and no-curriculum) is abrupt, making it harder to learn. We observe the same trend for other bits $m_2,m_3$ and $m_4$. Note that for $\polar(4,4)$, the reliability order is $m_1 < m_2 = m_3 < m_4$ and hence the L2R curriculum is same as N2C and R2L is same as C2N.

For a general $\polar(n,k)$, we can likewise compute the optimal MAP rules using the fact that the mapping $\mathrm{PlotkinTree}:\binary^n \to \binary^n$ is its own inverse, \ie $\bx = \mathrm{PlotkinTree}(\bma) \implies \bma = \mathrm{PlotkinTree}(\bx)$.

\begin{figure*}[ht]

\subfigure[Polar$(32,16)$]{
  \centering
 \includegraphics[width=\columnwidth]{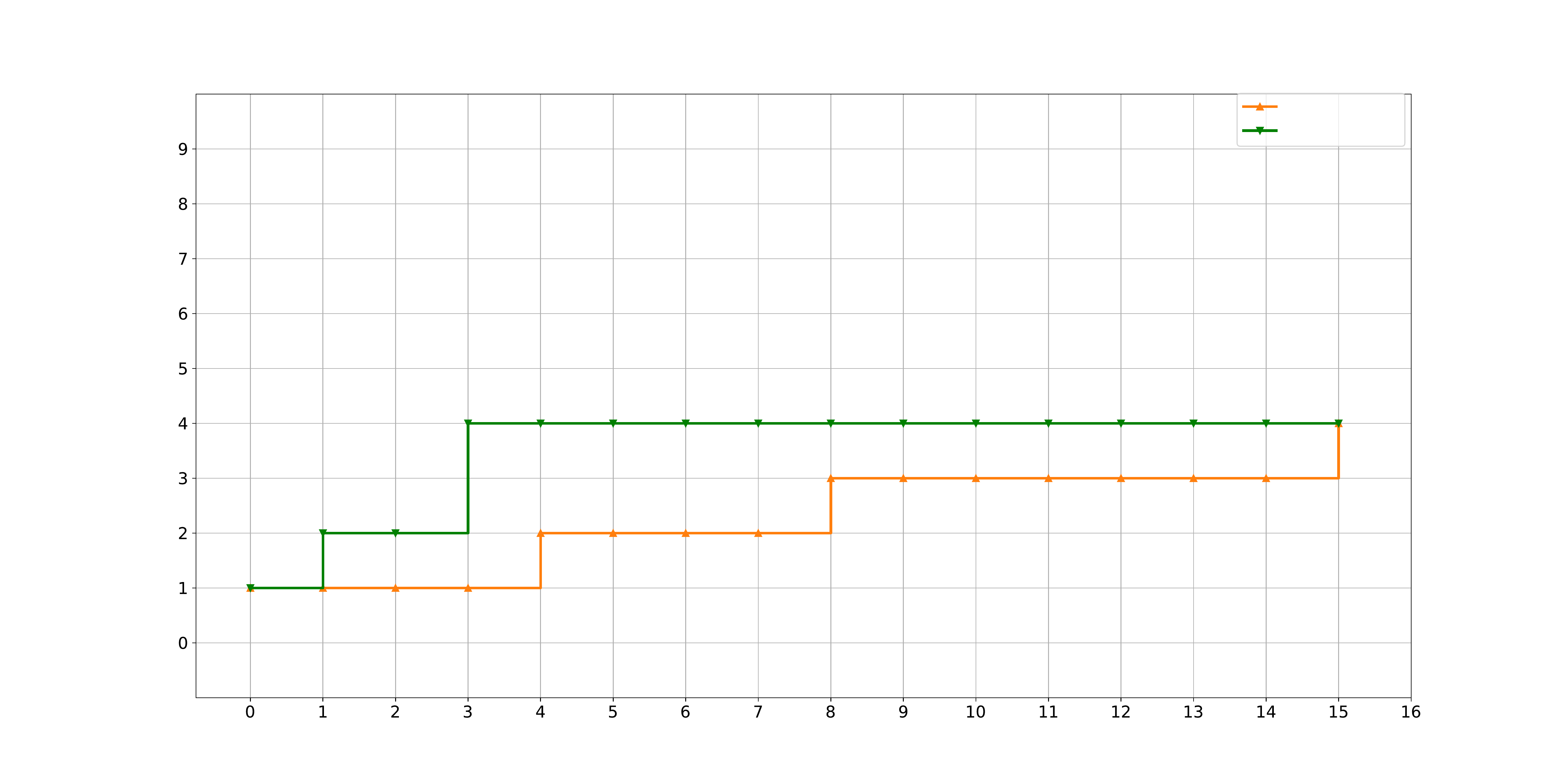}
 \put(-43,99.5){\fontsize{3}{10}\selectfont L2R curriculum}
    \put(-43,96){\fontsize{3}{10}\selectfont R2L curriculum}
    \put(-170,2){\footnotesize Training iterations}
    \put(-220,50){\rotatebox[origin=t]{90}{\footnotesize Learning Difficulty}}
  \label{fig:graph_maxlearndifficulty1632}
}
\hfill
\subfigure[Polar$(64,22)$]{
  \centering
  \hspace*{-0.3in}
 \includegraphics[width=\columnwidth]{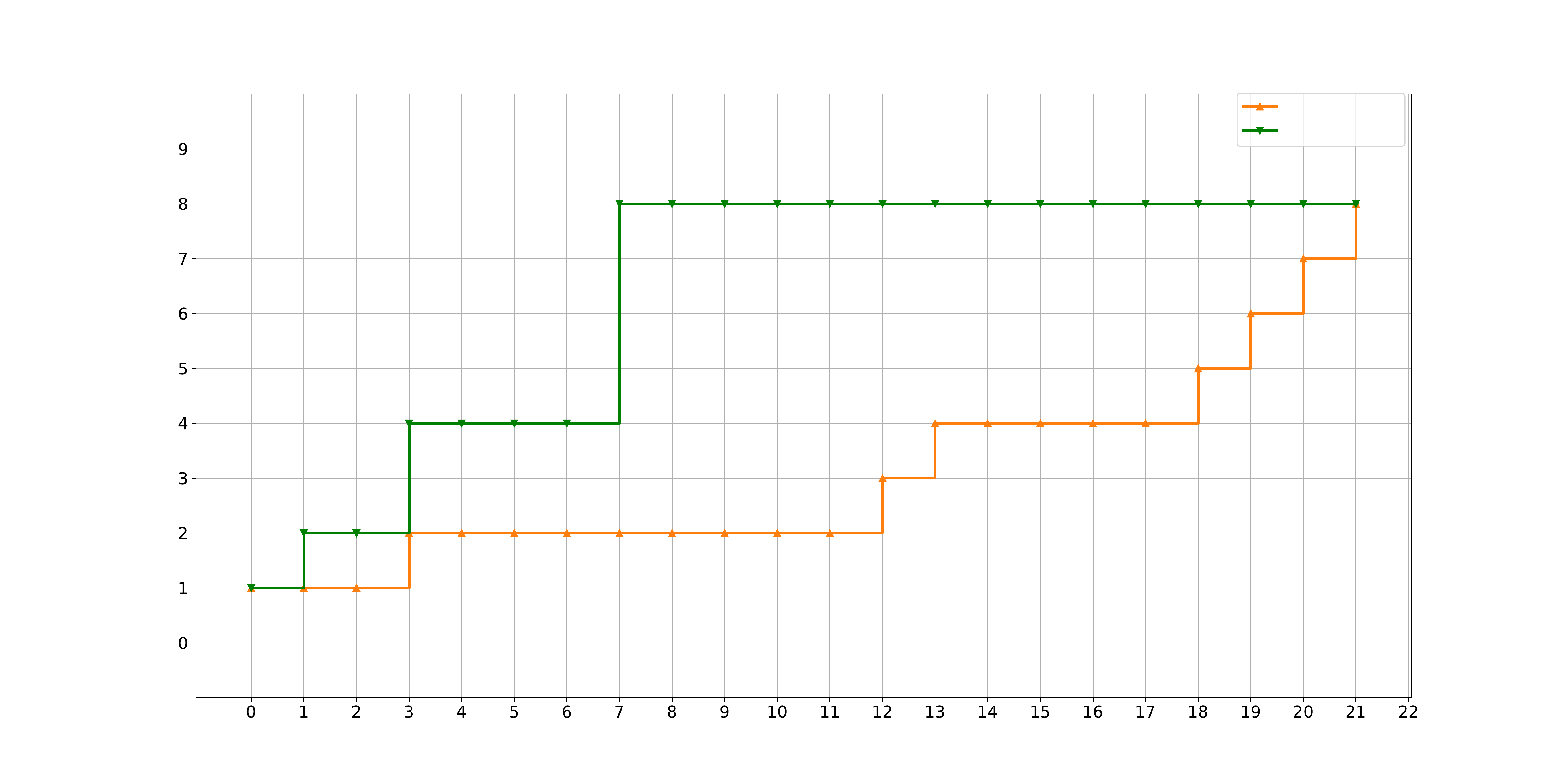}
 \put(-43,99.5){\fontsize{3}{10}\selectfont L2R curriculum}
    \put(-43,96){\fontsize{3}{10}\selectfont R2L curriculum}
    \put(-170,2){\footnotesize Training iterations}
    \put(-220,50){\rotatebox[origin=t]{90}{\footnotesize Learning Difficulty}}
  \label{fig:graph_maxlearndifficulty2264}
}

\caption{Evolution of the learning difficulty for L2R and R2L.}\label{fig:graph_maxlearndifficulty}
\end{figure*}

To concretely compare different curricula, we define the notion of \textit{learning difficulty} for a bit: the number of codeword bits multiplied in its optimal decoding rule. This metric roughly captures the number of multiplication operations a model has to learn at any curriculum step. For example, for $\polar(4,4)$, the learning difficulty for $m_1$ evloves as $1\to 2 \to 3 \to 4$ for the L2R curriculum and as $0 \to 0 \to 0 \to 4$ for the R2L curriculum. \prettyref{fig:graph_maxlearndifficulty} illustrates the evolution of learning difficulty (taking maximum over all bits) for Polar$(32,16)$ and Polar$(64,22)$ codes. We observe here that the jumps in the learning difficulty are larger for R2L, thus indicating a harder transfer than L2R, where it increases smoothly (at most one bit per step).

\begin{figure*}[ht]

\subfigure[$\polar(32,16)$]{
  \centering
  \includegraphics[width=\columnwidth]{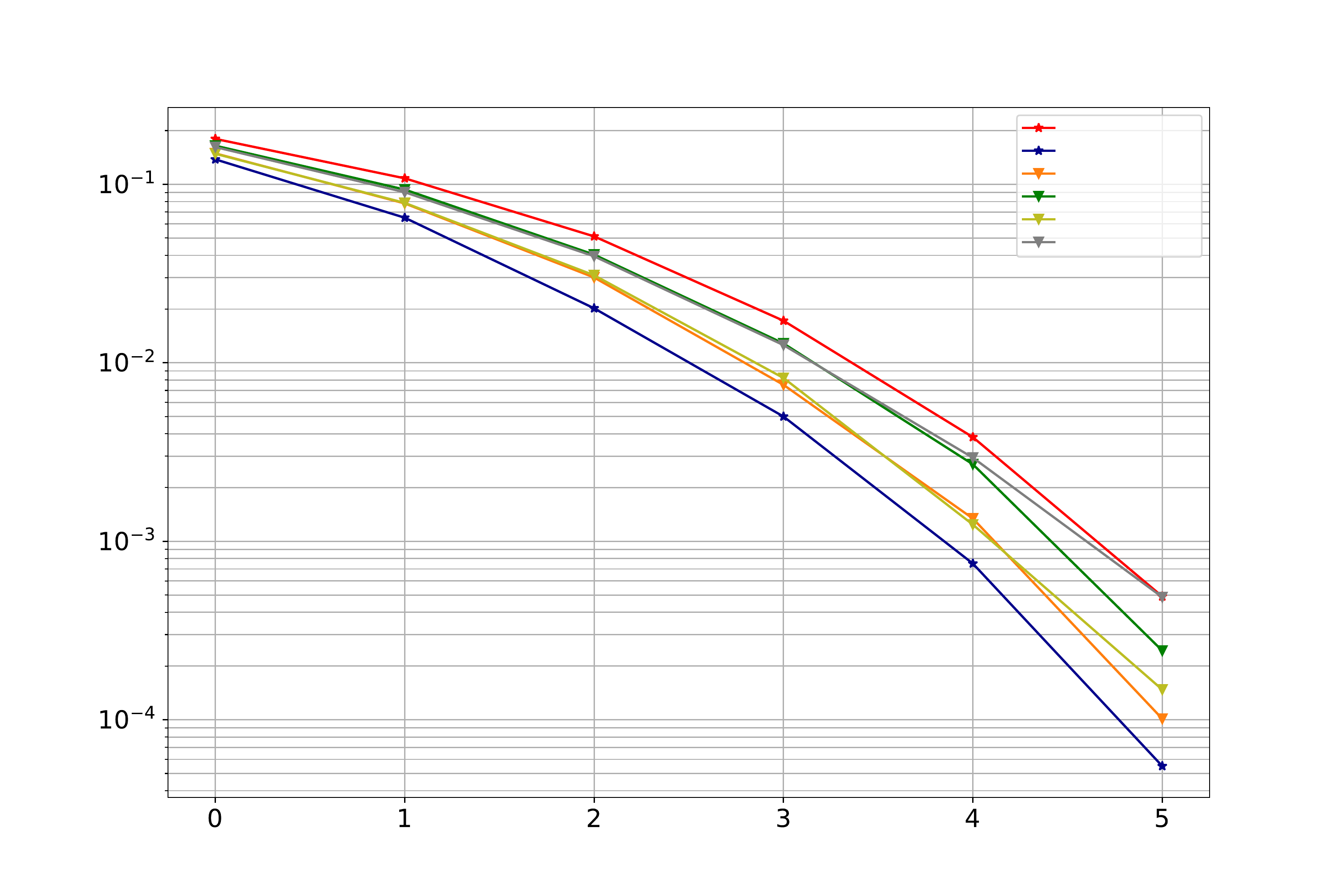}
    \put(-44,133){\fontsize{3}{5}\selectfont SC}
    \put(-44,129){\fontsize{3}{5}\selectfont SC-List, L=32}
    \put(-44,125){\fontsize{3}{5}\selectfont L2R (CRISP)}
    \put(-44,121){\fontsize{3}{5}\selectfont R2L}
    \put(-44,117){\fontsize{3}{5}\selectfont N2C}
    \put(-44,113.5){\fontsize{3}{5}\selectfont C2N}
    \put(-170,2){\footnotesize Signal-to-noise ratio (SNR) [dB]}
    \put(-230,70){\rotatebox[origin=t]{90}{\footnotesize Bit Error Rate}}
  \label{fig:plot_all_curricula1632}
}
\hfill
\subfigure[$\polar(64,22)$]{
  \centering
  \hspace*{-0.3in}
  \includegraphics[width=\columnwidth]{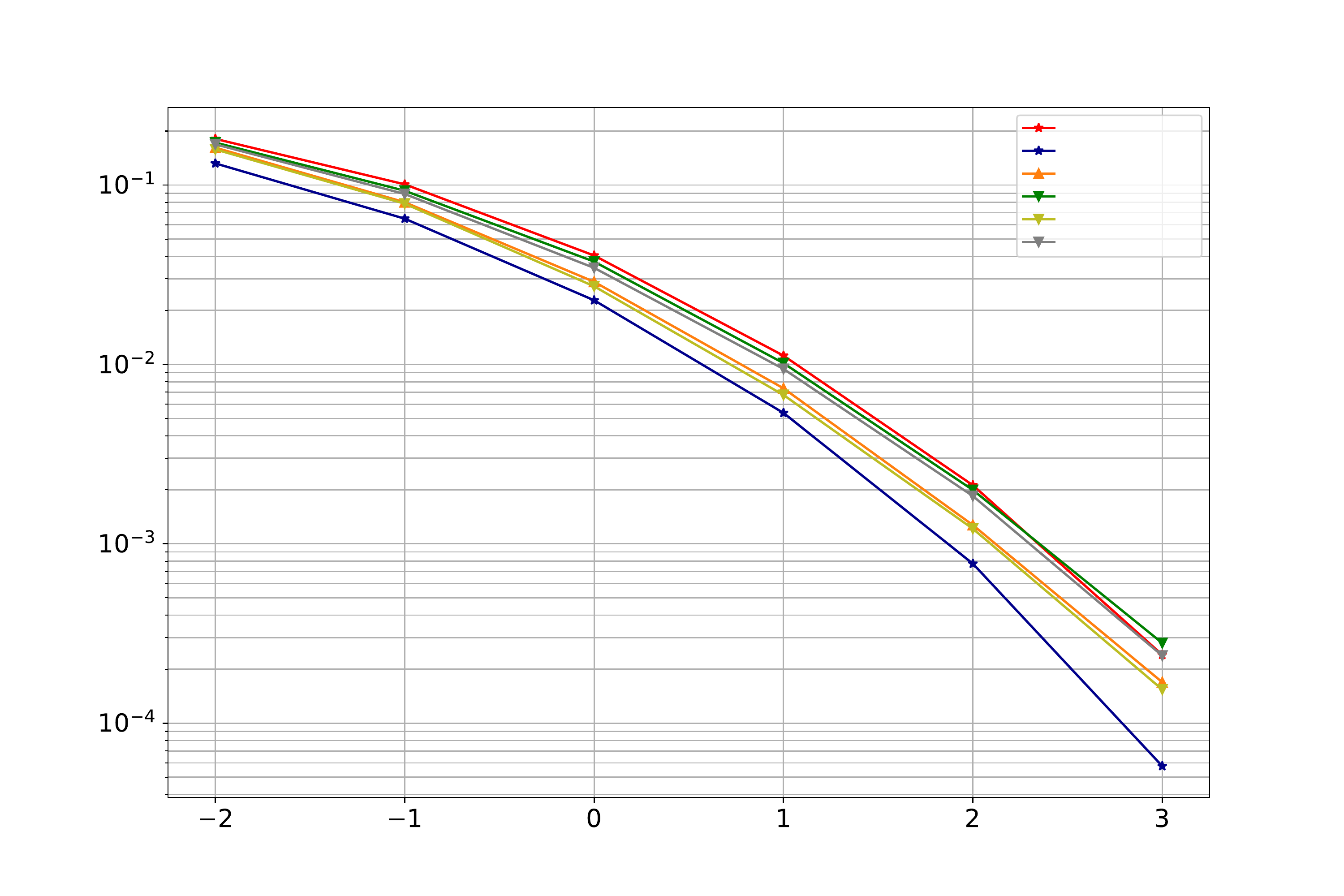}
    \put(-44,133){\fontsize{3}{5}\selectfont SC}
    \put(-44,129){\fontsize{3}{5}\selectfont SC-List, L=32}
    \put(-44,125){\fontsize{3}{5}\selectfont L2R (CRISP)}
    \put(-44,121){\fontsize{3}{5}\selectfont R2L}
    \put(-44,117){\fontsize{3}{5}\selectfont N2C}
    \put(-44,113.5){\fontsize{3}{5}\selectfont C2N}
    \put(-170,2){\footnotesize Signal-to-noise ratio (SNR) [dB]}
    \put(-230,70){\rotatebox[origin=t]{90}{\footnotesize Block Error Rate}}
  \label{fig:plot_all_curricula2264}
}

\caption{Choice of curriculum is crucial to obtain gains over SC. Information-theory guided curricula N2C and C2N are marginally better than the L2R and R2L schemes respectively.}\label{fig:plot_all_curricula}
\end{figure*}

\begin{figure*}[ht]

\subfigure[Noisiest bit]{
  \centering
 \includegraphics[width=\columnwidth]{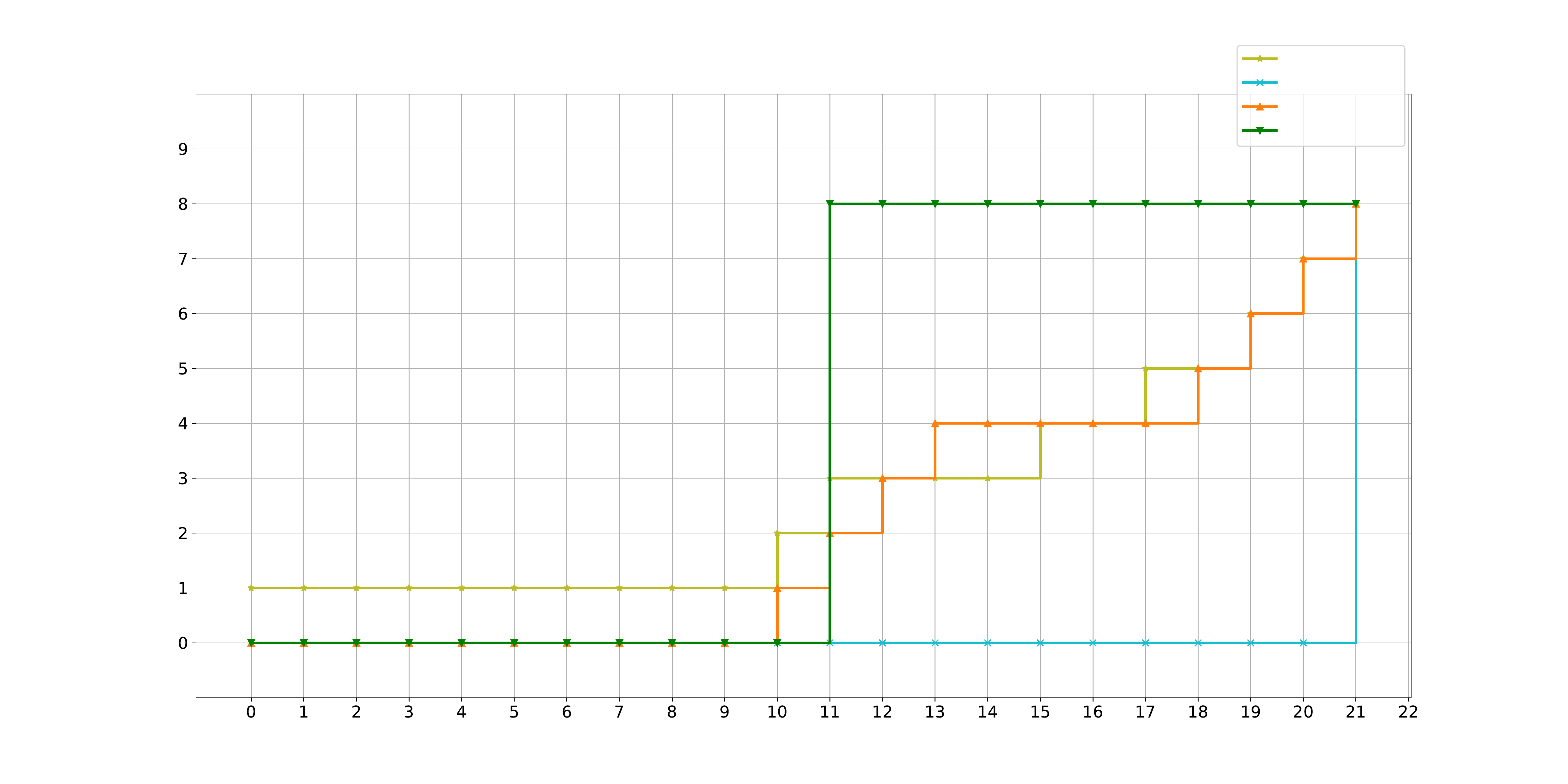}
  \put(-43,107){\fontsize{3}{10}\selectfont N2C curriculum}
    \put(-43,103.5){\fontsize{3}{10}\selectfont C2N curriculum}
 \put(-43,100){\fontsize{3}{10}\selectfont L2R curriculum}
    \put(-43,97){\fontsize{3}{10}\selectfont R2L curriculum}
    \put(-150,2){\footnotesize Training iterations}
    \put(-217,50){\rotatebox[origin=t]{90}{\footnotesize Learning Difficulty}}
  \label{fig:graph_learndifficulty_all1632}
}
\hfill
\subfigure[Maximum over all bits]{
  \centering
 \includegraphics[width=\columnwidth]{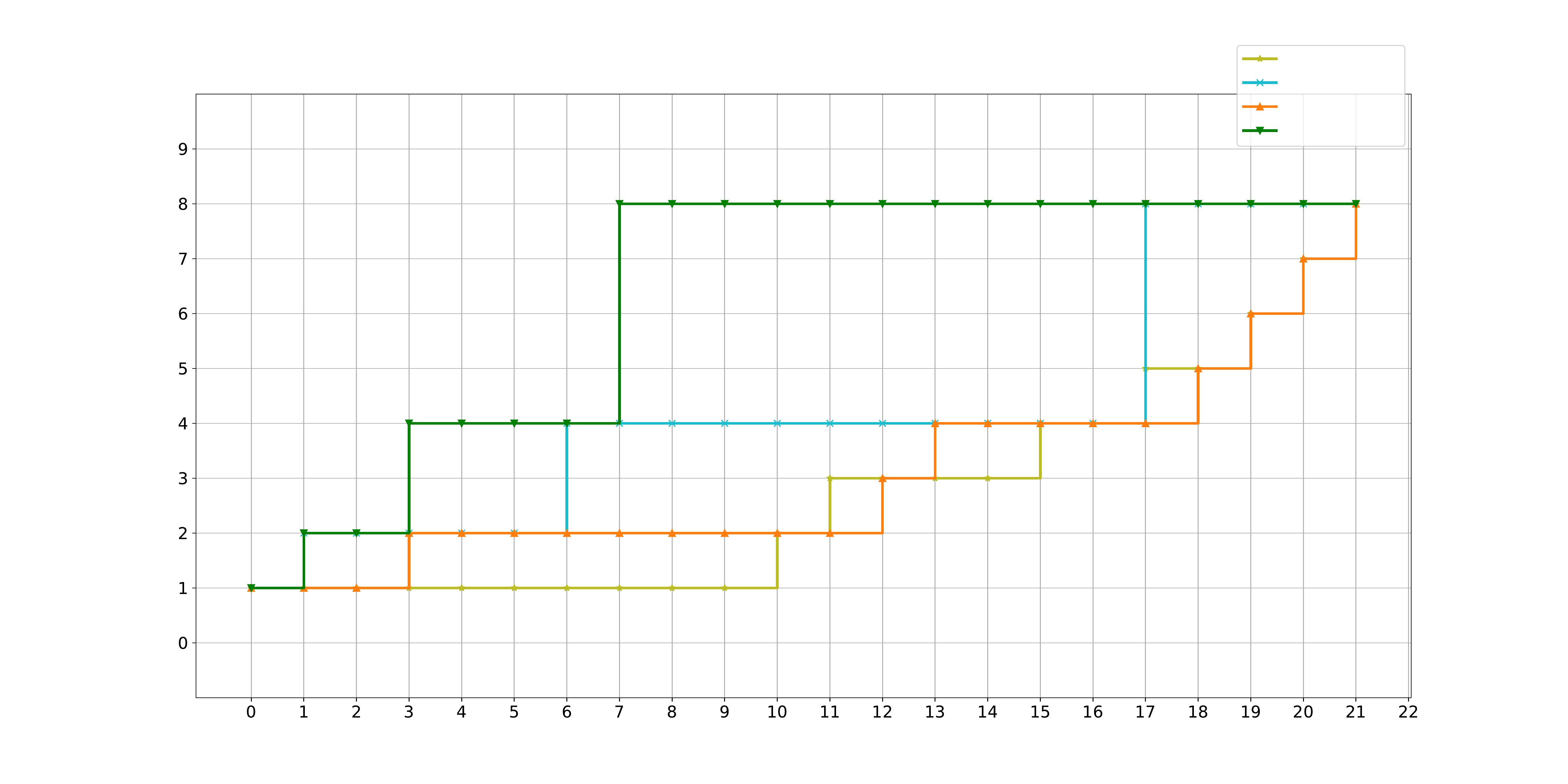}
 \put(-43,107){\fontsize{3}{10}\selectfont N2C curriculum}
    \put(-43,103.5){\fontsize{3}{10}\selectfont C2N curriculum}
 \put(-43,100){\fontsize{3}{10}\selectfont L2R curriculum}
    \put(-43,97){\fontsize{3}{10}\selectfont R2L curriculum}
    \put(-150,2){\footnotesize Training iterations}
    \put(-217,50){\rotatebox[origin=t]{90}{\footnotesize Learning Difficulty}}
  \label{fig:graph_learndifficulty_all2264}
}

\caption{Evolution of learning bit difficulty for different curricula for $\polar(64,22)$.}\label{fig:graph_learndifficulty_all}
\end{figure*}

\prettyref{fig:graph_learndifficulty_all} highlights a similar phenomenon for $\polar(64,22)$ for L2R, R2L, N2C and C2N curricula. We observe that the learning difficulties of the L2R and N2C curricula evolve smoothly while that of R2L and C2N are abrupt. Correspondingly, their final BER reliability performance follows the order N2C $\approx$ L2R < R2L $\approx$ C2N (\prettyref{fig:plot_all_curricula}).

\subsection{Error analysis}\label{app:error_analysis}
To interpret the CRISP decoder, we compare its bitwise error patterns against the SCL decoder. As shown in \prettyref{fig:condBitwise}, we plot the contribution of each bit to the block error rate; we condition on having no previous errors. We observe that the typical error events of CRISP, unlike CRISP\_CNN, closely resemble that of the SCL decoder. This aligns with our expectation since CRISP uses a sequential decoding paradigm similar to that of the successive cancellation framework.

\section{Ablation studies}
\label{app:ablation}

\begin{figure*}[ht]
\subfigure[]{
  \centering
  \includegraphics[width=\columnwidth]{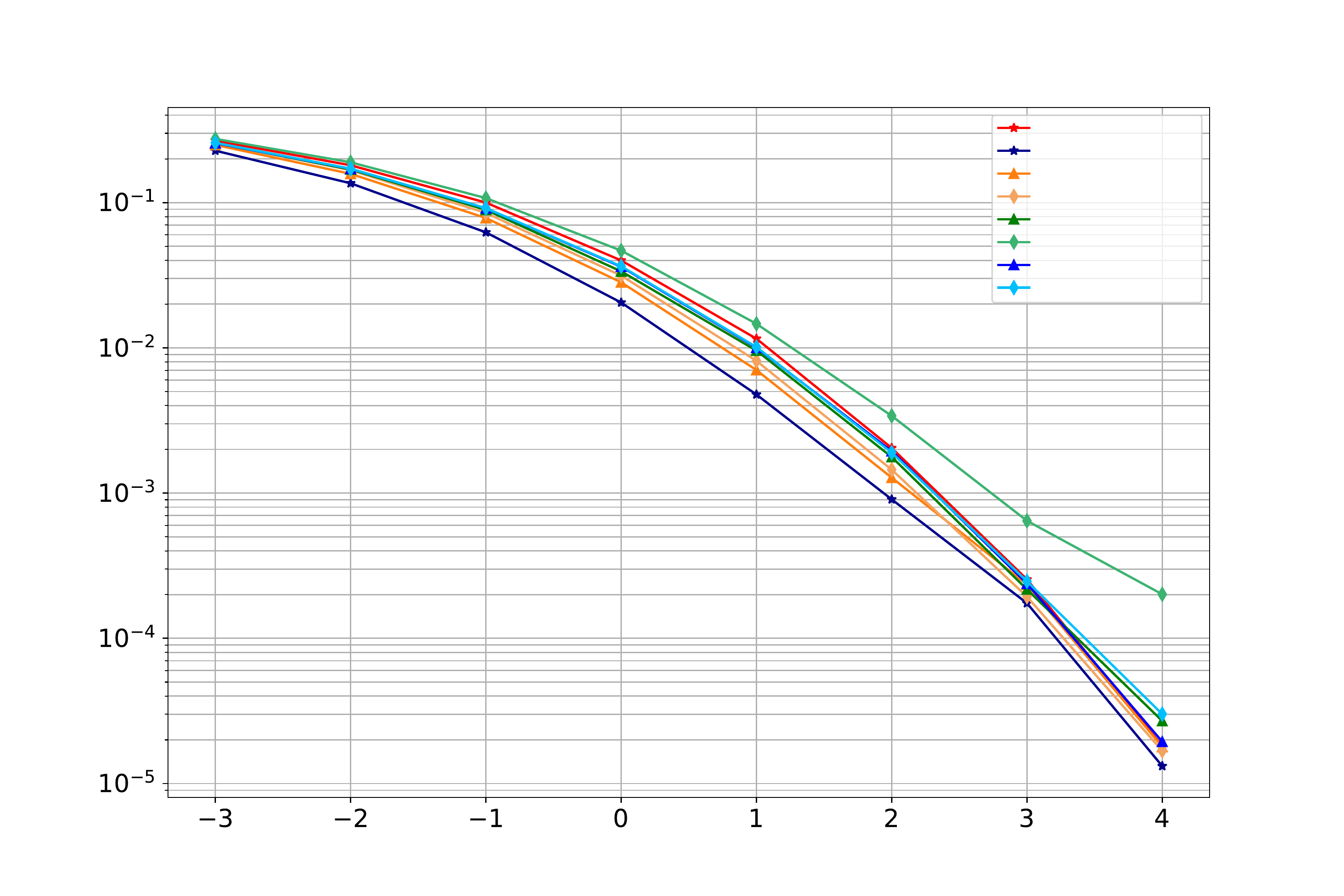}
    \put(-50,133){\fontsize{3}{5}\selectfont SC}
    \put(-50,129){\fontsize{3}{5}\selectfont SC-List, L=32}
    \put(-50,125){\fontsize{3}{5}\selectfont L2R, \quad h=512}
    \put(-50,121){\fontsize{3}{5}\selectfont L2R, \quad h=256}
    \put(-50,117){\fontsize{3}{5}\selectfont R2L, \quad h=512}
    \put(-50,113){\fontsize{3}{5}\selectfont R2L, \quad h=256}
    \put(-50,109.5){\fontsize{3}{5}\selectfont w/o C, \HS\HS h=512}
    \put(-50,105.5){\fontsize{3}{5}\selectfont w/o C, \HS\HS h=256}
    \put(-170,2){\footnotesize Signal-to-noise ratio (SNR) [dB]}
    \put(-230,70){\rotatebox[origin=t]{90}{\footnotesize Bit Error Rate}}
  \label{fig:modelsize_ablation}
}
\hfill
\subfigure[]{
  \centering
 \includegraphics[width=\columnwidth]{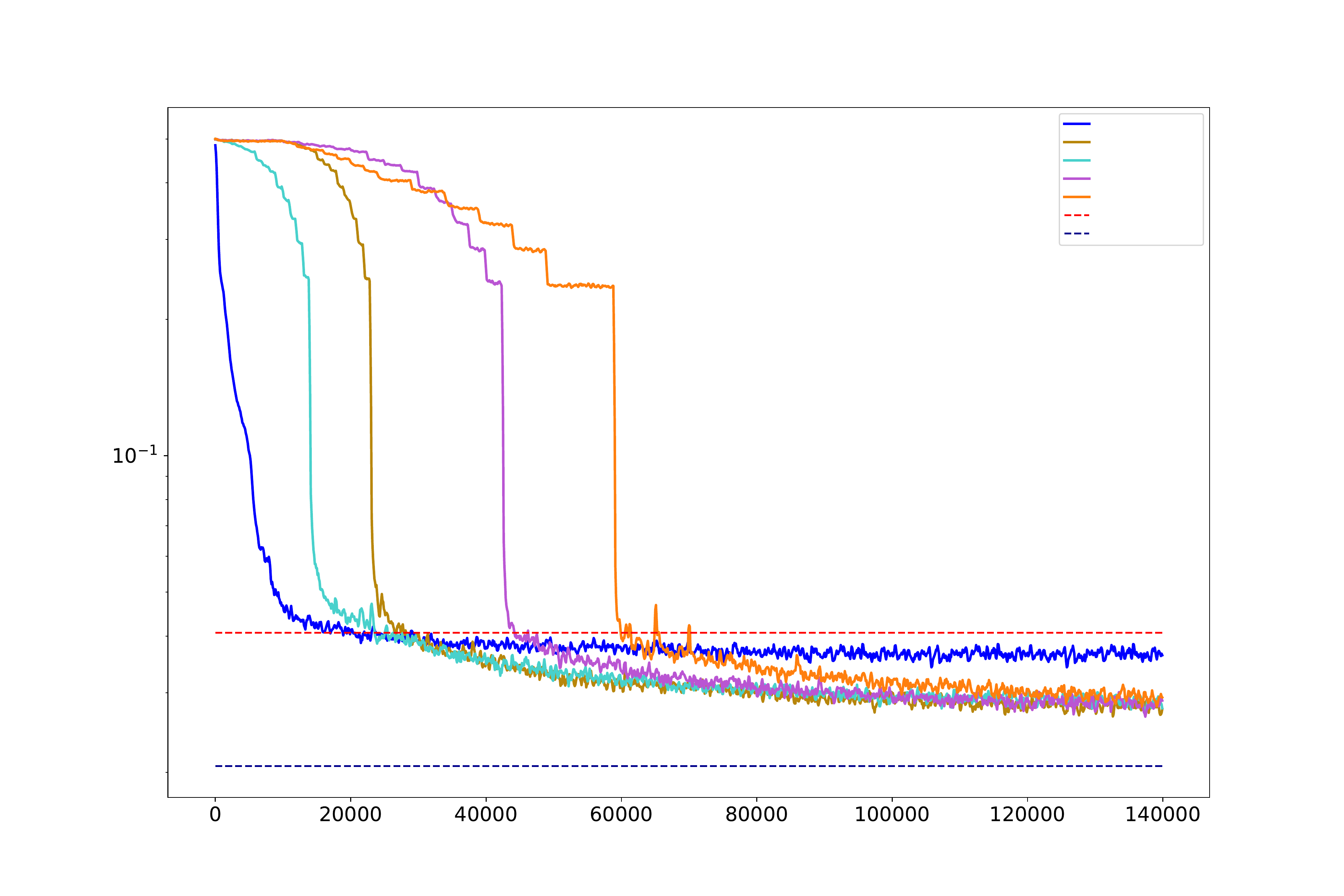}
 \put(-40,134){\fontsize{2}{5}\selectfont No curriculum}
    \put(-40,131){\fontsize{2}{5}\selectfont CRISP schedule 1}
    \put(-40,128){\fontsize{2}{5}\selectfont CRISP schedule 2}
    \put(-40,125){\fontsize{2}{5}\selectfont CRISP schedule 3}
    \put(-40,122){\fontsize{2}{5}\selectfont CRISP schedule 4}
    \put(-40,118.5){\fontsize{2}{5}\selectfont SC}
    \put(-40,115){\fontsize{2}{5}\selectfont SC-List (MAP)}
    \put(-150,0){\footnotesize Training iteration}
    \put(-230,70){\rotatebox[origin=t]{90}{\footnotesize Bit Error Rate}}
\label{fig:schedule}
}

\caption{Ablation plots: (a) Choosing the right curriculum is critical when model size is smaller, (b) The number of iterations to train CRISP on each subcode using the L2R/N2C curriculum are not critical to the final performance achieved.}
\end{figure*}

Recall that our CRISP decoder consists of the sequential RNN ($512$-dim hidden state) trained with the L2R curriculum. To understand the contribution of each of these components to its gains over SC, we did the following ablation experiments for $\polar(64,22)$ code.
\subsection{Effect of model size} We fix the decoder to be GRU and consider different model sizes via the hidden state size $h \in \{256,512\}$, and different curricula among $\{ \text{L2R, R2L, Without curriculum (w/o C)}\}$. \prettyref{fig:modelsize_ablation} demonstrates that the accuracy gains of the L2R curriculum are more pronounced for \emph{smaller} models ($h = 256$). On the other hand, we observe minimal reliabilty gains for L2R with large models ($h = 512$). We also tried other sequential architectures such as LSTMs \citep{lstm_original} and Transformers \citep{radford2019language}, but found GRUs to be the best (\prettyref{app:additional}).
\subsection{Sequential vs. block decoding}\label{app:block_decoding} The sequential GRU architecture for CRISP is inspired in part by the sequential SC algorithm. Alternatively, we also design CRISP\_CNN, a block decoder parameterized by $1$D Convolutional Neural Networks (CNNs). CRISP\_CNN estimates all the information bits $m_i$ in one shot given $\by$. 
Similar to sequential decoders, curriculum learning; in particular, the L2R scheme works the best for block decoding in achieving near-MAP reliability. 

\begin{figure*}[ht]
\subfigure[]{
  \centering
  \includegraphics[width=\columnwidth]{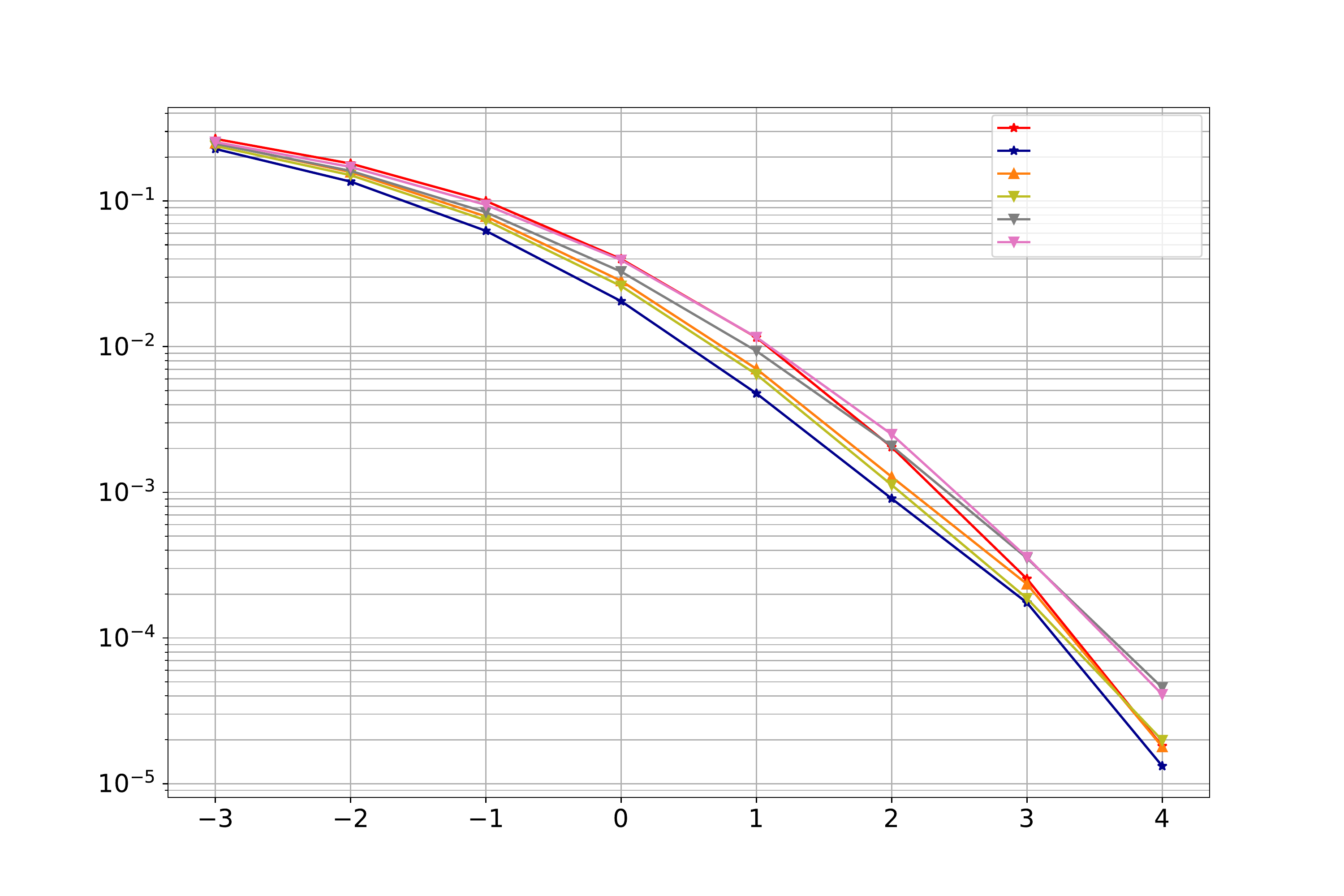}
    \put(-50,133){\fontsize{3}{5}\selectfont SC}
    \put(-50,129){\fontsize{3}{5}\selectfont SC-List, L=32}
    \put(-50,125){\fontsize{3}{5}\selectfont CRISP}
    \put(-50,121){\fontsize{3}{5}\selectfont CNN - L2R}
    \put(-50,117){\fontsize{3}{5}\selectfont CNN - R2L}
   \put(-50,113){\fontsize{3}{5}\selectfont CNN - w/o C}
    \put(-170,2){\footnotesize Signal-to-noise ratio (SNR) [dB]}
    \put(-230,70){\rotatebox[origin=t]{90}{\footnotesize Bit Error Rate}}
  \label{fig:seq_block}
}
\hfill
\subfigure[]{
  \centering
  \hspace*{-0.3in}
  \includegraphics[width=\columnwidth]{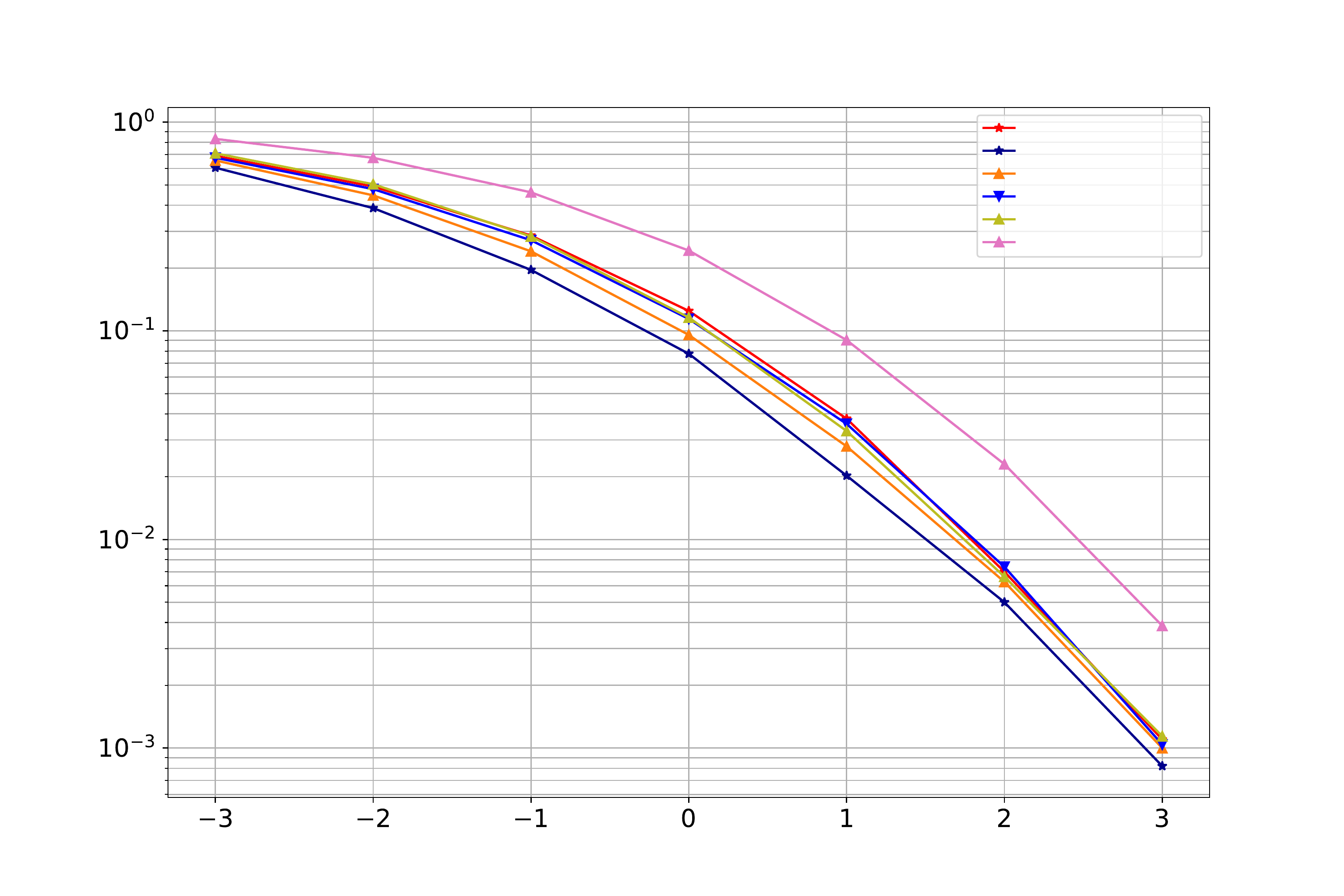}
    \put(-53,133){\fontsize{3}{5}\selectfont SC}
    \put(-53,129){\fontsize{3}{5}\selectfont SC-List, L=32}
    \put(-53,125){\fontsize{3}{5}\selectfont CRISP}
    \put(-53,121){\fontsize{3}{5}\selectfont RNN - No curriculum}
    \put(-53,117){\fontsize{3}{5}\selectfont CNN - L2R}
    \put(-53,113){\fontsize{3}{5}\selectfont CNN - No curriculum}
    \put(-170,2){\footnotesize Signal-to-noise ratio (SNR) [dB]}
    \put(-230,70){\rotatebox[origin=t]{90}{\footnotesize Block Error Rate}}
    \label{fig:bler_seq_block}
}

\caption{a) CNN decoder achieves near-MAP BER performance with L2R curriculum. \\ b) CRISP achieves near-MAP BLER for Polar$(64,22)$. CNN is slightly worse. }
\end{figure*}

\begin{figure*}[ht]

\subfigure[]{
  \centering
  \includegraphics[width=\columnwidth]{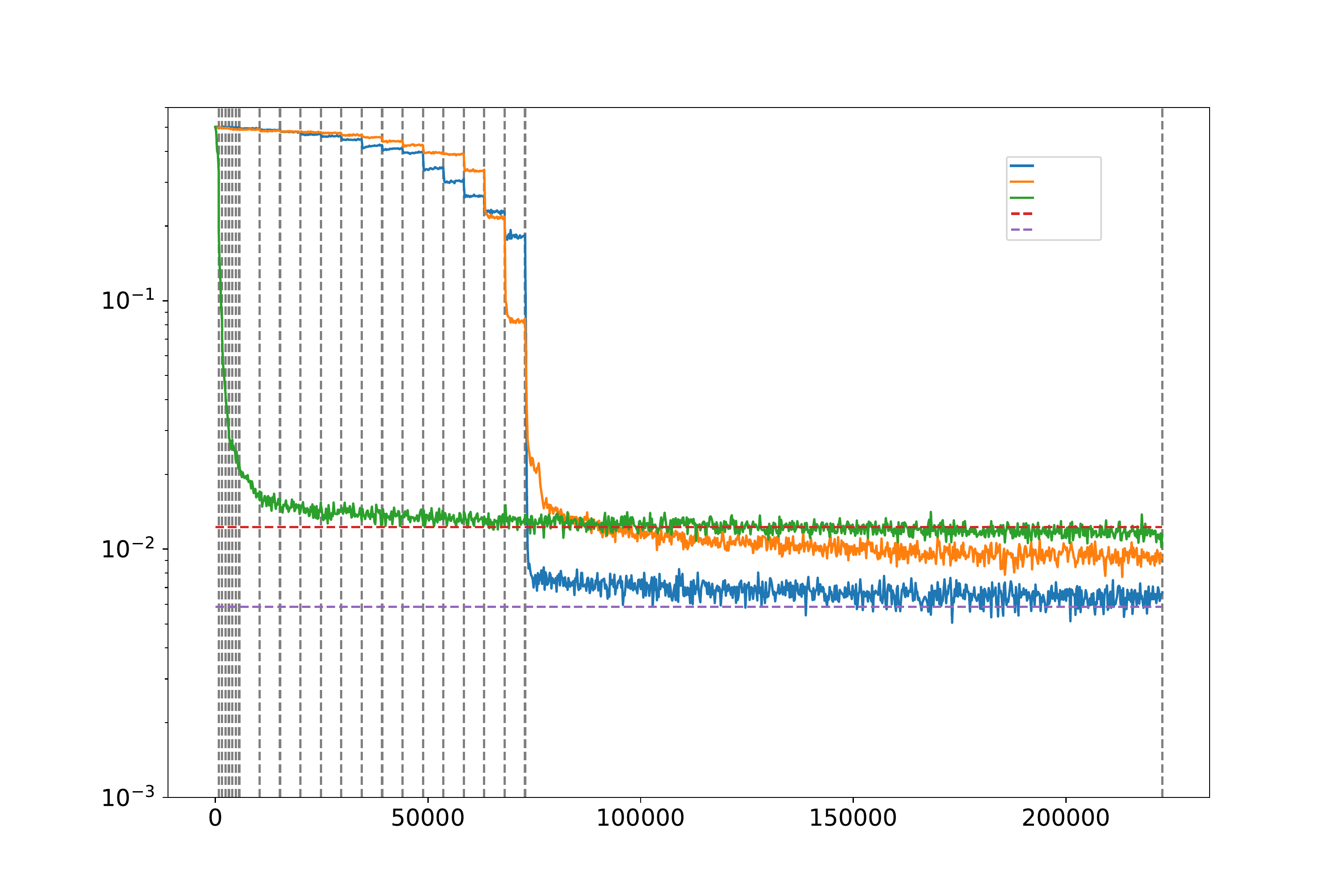}
 \put(-53,127){\fontsize{1.5}{5}\selectfont N2C curriculum}
    \put(-53,124){\fontsize{1.5}{5}\selectfont C2N curriculum}
    \put(-53,121){\fontsize{1.5}{5}\selectfont No curriculum}
    \put(-53,118.5){\fontsize{1.5}{5}\selectfont SC}
    \put(-53,116){\fontsize{1.5}{5}\selectfont SC-List (MAP)}
    \put(-150,0){\footnotesize Training iteration}
    \put(-230,70){\rotatebox[origin=t]{90}{\footnotesize Bit Error Rate}}
    \label{fig:cnn_plots}
}
\hfill
\subfigure[]{
  \centering
  \includegraphics[width=\columnwidth]{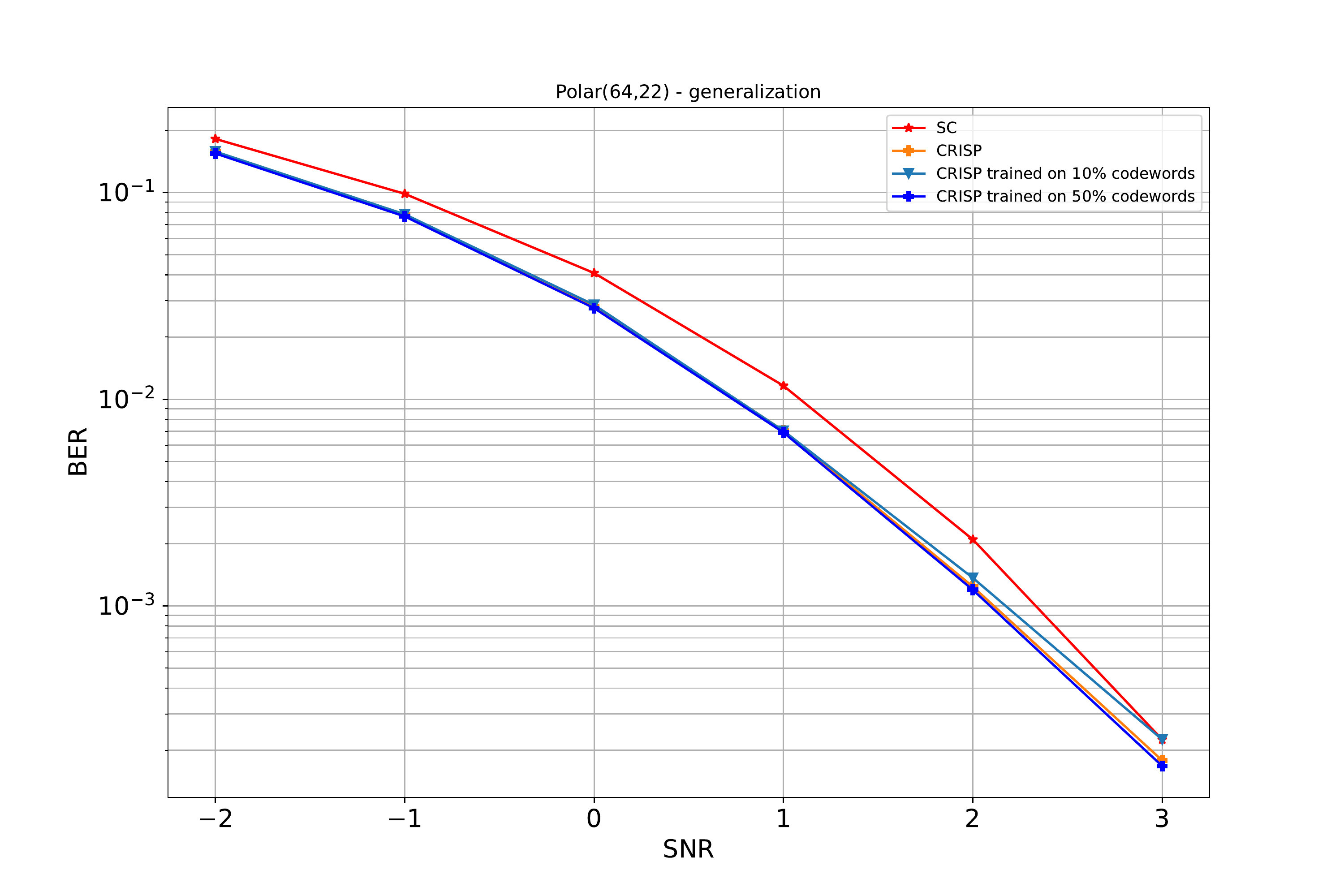}
    \label{fig:crisp_generalization}
}
\caption{(a) L2R curriculum helps CNN to achieve near-optimal reliability}

\end{figure*}

\prettyref{fig:bler_seq_block} compares RNNs and CNNs in terms of BLER for $\polar(64,22)$ with L2R and R2L curricula. We observe that RNN-based decoders (CRISP) are more reliable in terms of BLER than CNNs; in contrast, RNNs and CNNs achieve similar BER performance (\prettyref{fig:seq_block}). Further, we observe that the error patterns corresponding to bitwise contribution to the total BLER for the RNN model resemble that of SC-List, as opposed to CNN models (\prettyref{fig:condBitwise}).
 We show the evolution of validation BER for CNN training in \prettyref{fig:cnn_plots}. We see that the C2N curriculum performs worse than the N2C curriculum. 
\looseness=-1

\section{Additional results}
\label{app:additional}
We present our additional results on the Polar code family with various decoding architectures such as CNNs and transformers, with BLER reliability, and for longer blocklengths ($n=128$). Recall that the CRISP decoder uses the GRU-based RNN (\prettyref{fig:crisp_arch}) trained with the L2R curriculum.

\subsection{Additional results for polar codes}
\label{app:additional_polar}

\subsubsection{Robustness to non-AWGN noise}\label{app:robustness}
\begin{figure*}[ht]
\subfigure[$\polar(64,22)$]{
  \centering
  \includegraphics[width=\columnwidth]{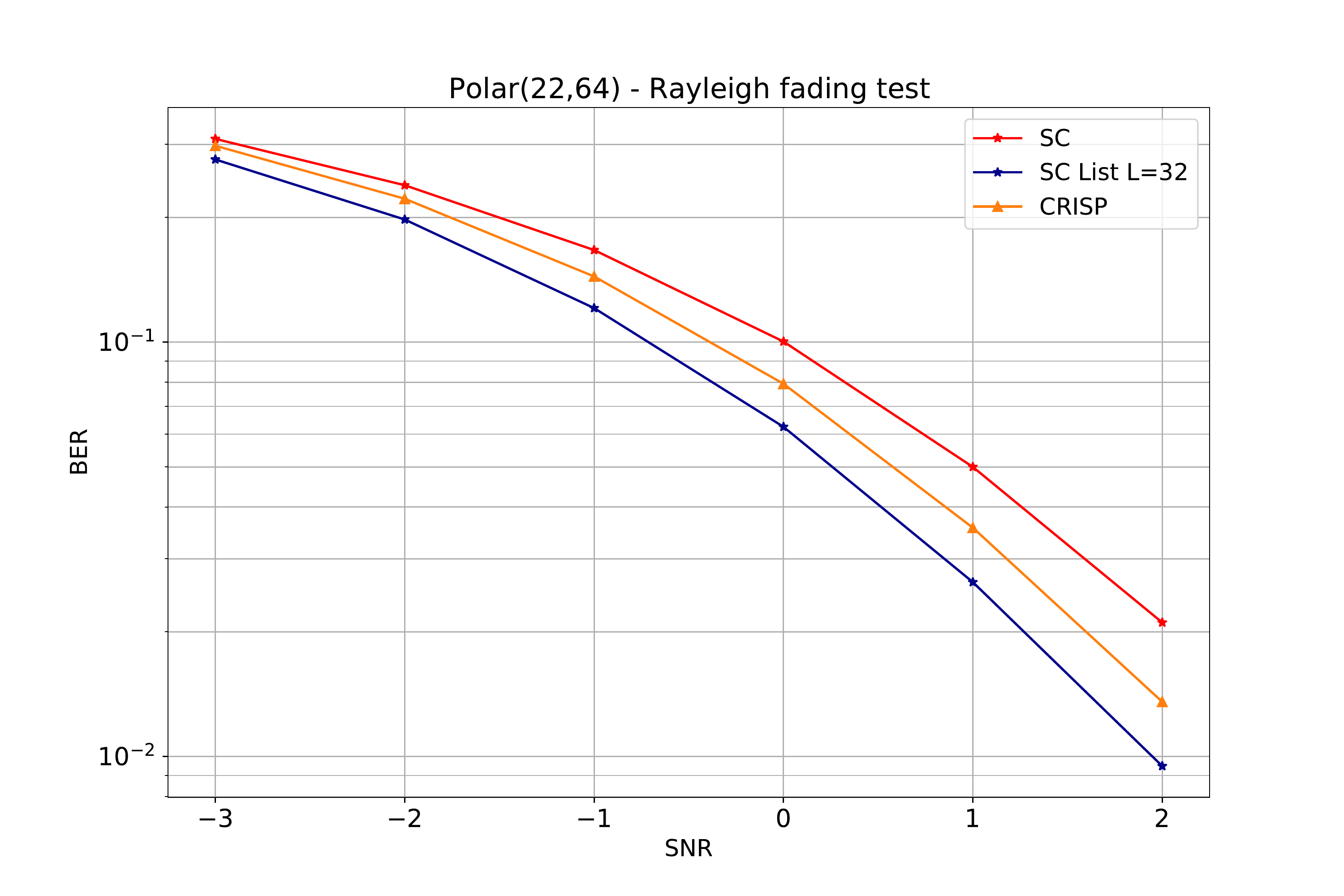}
  \label{fig:fading2264}
}
\hfill
\subfigure[$\polar(32,16)$]{
  \centering
  \includegraphics[width=\columnwidth]{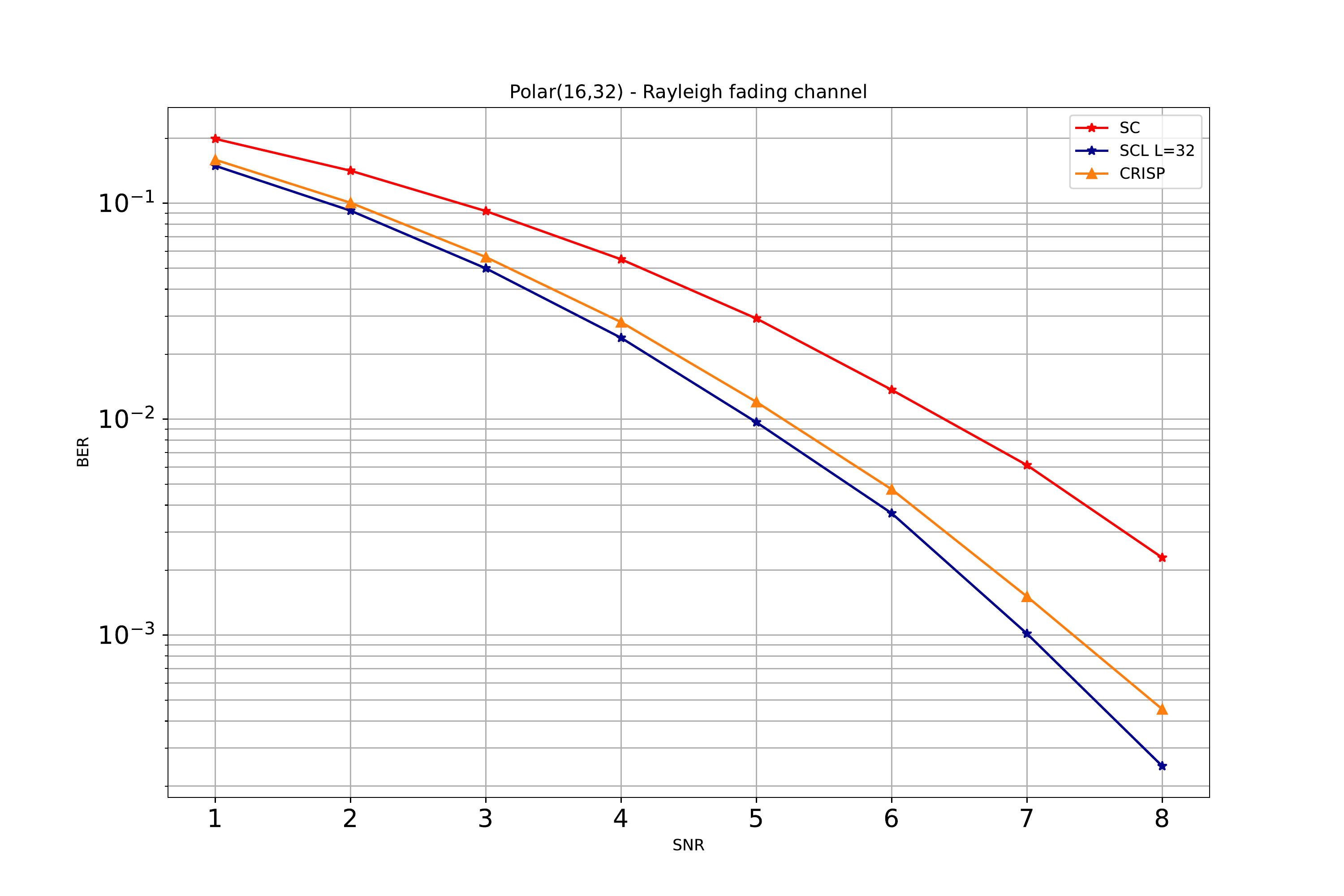}
  \label{fig:fading1632}
}
\caption{CRISP achieves good reliability on Rayleigh fading channels}\label{fig:fading}
\end{figure*}

\begin{figure*}[ht]
\subfigure[$\polar(64,22)$]{
  \centering
  \includegraphics[width=\columnwidth]{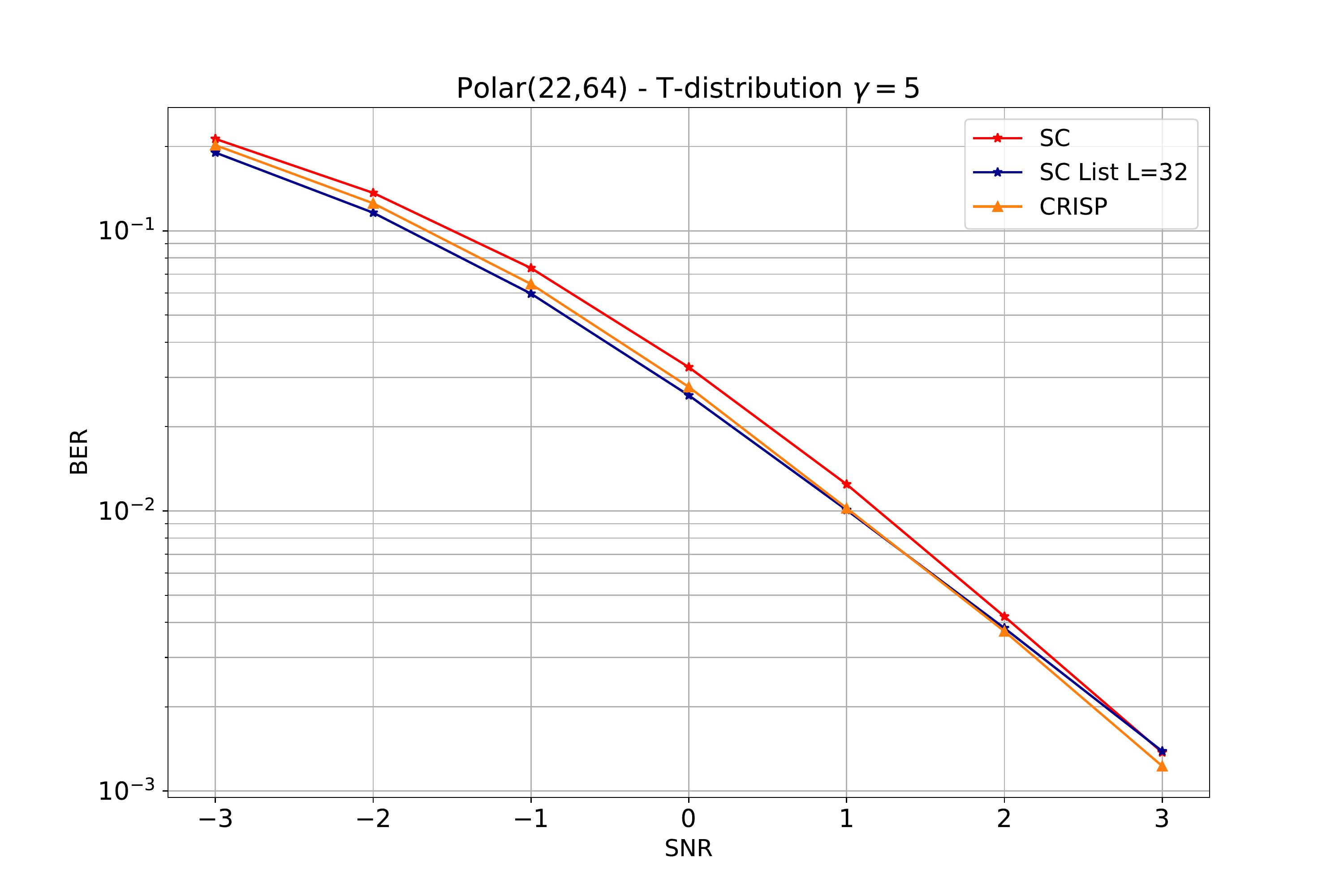}
  \label{fig:tdist2264}
}
\hfill
\subfigure[$\polar(32,16)$]{
  \centering
  \includegraphics[width=\columnwidth]{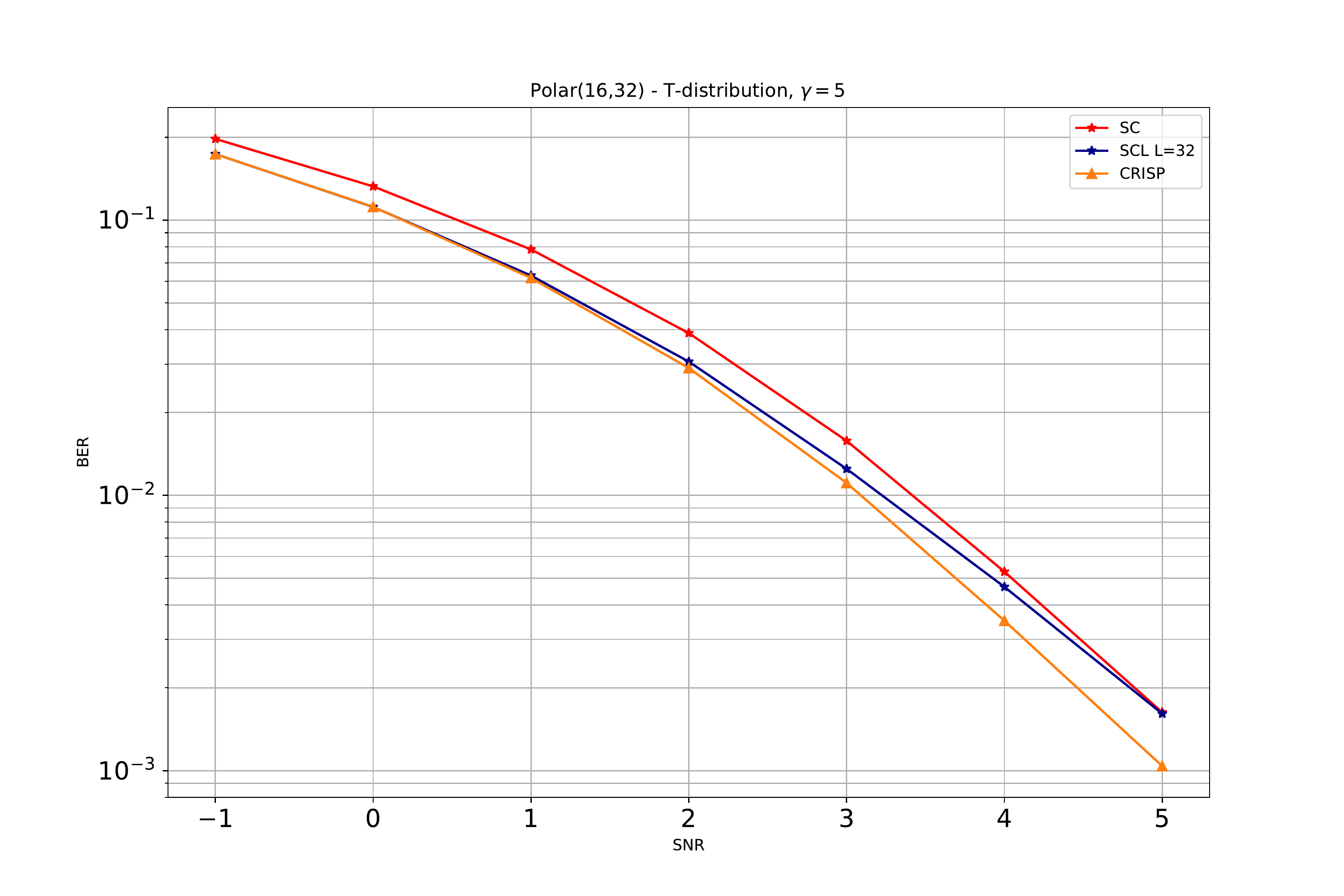}
  \label{fig:tdist1632}
}
\caption{CRISP matches SCL reliability on T-distributed channels}\label{fig:tdist}
\end{figure*}

In this section we evaluate CRISP trained on AWGN on non-AWGN settings. 
We test CRISP on a Rayleigh fading channel, and T-distributed noise.
As shown in \prettyref{fig:fading}, CRISP retains its gains when tested on a Rayleigh fading channel. Further, as demonstrated in \prettyref{fig:tdist}, CRISP is very robust to T-distributed noise and marginally outperforms SCL at higher SNRs. These experiments suggest that the CRISP decoder inherits the robustness inherent to nearest neighbor decoding even though this was not explicitly featured in the training – this intuition is further justified by our experiments showing that the typical error events of CRISP match that of the optimal SCL (MAP) decoder (\prettyref{app:error_analysis})

\subsubsection{Generalization to unseen codewords}

We emphasize that the CRISP decoder does not rely on memorizing codewords to achieve its high performance. We demonstrate this on a $\polar(64,22)$ codebook consisting of $2^22$
 codewords, where we randomly selected subsets comprising of the codebook as training data, and held out the remaining codewords for evaluation. We observe in \prettyref{fig:crisp_generalization} that the CRISP decoder trained on a limited set of codewords does not lead to performance deterioration. This shows that our training method learns the structural patterns inherent to Polar and PAC codes, rather than just memorizing the codewords.

\subsubsection{CRISP for CRC-Polar codes}
\begin{figure*}[ht]
\subfigure[$\polar(64,22)$]{
  \centering
  \includegraphics[width=\columnwidth]{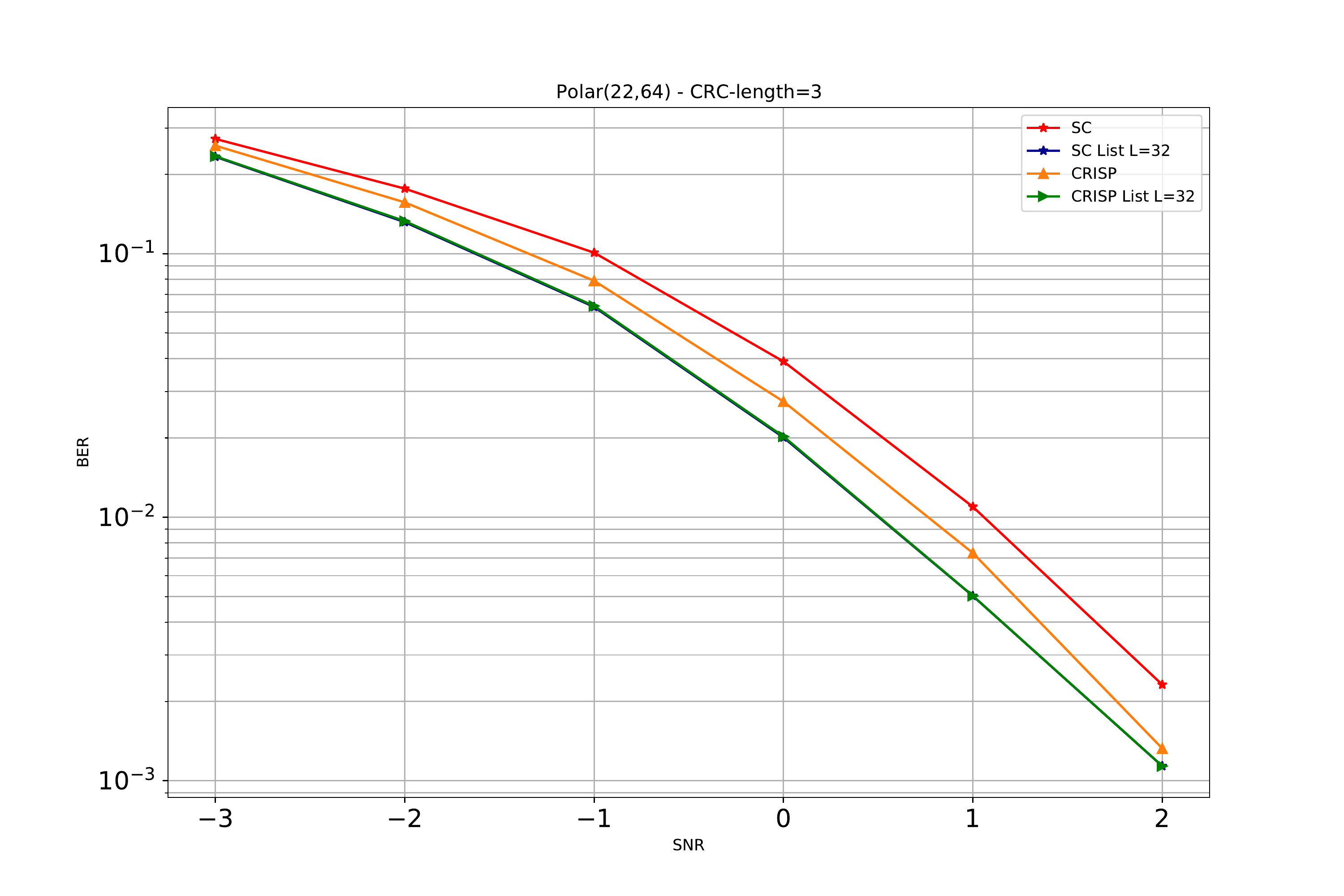}
  \label{fig:crc3}
}
\hfill
\subfigure[$\polar(32,16)$]{
  \centering
  \includegraphics[width=\columnwidth]{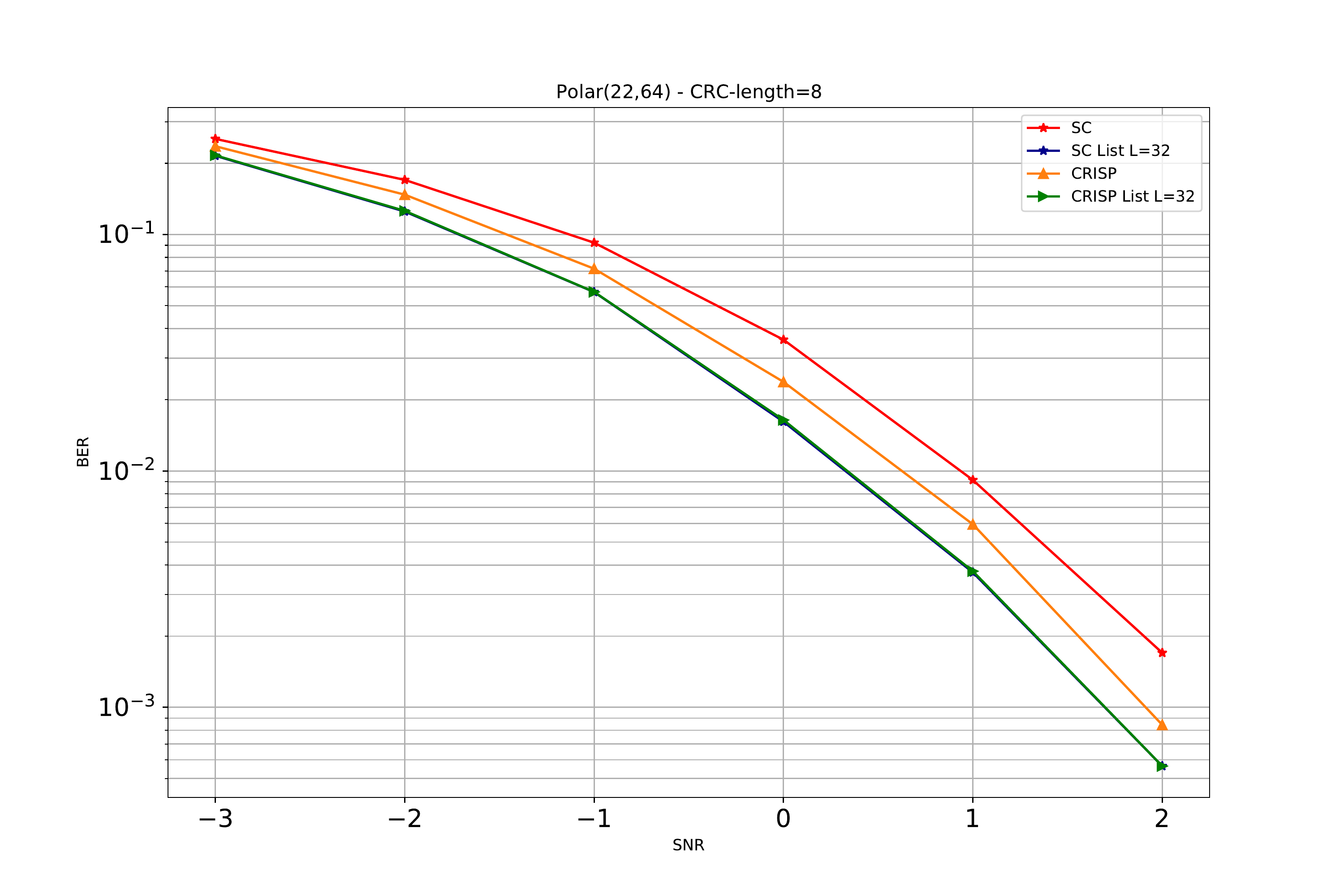}
  \label{fig:crc8}
}
\caption{CRISP performs well on CRC-Polar code}\label{fig:crc}
\end{figure*}

In practice, polar codes with successive cancellation list decoding is used in conjunction with a cyclic redundancy check (CRC) outer code. The message $u \in \{0, 1\}^{k_m}$ is encoded by a systematic cyclic code of rate $\frac{k_m}{k}$ to obtain a vector $m \in \{0, 1\}^{k}$. We obtain the codewords via the normal polar encoding procedure on $m$.
CRISP can be used to decode such CRC-Polar codes by considering $m$ as the input to the polar code block.
As shown in \prettyref{fig:crc}, CRISP achieves near-MAP reliability when CRCs of length $3$ and $8$ are used for a $\polar(64,22)$ code.

\subsubsection{Scaling to larger codes} 
\begin{figure*}[ht]

\subfigure[BERs Polar$(128,22)$]{
  \centering
  \includegraphics[width=\columnwidth]{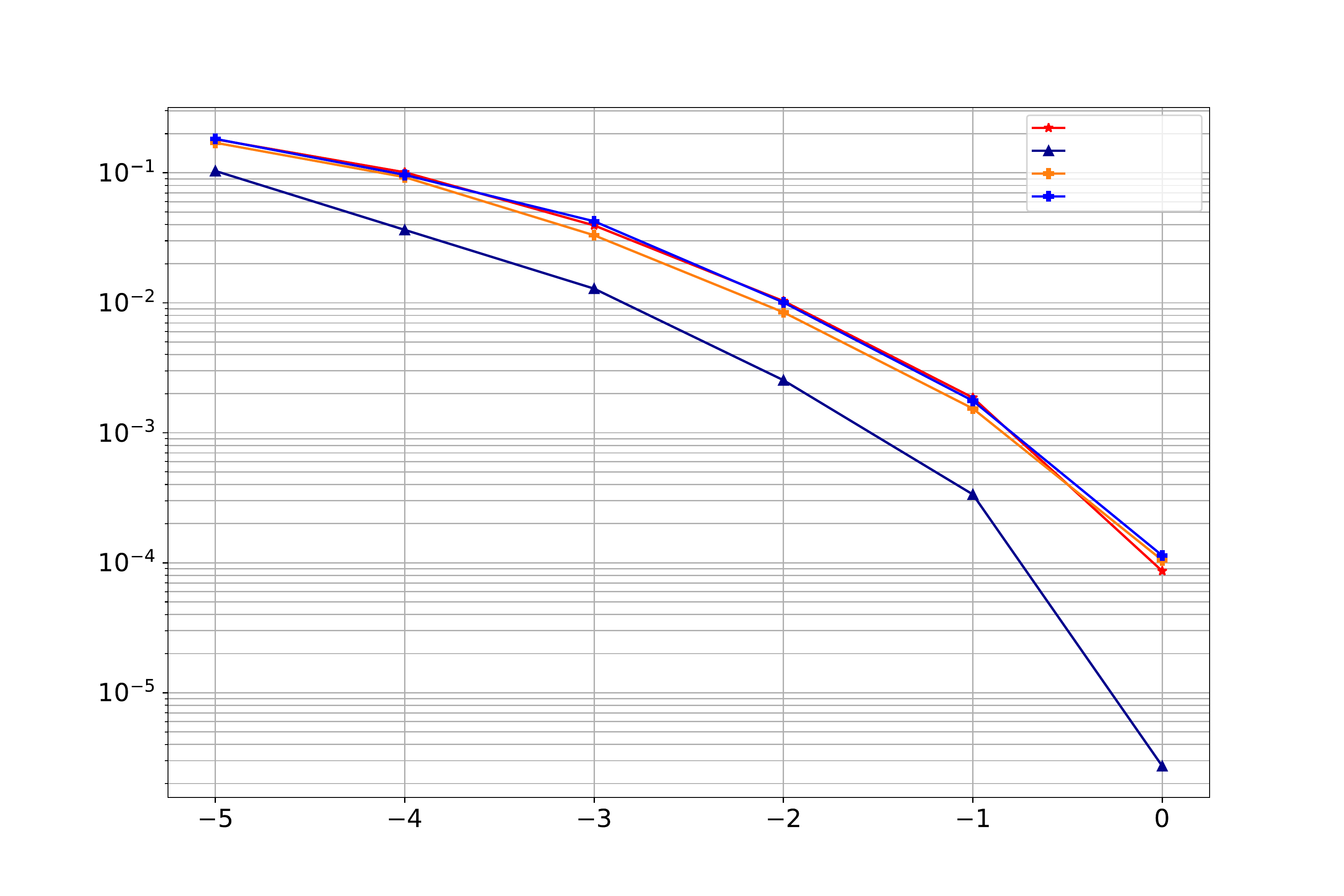}
    \put(-44,133){\fontsize{3}{5}\selectfont SC}
    \put(-44,129){\fontsize{3}{5}\selectfont SC-List, L=32}
    \put(-44,125){\fontsize{3}{5}\selectfont CRISP}
    \put(-44,121){\fontsize{3}{5}\selectfont No curriculum}
    \put(-170,2){\footnotesize Signal-to-noise ratio (SNR) [dB]}
    \put(-230,70){\rotatebox[origin=t]{90}{\footnotesize Bit Error Rate}}
  \label{fig:plot_bler_1281}
}
\hfill
\subfigure[BLERs Polar$(128,22)$]{
  \centering
  \includegraphics[width=\columnwidth]{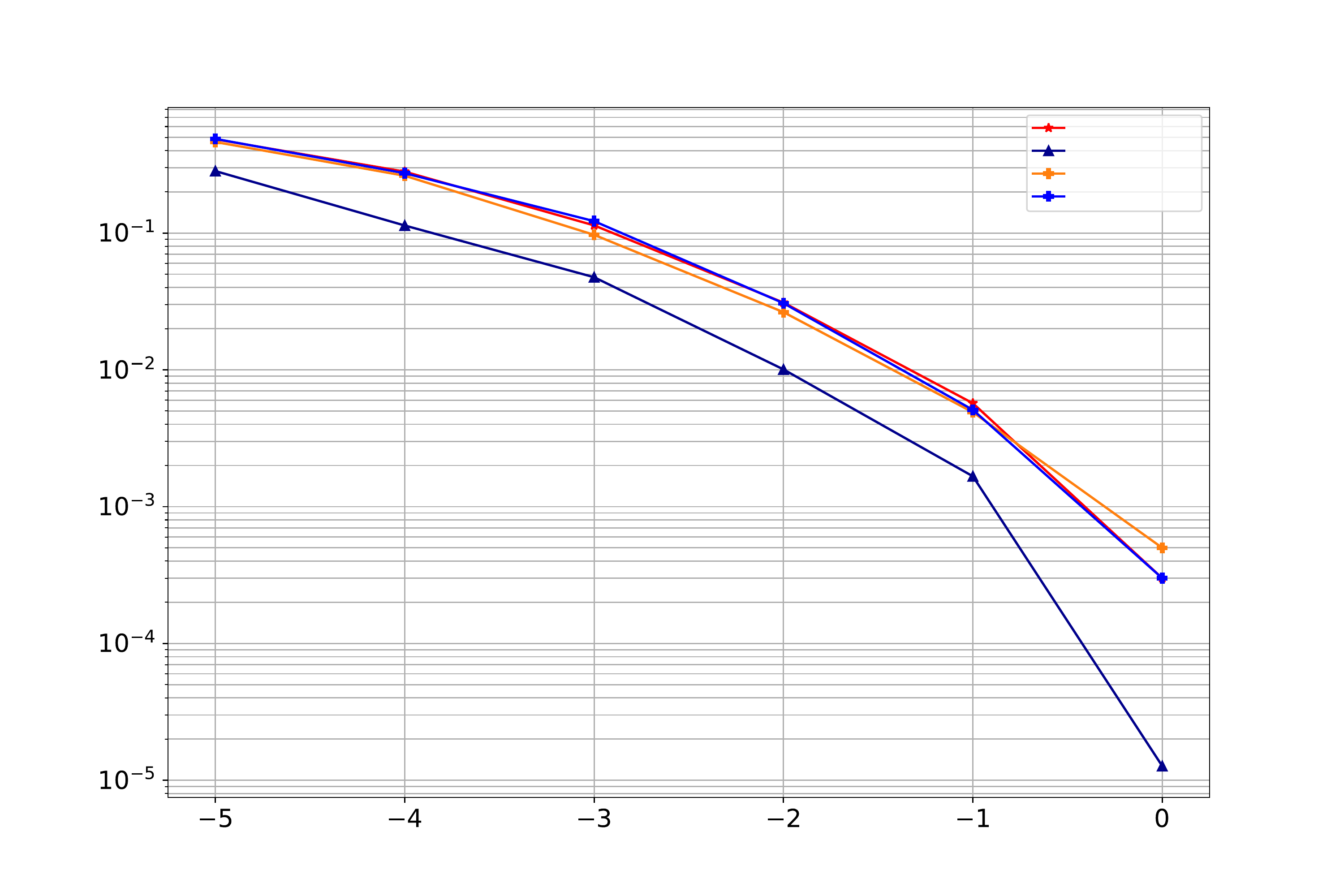}
    \put(-44,133){\fontsize{3}{5}\selectfont SC}
    \put(-44,129){\fontsize{3}{5}\selectfont SC-List, L=32}
    \put(-44,125){\fontsize{3}{5}\selectfont CRISP}
    \put(-44,121){\fontsize{3}{5}\selectfont No curriculum}
    \put(-170,2){\footnotesize Signal-to-noise ratio (SNR) [dB]}
    \put(-230,70){\rotatebox[origin=t]{90}{\footnotesize Block Error Rate}}
  \label{fig:plot_bler_1282}
}
\caption{CRISP matches SC reliabilty on $\polar(128,22)$.}\label{fig:polar128}
\end{figure*}

Curriculum training can be used to train even larger codes and obtain gains over naive training methods. However, we observed that our models were only able to achieve a reliability marginally better than SC. As shown in \prettyref{fig:polar128}, CRISP performs similar to SC decoding on the $\polar(128,22)$ code. We believe that it is possible to close the gap with MAP with more training tricks.

\subsubsection{Results with transformers} 
\begin{figure*}[ht]

\subfigure[GPT-based architecture (sequential decoding)]{
  \centering
 \includegraphics[width=\columnwidth]{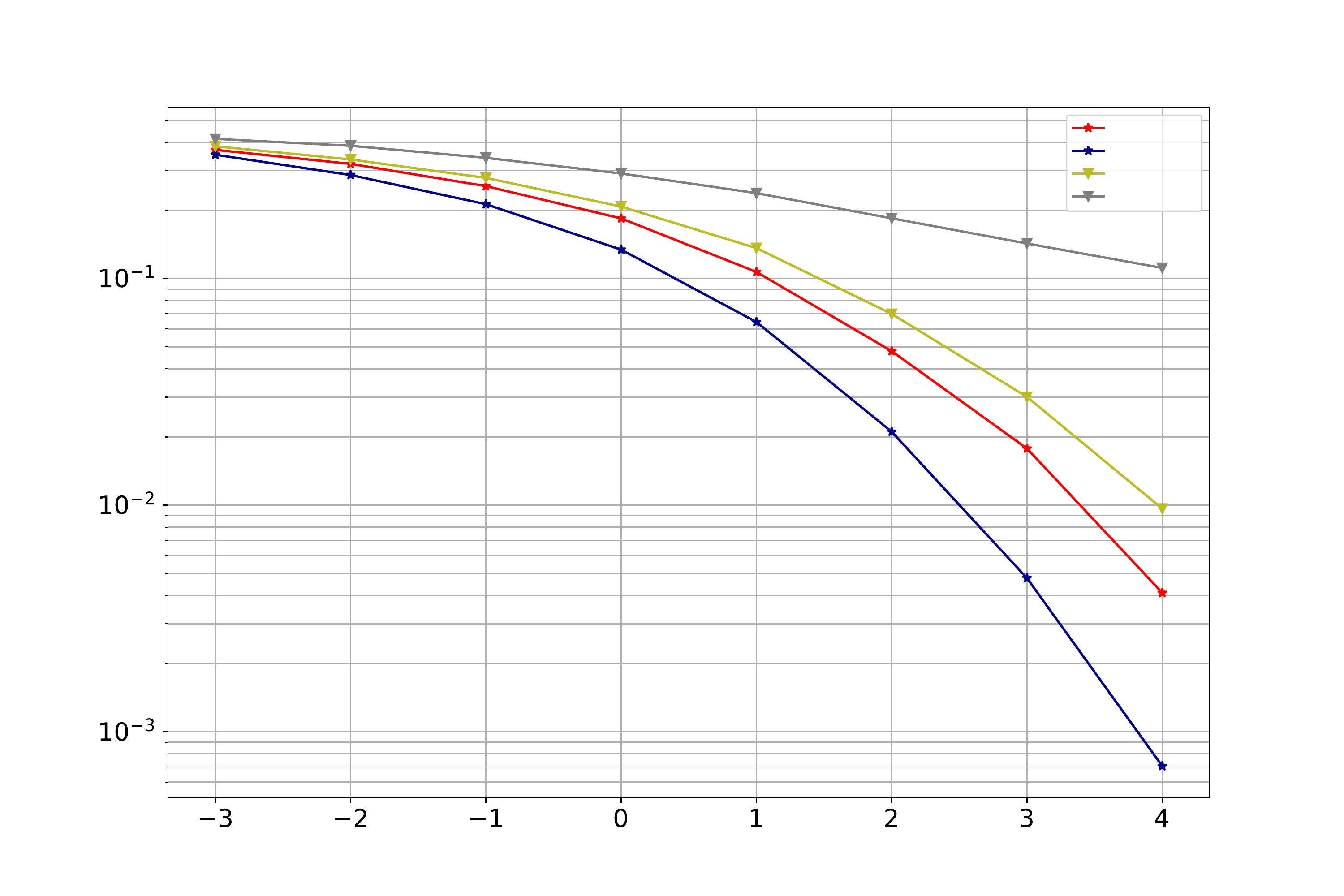}
 \put(-39,133){\fontsize{3}{5}\selectfont SC}
    \put(-39,129.5){\fontsize{3}{5}\selectfont MAP}
    \put(-39,126){\fontsize{3}{5}\selectfont GPT - L2R}
    \put(-39,122.5){\fontsize{3}{5}\selectfont GPT - R2L}
    \put(-170,2){\footnotesize Signal-to-noise ratio (SNR) [dB]}
    \put(-230,70){\rotatebox[origin=t]{90}{\footnotesize Bit Error Rate}}
  \label{fig:transformer_perf1}
}
\hfill
\subfigure[BERT-based architecture (block decoding)]{
  \centering
 \includegraphics[width=\columnwidth]{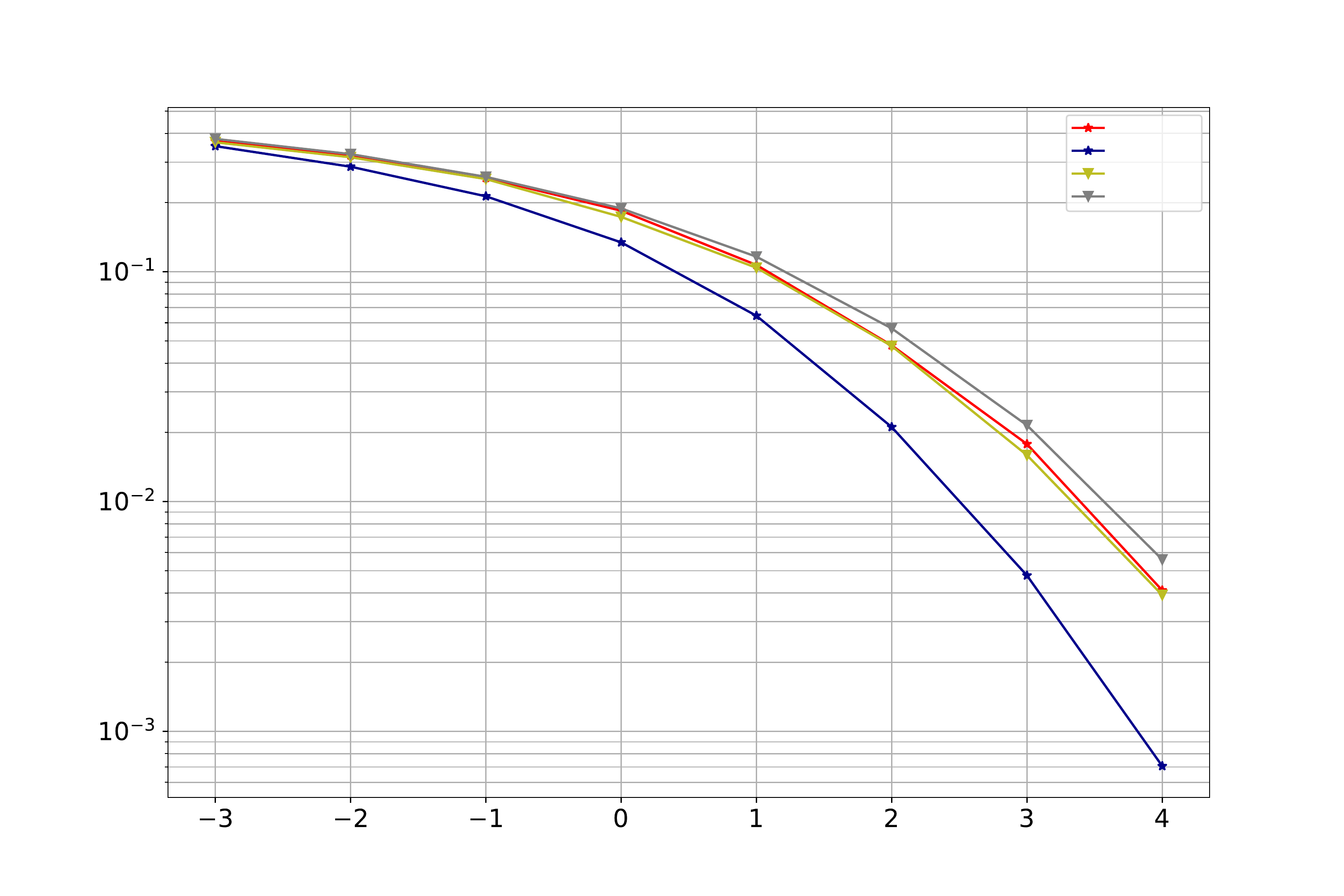}
 \put(-39,133){\fontsize{3}{5}\selectfont SC}
    \put(-39,129.5){\fontsize{3}{5}\selectfont MAP}
    \put(-39,126){\fontsize{3}{5}\selectfont BERT - L2R}
    \put(-39,122){\fontsize{3}{5}\selectfont BERT - R2L}
    \put(-170,2){\footnotesize Signal-to-noise ratio (SNR) [dB]}
    \put(-230,70){\rotatebox[origin=t]{90}{\footnotesize Bit Error Rate}}
  \label{fig:transformer_perf2}
}

\caption{Transformer performance on $\polar(32,16)$.}\label{fig:transformer_perf}
\end{figure*}

We also experimented with transformer-based architecures \cite{vaswani2017attention} for our decoder. In particular, we tried an autoregressive transformer-decoder network (similar to GPT \citep{brownGPT} that does sequential decoding) and the transformer-encoder network (similar to BERT \citep{devlin18} that does block decoding). Preliminary results indicate that these transformer-based models are less reliable compared to RNNs and CNNs (\prettyref{fig:transformer_perf}). In addition, these models take a greater number of iterations (\ref{app:exp_details}) to train on each of the subcodes than RNNs and CNNs during curriculum training. Transformer training is sensitive to architectural and hyperparameter choices and is computationally expensive. We believe that with the right training tricks, transformer-based models can be used to decode larger codes. This is ongoing work. 

\subsection{Additional results for PAC codes}
\label{app:additional_pac}

\begin{figure*}[ht]

\subfigure[BERs PAC$(32,16)$]{
  \centering
  \includegraphics[width=\columnwidth]{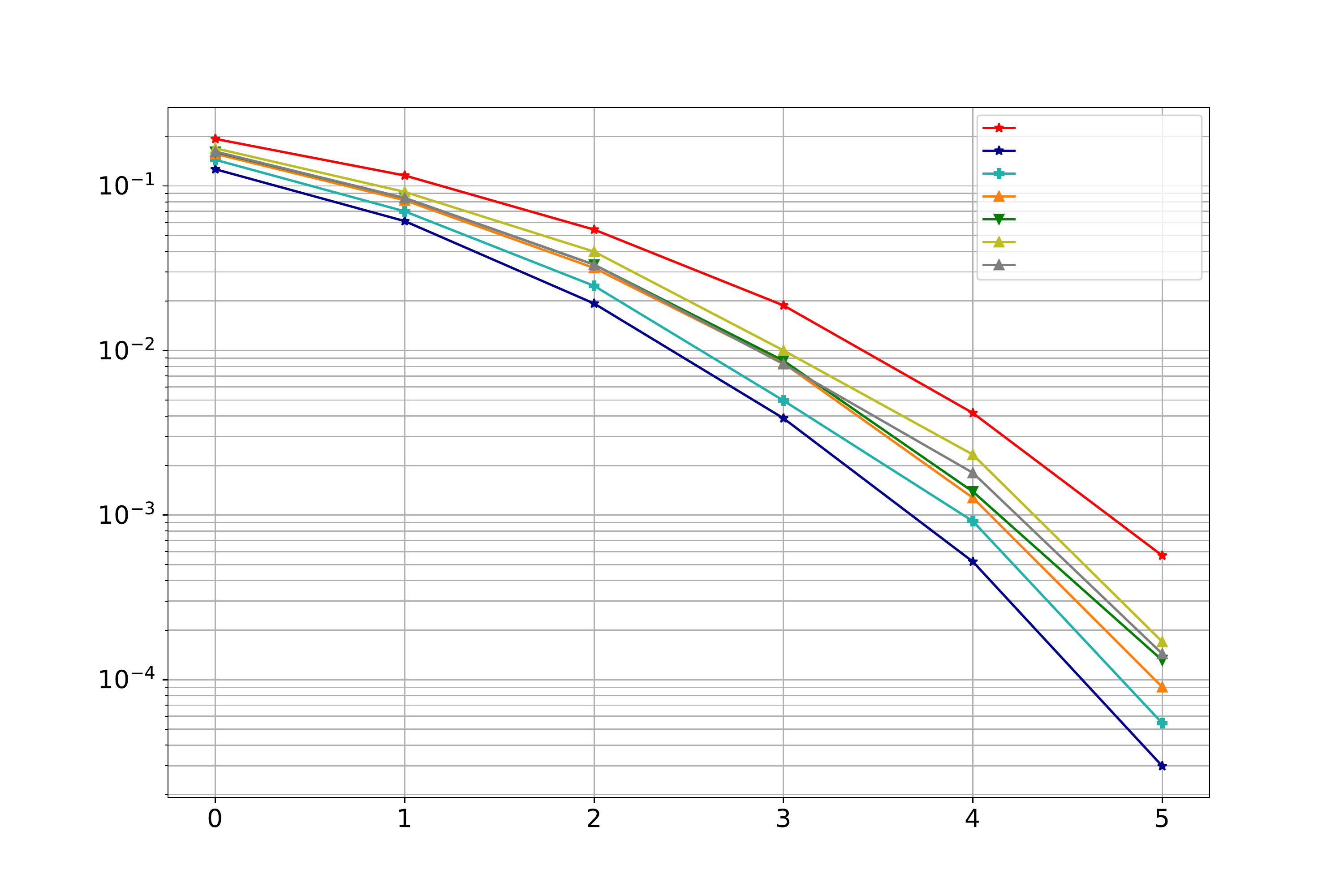}
    \put(-50,133){\fontsize{3}{5}\selectfont SC}
    \put(-50,129){\fontsize{3}{5}\selectfont SC-List, L=32}
    \put(-50,125){\fontsize{3}{5}\selectfont Fano}
    \put(-50,121){\fontsize{3}{5}\selectfont RNN - L2R}
    \put(-50,117){\fontsize{3}{5}\selectfont RNN - R2L}
    \put(-50,113){\fontsize{3}{5}\selectfont CNN - L2R}
    \put(-50,109.5){\fontsize{3}{5}\selectfont CNN - R2L}
    \put(-170,2){\footnotesize Signal-to-noise ratio (SNR) [dB]}
    \put(-230,70){\rotatebox[origin=t]{90}{\footnotesize Bit Error Rate}}
  \label{fig:plot_pac_ber1632}
}
\hfill
\subfigure[BLERs PAC$(32,16)$]{
  \centering
  \includegraphics[width=\columnwidth]{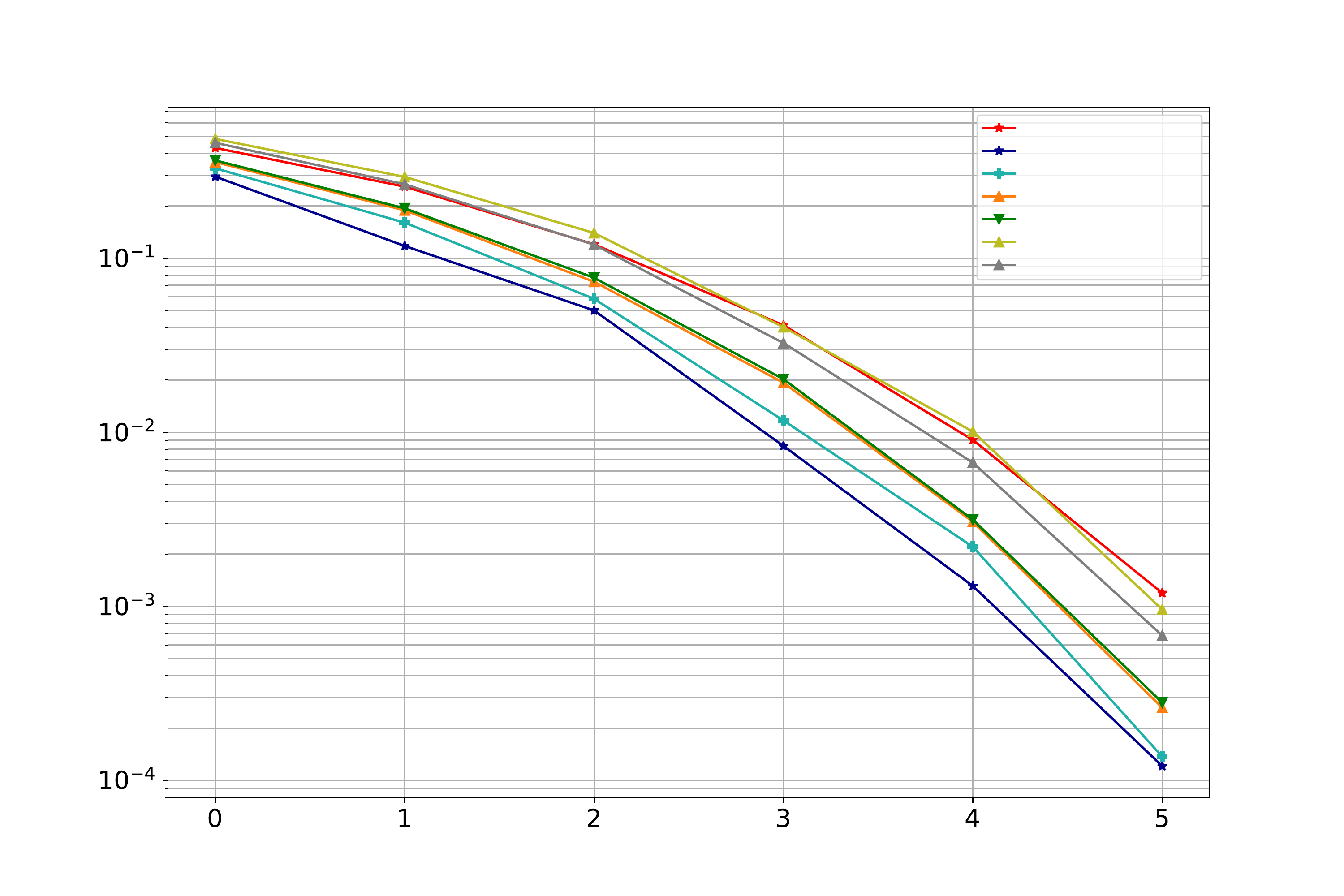}
    \put(-50,133){\fontsize{3}{5}\selectfont SC}
    \put(-50,129){\fontsize{3}{5}\selectfont SC-List, L=32}
    \put(-50,125){\fontsize{3}{5}\selectfont Fano}
    \put(-50,121){\fontsize{3}{5}\selectfont RNN - L2R}
    \put(-50,117){\fontsize{3}{5}\selectfont RNN - R2L}
    \put(-50,113.5){\fontsize{3}{5}\selectfont CNN - L2R}
    \put(-50,109.5){\fontsize{3}{5}\selectfont CNN - R2L}
    \put(-170,2){\footnotesize Signal-to-noise ratio (SNR) [dB]}
    \put(-230,70){\rotatebox[origin=t]{90}{\footnotesize Block Error Rate}}
  \label{fig:plot_pac_bler1632}
}
\caption{With correct choice of curriculum, CNNs match the BER performance of CRISP on PAC(32,16). However, they are sub-optimal in BLER.}\label{fig:bler_seq_block_pac}
\end{figure*}

CRISP maintains its good performance even in block error rate, as we show in \prettyref{fig:plot_pac_bler1632}. \prettyref{fig:plot_pac_ber1632} compares RNNs and CNNs in terms of BER for PAC$(32,16)$ code with L2R and R2L curricula. We observe that while both RNNs and CNNs outperform SC, RNNs achieve slightly better BER reliability than CNNs. On the other hand, \prettyref{fig:plot_pac_bler1632} highlights that CNNs achieves an SC-like BLER.

\section{Experimental details}
\label{app:exp_details}

We provide our code at the following \href{https://github.com/hebbarashwin/neural_polar_decoder}{link}.

{\bf Data generation.} Note that for any $\polar(n,k)$ or $\text{PAC}(n,k))$ code, the input message $m$ is chosen uniformly at random from $\binary^k$. We simulate this by drawing $k$ \iid Bernoulli random variables with probability $1/2$. We follow a similar procedure to generate a batch of message blocks (in $  \binary^{B \times k}$) with batch size $B$, both during training and inference. For the AWGN channel, the batch noise (in $ \reals^{B \times n}$) is accordingly generated by drawing \iid Gaussian samples from $\calN(0,\sigma^2)$.

{\bf Hyper-parameters.} For training our models (both sequential and block decoders), we use AdamW optimizer \citep{loshchilov2017decoupled} with a learning rate of $10^{-3}$. At each curriculum step, corresponding to training a subcode, we choose the SNR corresponding to which the optimal decoder for that subcode has BER in the range of $10^{-2} \sim 10^{-1}$ \citep{kim2018communication}. This ensures that a significant portion of training examples lie close to the decision boundary. It is well known that using a large batch size is essential to train a reliable decoder \citep{jiang2019deepturbo}; we use a batch size of $4096$ or $8192$. 

\subsection{Sequential decoders}

We present the architectures and training details for our sequential decoders. We consider two popular choices for our sequential models: RNNs and GPT. We also note that it is a standard practice to use \emph{teacher forcing} to train sequential models \citep{teacherForcing}: during training, as opposed to feeding the model prediction $\mhat_i$ as an input for the next time step, the ground truth message bit $m_i$ is provided as an input to the model instead (\prettyref{fig:crisp_arch}). \emph{Student forcing} refers to using the same $\mhat_i$ as an input.   

\subsubsection{RNNs}

\begin{figure*}[ht]

\subfigure[]{
  \centering
  \includegraphics[width=\columnwidth]{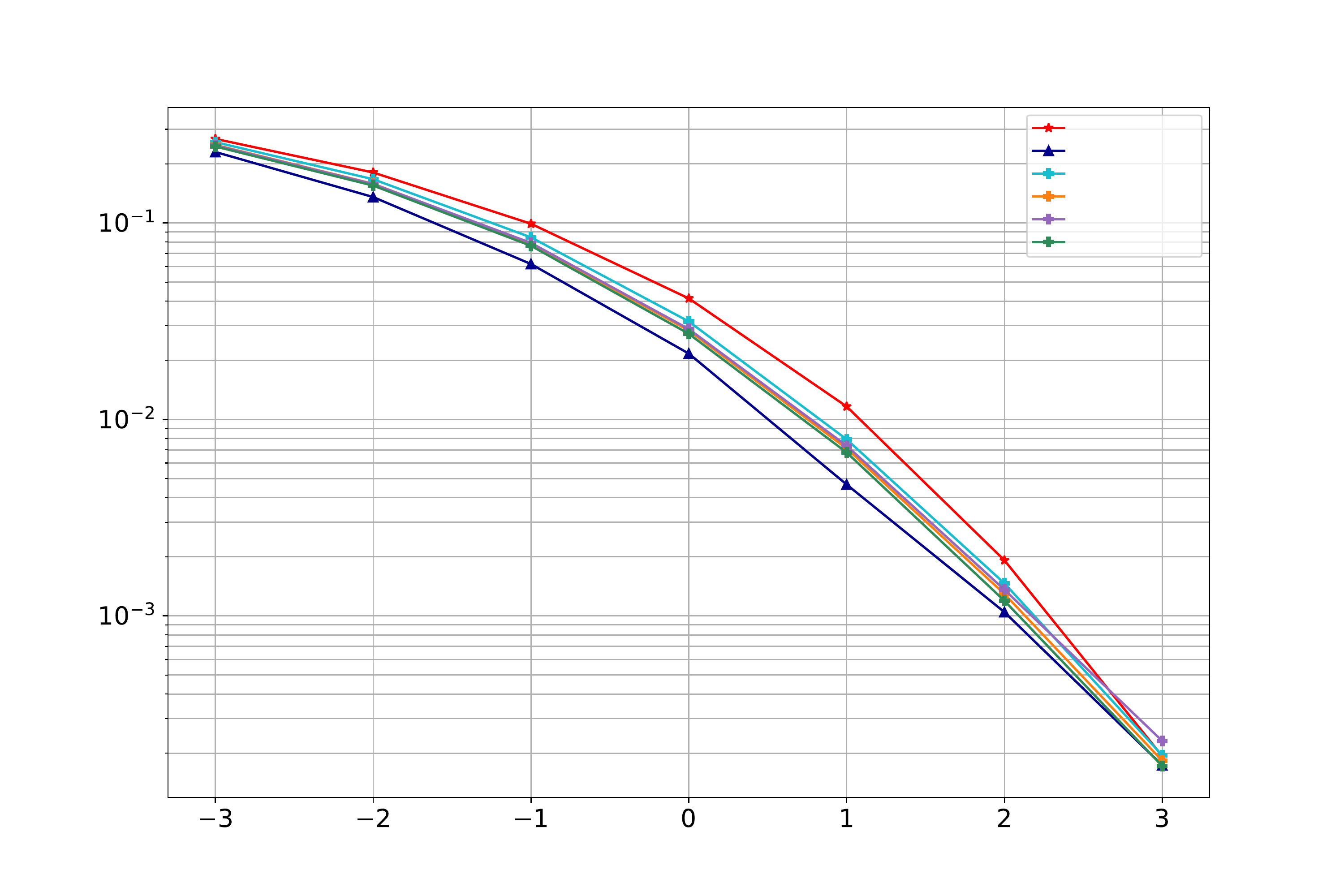}
    \put(-46,133){\fontsize{3}{5}\selectfont SC}
    \put(-46,129){\fontsize{3}{5}\selectfont SC-List, L=32}
    \put(-46,125){\fontsize{3}{5}\selectfont GRU -  depth=1}
    \put(-46,121){\fontsize{3}{5}\selectfont GRU -  depth=2}
    \put(-46,117){\fontsize{3}{5}\selectfont GRU -  depth=3}
    \put(-46,113.5){\fontsize{3}{5}\selectfont LSTM - depth=2}
    \put(-170,2){\footnotesize Signal-to-noise ratio (SNR) [dB]}
    \put(-230,70){\rotatebox[origin=t]{90}{\footnotesize Bit Error Rate}}
  \label{fig:fig:ber_polar_othermodels}
}
\hfill
\subfigure[]{
  \centering
  \includegraphics[width=\columnwidth]{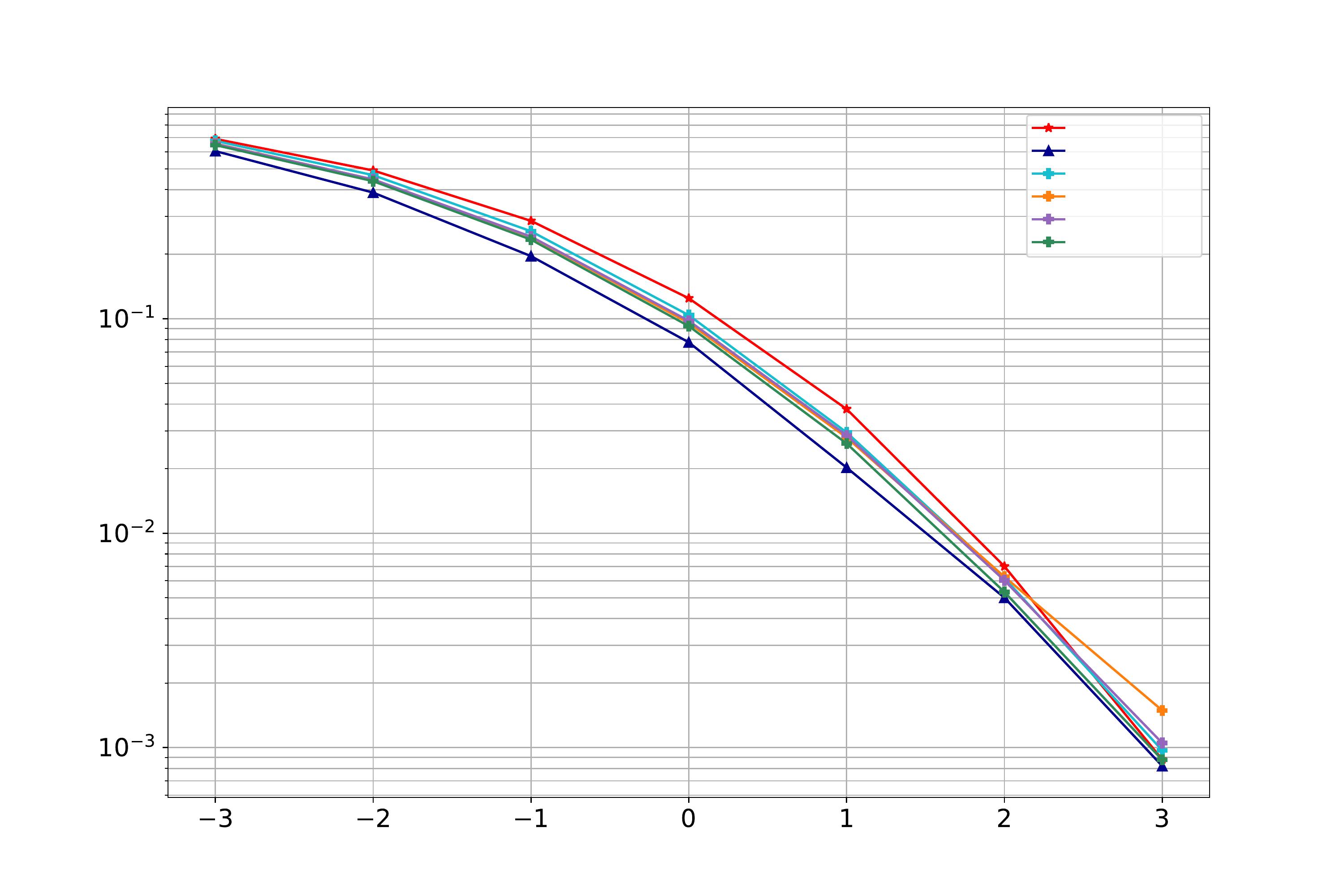}
    \put(-44,133){\fontsize{3}{5}\selectfont SC}
    \put(-44,129){\fontsize{3}{5}\selectfont SC-List, L=32}
    \put(-44,125){\fontsize{3}{5}\selectfont GRU - depth=1}
    \put(-44,121){\fontsize{3}{5}\selectfont GRU - depth=2}
    \put(-44,117){\fontsize{3}{5}\selectfont GRU - depth=3}
    \put(-44,113.5){\fontsize{3}{5}\selectfont LSTM - depth=2}
    \put(-170,2){\footnotesize Signal-to-noise ratio (SNR) [dB]}
    \put(-230,70){\rotatebox[origin=t]{90}{\footnotesize Block Error Rate}}
  \label{fig:plot_bler2264_polar}
}
\caption{Polar$(64,22)$: LSTMs and GRUs achieve similar reliability.}\label{fig:bler_polar_othermodels}
\end{figure*}

{\bf Architecture.}
We use a $2$-layer GRU with a hidden state size of $512$. The output at each timestep is obtained through a fully connected layer (as shown in \prettyref{fig:crisp_arch}). The network has $2.5$M and $600$K parameters for block lengths 64 and 32. As shown in Figure \ref{fig:bler_polar_othermodels}, $2$-layer-LSTM and $3$-layer-GRU models achieve similar performance. We choose a $2$-layer GRU for our experiments since it allows for faster training and has fewer parameters.

\begin{figure*}[ht]

\subfigure[]{
  \centering
  \includegraphics[width=\columnwidth]{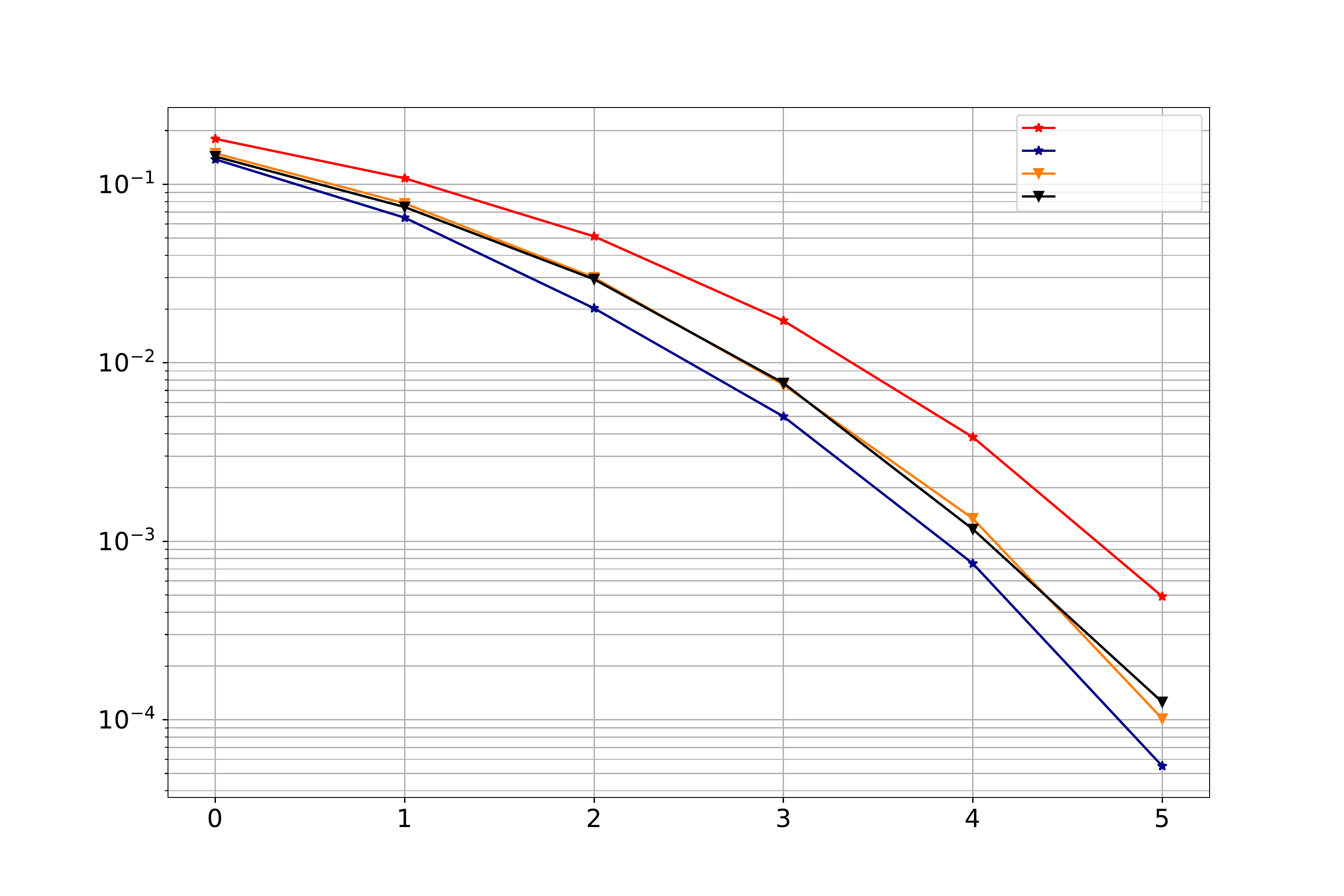}
    \put(-46,133){\fontsize{3}{5}\selectfont SC}
    \put(-46,129){\fontsize{3}{5}\selectfont SC-List, L=32}
    \put(-46,125){\fontsize{3}{5}\selectfont Teacher-forcing}
    \put(-46,121){\fontsize{3}{5}\selectfont Student-forcing}
    \put(-170,0){\footnotesize Signal-to-noise ratio (SNR) [dB]}
    \put(-230,70){\rotatebox[origin=t]{90}{\footnotesize Bit Error Rate}}
  \label{fig:plot_ber1632_st}
}
\hfill
\subfigure[]{
  \centering
  \includegraphics[width=\columnwidth]{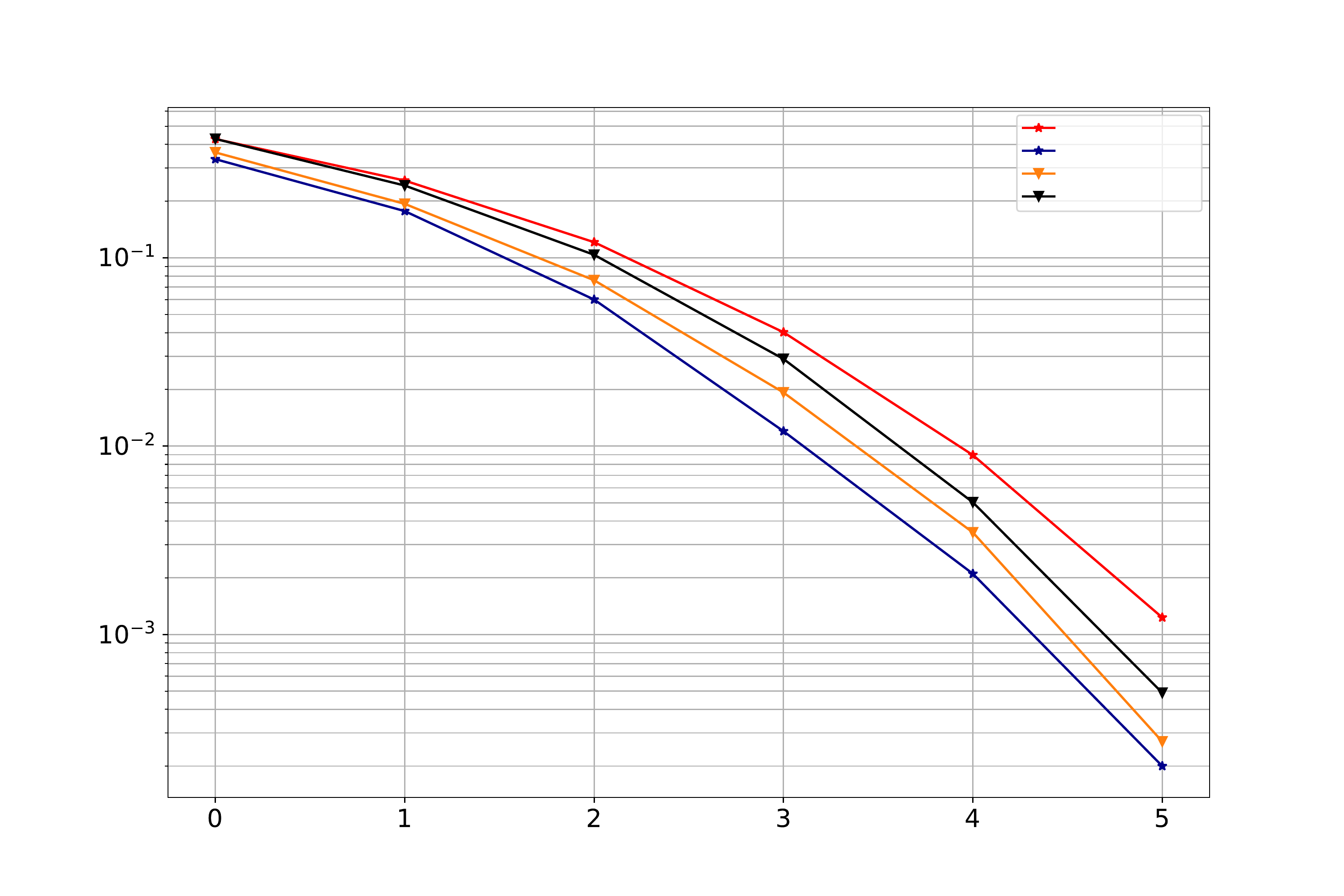}
    \put(-46,133){\fontsize{3}{5}\selectfont SC}
    \put(-46,129){\fontsize{3}{5}\selectfont SC-List, L=32}
    \put(-46,125){\fontsize{3}{5}\selectfont Teacher-forcing}
    \put(-46,121){\fontsize{3}{5}\selectfont Student-forcing}
    \put(-170,0){\footnotesize Signal-to-noise ratio (SNR) [dB]}
    \put(-230,70){\rotatebox[origin=t]{90}{\footnotesize Block Error Rate}}
  \label{fig:plot_bler1632_st}
}

\caption{Training CRISP using student forcing results in sub-optimal BLER.}\label{fig:bler_st}
\end{figure*}

{\bf Training.} 
We use the teacher forcing mechanism to train our models. We found that teacher forcing gives a better final performance in terms of both BER and BLER, whereas student forcing only provides gains in the BER reliability (\prettyref{fig:bler_st}). We observed that student forced training achieved sub-optimal performance for larger block lengths.
Empirically we observed that the number of iterations spent on training each intermediate subcode of the curriculum is not critical to the performance of the final model (\prettyref{fig:schedule}). To train CRISP for Polar(64,22), we use the following curriculum schedule: Train each subcode for $2000$ iterations, and finally train the full code until convergence with a decaying learning rate. This training schedule required 13-15 hours of training on a GTX 1080Ti GPU.

\subsubsection{GPT}

{\bf Architecture.}
The model consists of $6$ transformer blocks with masked self-attention and GELU activation. The multiheaded attention unit has $8$ heads in each block, and an embedding/hidden size of $64$ is used throughout the network. The output vectors of the final transformer block are passed through a linear layer to estimate each bit sequentially. The model has 350K parameters for blocklength 32.

{\bf Training.} 
For training the GPT-based transformer, we use a teacher forcing mechanism. Here, we observed that the decoder takes a greater number of iterations ($40,000$) to train on each of the subcodes than RNNs and CNNs ($2,000-10,000$) during curriculum training of $\polar(32,16)$. For a fixed batch size, GPT also takes significantly longer to train ($12$ hours) compared to CNNs ($3$ hours) and RNNs ($4$ hours) on GTX 1080 Ti GPU.

\subsection{Block decoders}

\subsubsection{CNNs}

{\bf Architecture.} 
For block decoding using Convolutional Neural Networks (CNNs), we use a ResNet-like architecture \citep{resnets}, with the primary difference being the use of 1D convolutions instead of 2D. The model has $10$ 1D-convolutional layers with residual connections skipping every two consecutive layers. Each convolutional layer has $64$ channels, which are flattened at the penultimate layer and fed as an input to a fully-connected neural network with one hidden layer. We use the GELU \citep{gelu} activation function throughout the network. The model has 2.5M parameters for blocklength 64.

{\bf Training.}
We train the CNN model for $5,000$ iterations for each intermediate subcode of the curriculum. In the last step of the curriculum, we train it for $100,000$ iterations with a decaying cosine annealing schedule for the learning rate \citep{cosineAnneal}.

\subsubsection{BERT}

{\bf Architecture.}
The model consists of $6$ transformer blocks with unmasked self-attention and GELU activation. In each block, the multiheaded attention unit has $8$ heads, and an embedding/hidden size of $64$ is used throughout the network. The output vectors of the final transformer block are passed through a linear layer to estimate all the bits in one shot. The model has 350K parameters for blocklength 32.

{\bf Training.}
 We train this model on each intermediate subcode for around $10,000-20,000$ steps. Thus the BERT-based decoder achieves better reliability than its GPT counterpart despite fewer training iterations (\prettyref{fig:transformer_perf}).

\section{Reliability-complexity comparison}
\label{app:complexity}
Two important metrics in evaluating a decoding algorithm are the decoding reliability and complexity. In this paper, we focus on optimizing the BER performance; the main goal of our paper is to design a curriculum based decoder for Polar and PAC codes that can achieve near-optimal reliability performance as opposed to the current data-driven approaches that only match the SC. In \prettyref{sec:complexity_compar}, we demonstrated that CRISP achieves excellent inference throughput on GPUs. We also see that the decoding complexity of CRISP can be further improved with a hardware-aware neural architecture. 

We believe that neural decoders, coupled with the recent advances in distillation \cite{sanh19} and pruning of neural networks \cite{hinton15distil,Wang_2020,anwar15} far larger than ours (E.g., 110M for BERT vs. 2.5M for CRISP), can achieve even better runtimes. For instance, TinyBERT (\cite{jiao19}) uses knowledge distillation to learn a model 9.4x faster on inference compared to the parent BERT.
Coupled with efficient GPU implementations, which are optimized for vector-matrix multiplications, and the aforementioned compression techniques, we believe neural decoders offer a great potential for fast and reliable channel decoding. 
{It is important to note that inference throughput is hardware and software dependant. In \prettyref{tab:complexity}, we report throughput numbers of the optimized C++ multithreaded implementation of SC/SCL decoding on CPU using the aff3ct toolbox (\cite{Cassagne2019a}). There has been progress in developing GPU implementations of SCL (\cite{cammerer2017combining, han2017successive}). Since we could not find publicly available implementations of these works, we report throughput numbers of our implementation.}

\end{document}